\documentclass[twocolumn]{aastex7}

\usepackage{amsmath}
\begin{document}
%%\begin{CJK*}{UTF8}{gbsn} 
	
%\title{ARE CL-AGN change their look temporary?}
\title{22 Newly Identified Repeating Changing-look AGNs and Evidence for Extreme Broad-line Region Breathing}

\correspondingauthor{Litao Zhu, Zhongxiang Wang}

\author[orcid=0009-0005-9652-8946]{Litao Zhu}
\affiliation{Department of Astronomy, School of Physics and Astronomy, Key Laboratory of Astroparticle Physics of Yunnan Province, Yunnan, China}
\email{litaozhu6@gmail.com}

\author[orcid=0000-0003-1984-3852]{Zhongxiang Wang}
\affiliation{Department of Astronomy, School of Physics and Astronomy, Key Laboratory of Astroparticle Physics of Yunnan Province, Yunnan, China}
%\affiliation{Shanghai Astronomical Observatory, Chinese Academy of Sciences, 80 Nandan Road, Shanghai 200030, China}
\email{wangzx20@ynu.edu.cn}

\author[0000-0002-9331-4388]{Alok C. Gupta}
\affiliation{Aryabhatta Research Institute of Observational Sciences (ARIES), Manora Peak, Nainital-263001, India}
\email{acgupta30@gmail.com}

\author[0009-0001-0943-8602]{Yiyang Zeng}
\affiliation{Department of Astronomy, School of Physics and Astronomy, Key Laboratory of Astroparticle Physics of Yunnan Province, Yunnan, China}
\email{zengyiyang@stu.ynu.edu.cn}

\author[0000-0003-3337-4861]{P. U. Devanand}
\affiliation{Aryabhatta Research Institute of Observational Sciences (ARIES), Manora Peak, Nainital-263001, India}
\email{devanandullas@gmail.com}
	
\author[0009-0007-3214-602X]{Karan Dogra}
\affiliation{Aryabhatta Research Institute of Observational Sciences (ARIES), Manora Peak, Nainital-263001, India}
\email{karandogra987@gmail.com}
	
\author[0000-0002-7658-0350]{Riya Bhowmick}
\affiliation{Aryabhatta Research Institute of Observational Sciences (ARIES), Manora Peak, Nainital-263001, India}
\email{riyabhowmickmalda@gmail.com}	

\author[]{Ju-Jia Zhang}
\affiliation{Yunnan Observatories, Chinese Academy of Sciences, Kunming 650216, China}
\affiliation{Key Laboratory for the Structure and Evolution of Celestial Objects, Chinese Academy of Sciences, Kunming 650216, China}
\email{jujia@ynao.ac.cn}

\begin{abstract}
	Changing-look active galactic nuclei (CL AGNs) show the appearance or 
	disappearance of broad emission lines on timescales of years. 
	Among them, the repeating CL (RCL) AGNs may provide a clear clue for
	our understanding of the CL transitions because the same nucleus
	crosses some physical boundaries more than once. 
We search for RCL AGNs in known CL-AGN samples using long-term multi-band
	light curves, and selected 34 candidates for spectroscopic
follow-up. We confirm 25 RCL AGNs, including 22 newly identified cases. 
	Properties of these RCL AGNs are analyzed.	
The observed rest-frame intervals of the second transitions are mostly
3--4 yr, while the variations of the optical light curves suggest that some 
transitions might occur on timescales of several months.  The latest spectra 
show that the on/off states correspond to higher/lower Eddington-ratios,
	in the expected direction relative to the parent CL-AGN samples. 
	As Seven RCL AGNs are well covered by nearly continuous single-band 
	light curves, their on/off states can be found to follow multi-year 
	optical excursions, and their Eddington ratios vary consistently with 
	the photometric changes.  We also find that the H$\beta$-only 
	transitions occur at higher Eddington ratios than the transitions 
	involving both H$\alpha$ and H$\beta$, suggesting a line-dependent 
	Broad-Line-Region (BLR) visibility threshold.  These results support a 
	picture in which different accretion-flow processes drive 
	reversible changes in the central ionizing emissions, while the 
	observed RCL transitions are produced by the BLR breathing across 
	line-dependent visibility thresholds.
\end{abstract}
	

\section{introduction} 
The optical spectral type of an Active Galactic Nucleus (AGN) is not always 
a fixed property. Although the classical unification model for AGNs explains 
much of their Type~1/Type~2 difference as orientation-dependent 
obscuration due to the dusty torus
\citep[e.g.,][]{law87, ant93, up95, tad08}, multi-epoch spectroscopy has 
shown that some AGNs can change their broad-line appearance within only a few 
years.  These changing-look (CL) AGNs show the emergence or disappearance 
of broad Balmer lines, which are often accompanied by large continuum 
variations \citep[e.g.,][]{to76, crp+86, sbw93, ajk+99, eh01, ddc+14, spg+14, lcm+15}.
Early individual discoveries and small spectroscopic searches have indicated
the CL phenomenon
\citep[e.g.,][]{rac+16,rcr+16,ghc+17}, and now large time-domain photometric 
and
spectroscopic surveys have expanded CL AGNs into statistical samples
\citep[e.g.,][]{gzf+24,gzg+25,dzg+25}.
The short-timescale CL phenomenon is obviously not described in the unification
model and points instead to rapid changes in the
central radiation field, variable obscuration, or both
\citep[e.g.,][]{eli12, nd18, rae+19,lwl+19,ady+20,grd+25}. In this context,
some off-state CL AGNs may be related to true or naked Type~2 AGNs, where the
broad lines are intrinsically weak or absent rather than hidden
\citep[e.g.,][]{ygm+23,lyp+14,jrt+25}.

Most CL-AGN studies have focused on identifying a CL transition from two 
or a few
spectroscopic epochs, through either direct spectral comparisons
\citep[e.g.,][]{ywf+18,gzf+24,gzg+25,dzg+25,gpa+22,zte+24,cjg+25,lkc+26},
or variability-selected follow-up observations
\citep[e.g.,][]{mrl+16,fgg+19,grs+20,swj+20, lmb+22, wzb+23,lsa+23,wwg+24,zlw+24,zwd+25,zhu+26}. 
Such studies have established the diversity of CL transitions, but a single
on/off event cannot determine whether the broad-line-region (BLR) gas is 
physically removed and
rebuilt or whether the broad-line emission falls below the visibility
threshold as the accretion power decreases \citep{eh09,eht14}. Repeating
CL (RCL) AGNs provide a test probe, because the same nucleus needs to
cross the same physical condition more than once, producing on--off--on or 
off--on--off
sequences \citep[e.g.,][]{wzx+22,lwp+25,kgw+26,lnn+25}. The repetition also
requires a reversible and repeatable driver around the same accreting
super-massive black hole (SMBH). Thus,
compared with ordinary CL events, RCL AGNs can limit the cause of the CL to 
changes in
the central engine and the BLR response, rather than changes among different
systems.

Recent multi-object searches have further shown that RCL
transitions are not isolated cases and may occur on year-long timescales
\citep{wwg+25,dzg+26,gcw+26,cyy+26}. However, most current RCL-AGN studies
still focus on identification, classification, and transition timescales,
while the connection between the spectral state and the contemporaneous
continuum evolution remains less explored. This connection is essential,
because ordinary CL AGNs are often selected through large optical variability
or found to show strong optical and mid-infrared (MIR) flux changes associated 
with the transition
\citep[e.g.,][]{mrl+16,fgg+19,grs+20,swj+20,wzb+23,rsm+18,rwc+22}. If the MIR
emissions follow the optical changes, the transition is more likely driven by
an intrinsic change in the central radiation field (as in the case of naked
Type 2 AGNs) rather than by simple line-of-sight 
obscuration \citep{swj+17,ygm+23}. 
In this case, different responses of H$\beta$ and
H$\alpha$ can be used to test whether the apparent type change reflects a
line-dependent BLR visibility threshold, as expected from BLR breathing and
line responsivity \citep[e.g.,][]{kg04,ch06,bdg+13,gk14}. 

In this work, we have searched for new RCL transitions from known CL-AGN 
samples and identified 25 RCL-AGNs.  The multi-epoch spectra and long-term 
multi-band light curves of the AGNs are used to examine
whether the repeated broad-line changes are driven by reversible continuum
variations and BLR breathing. In Section~\ref{sec:sample}, we describe 
the CL-AGN samples we used for searching for RCL AGNs, the candidate selection,
and the spectroscopic identification. The properties of 25 identified RCL
AGNs are presented in Seciont~\ref{sec:prop}. In Section~\ref{sec:dis},
we discuss our identification and the implications for our understanding of
the CL phenomenon.

\section{Sample and Identification}
\label{sec:sample}

\subsection{Parent CL-AGN samples and light-curve construction}

We started from the spectroscopically selected CL-AGN samples reported in
\citet{gzf+24, gzg+25} and \citet{dzg+25}. The former two were built 
by comparing
spectra from the Sloan Digital Sky Survey (SDSS; \citealt{aaa+09}) and the
Dark Energy Spectroscopic Instrument (DESI; \citealt{desi16}), while the last
one
was based on comparisons between spectra from the SDSS and the Large Sky Area 
Multi-Object
Fiber Spectroscopic Telescope (LAMOST; \citealt{lamost12}). These samples
provide known CL transitions and spectral states at
different epochs, and were used as parent ones for searching for
RCL AGNs.

\begin{figure}[htbp]
	\centering
	%\includegraphics[width=0.49\textwidth]{on_delta_g_kr.eps}
	%\includegraphics[width=0.49\textwidth]{off_delta_g_kr.eps}
	%\caption{Schematic view of our photometric selection. The left panel shows the candidates selected from known turn-on CL-AGNs for possible repeating turn-off transitions, while the right panel shows the candidates selected from known turn-off CL-AGNs for possible repeating turn-on transitions. The background Type~1 and Type~2 AGNs are SDSS DR16 sources with S/N$>10$.}
	\includegraphics[width=0.49\textwidth]{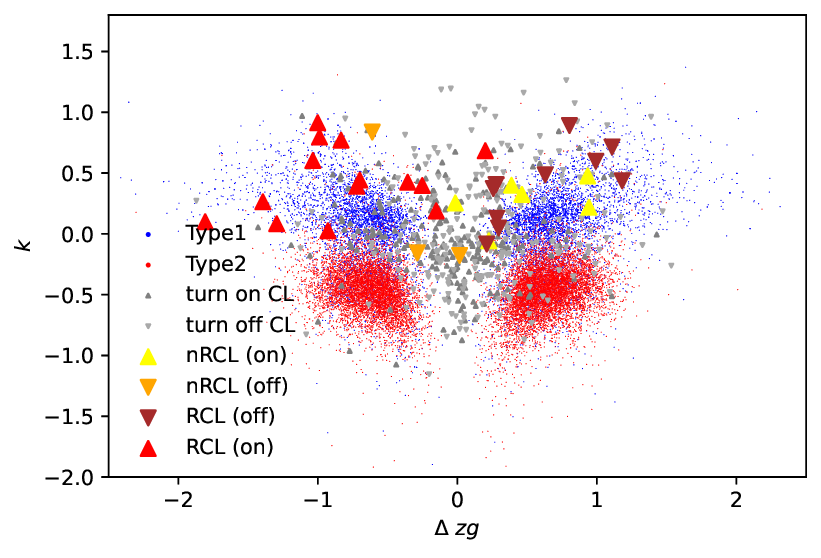}
	\caption{Distribution of the color--magnitude slope $k$ versus 
	the $zg$-band magnitude change $\Delta zg$ for the CL parent samples
	(gray triangles) and our selected 
		candidates (consisting of the confirmed RCL, marked with red or
		indian-red, and the non-confirmed RCL (nRCL), 
marked with yellow and orange). 
		The up- and down-pointing triangles denotes the on and off
		spectral state, respectively. The background Type~1 
		(blue dots) and 
		Type~2 AGNs (red dots) are SDSS DR16 sources with S/N$>10$.
		Among the candidates we selected, J1330 is not shown because 
		too few $zg$ measurements were available 
		(Figure~\ref{fig:crcl}).}
	%	, and J1518 is omitted as well because its current spectral state is uncertain (Section~\ref{sec:un}).
	\label{fig:select}
\end{figure}

For each source in the samples, we constructed long-term optical and MIR
light curves. The optical data were taken from the Zwicky Transient Facility
(ZTF; \citealt{bkg+19}), the Asteroid Terrestrial-impact Last Alert System
(ATLAS; \citealt{tdh+18}), and the Catalina Real-time Transient Survey
(CRTS; \citealt{ddm+09}). For the ZTF data, we used the $zg$- and $zr$-band
measurements after requiring \texttt{catflags}=0 and $\chi<4$, and then binned
the data in 3-day intervals. For ATLAS, we used the 1-day binned
stacked forced-photometry light curves in the $c$ and $o$ bands 
($ac$ and $ao$, respectively), applied basic
quality cuts to remove unreliable measurements, and converted the fluxes to
AB magnitudes. The MIR light curves were from WISE (\citealt{wem+10}) and
NEOWISE (\citealt{mbg+11,neowise}) at the W1 (3.4\,$\mu$m) and W2
(4.6\,$\mu$m) bands. The WISE data were binned by observing epochs. These
multi-band light curves allowed us to check whether the continuum emission had
changed significantly after the CL transition reported in the parent samples.

\subsection{Selection of RCL-AGN candidates}

The aim of our selection was to find sources that might have changed their
spectral type again after the known CL transition. We therefore inspected the
light curves of the CL AGNs in the parent samples and selected candidates by
visual examination, guided by their variability amplitude and
color--magnitude behaviour rather than by a strict criterion in parameter
space (such as those in \citealt{zwd+25}). The basic idea for the selection 
is illustrated in Figure~\ref{fig:select}.
The color--magnitude slope, $k$, was measured from a linear fit to
$zg-zr$ versus $zr$, where the magnitudes of the two bands were taken
from nearly simultaneous observations \citep{zwd+25}. A large positive $k$ 
indicates a strong bluer-when-brighter (BWB) trend, which is usually 
associated with enhanced AGN continuum variability. For known turn-on CL AGNs, 
we searched for sources that had faded after their turn-on transition. 
Thus, the recent $zg$ magnitudes were expected to be larger than that around 
the previous on state, corresponding to $\Delta zg>0$ (right side of
Figure~\ref{fig:select}). For known turn-off CL AGNs, we correspondingly
searched for sources that had brightened after their turn-off transition, 
for which $\Delta zg<0$ (left side of Figure~\ref{fig:select}) was expected.
In practice, we gave priority to sources with relatively large $|\Delta zg|$
and large positive $k$, and then checked their light curves for validation.

\begin{figure}[htbp]
	\centering
	\includegraphics[width=0.49\textwidth]{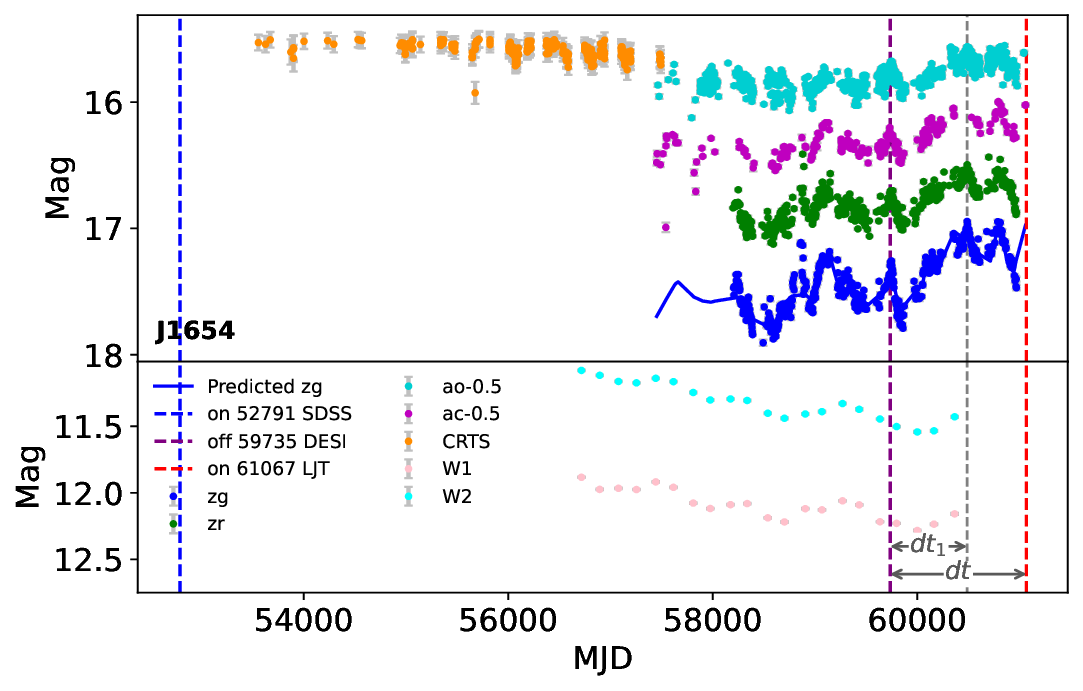}
	\caption{An example for the $zg$-like light curve 
	(the blue solid curve) of J1654 constructed from the $ac$-band data.
	The observed long-term optical 
	and MIR light curves of the source are also shown. The vertical 
	dashed color lines mark the epochs of the SDSS, DESI, and LJT 
	spectroscopic observations. 
	The timescales $dt$ and $dt1$ are marked, which are defined 
	in Section~\ref{subsec:basic}.}
	\label{fig:J1654}
\end{figure}

Since the public ZTF light curves are not released in real time, we also used
the real-time ATLAS forced-photometry data to check the recent brightnesses 
of the sources. Given that the $ac$ band largely overlaps in wavelength
with the $zg$ band \citep{tdh+18,bkg+19}, we constructed a $zg$-like light 
curve from the
$ac$-band data for each source. Specifically, during the time intervals
covered by both data, we interpolated the $ac$-band light curve to the
epochs of the $zg$-band measurements and fitted the paired magnitudes with
a linear relation using Huber regression \citep{hub64}. This source-specific
relation was then applied to the full $ac$-band light curve, and the
predicted $zg$-like light curve was smoothed with LOWESS \citep{cle79}. 
In Figure~\ref{fig:J1654}, an example (source J1654) of the constructed 
$zg$-like light curve is shown.

We defined MJD2 as the spectroscopic epoch when a turn-on or turn-off CL AGN
was identified. The $k$ values (shown in Figure~\ref{fig:select})
were derived using only the ZTF data
obtained from MJD2 to the end epoch of the latest ZTF data release,
and $\Delta zg$ were calculated from the end-epoch $zg$ magnitude minus that
matched to MJD2. As can be noted in Figure~\ref{fig:select}, the only 
up-pointing (red) triangle in the positive $\Delta zg$ region
is J1654, which was caused by the relatively large $zg$ magnitude at
the end epoch of the ZTF data. However, it had a more recent brightening 
caught by the ATLAS light curves, and thus its $zg$-like light curve 
(Figure~\ref{fig:J1654}) led to our selection of it as a candidate, 
which was confirmed as an on--off--on RCL AGN.

We would like to mention that for the background Type~1
and Type~2 AGNs in Figure~\ref{fig:select}, both parameters were derived
from the full ZTF light curves; for example,
$\Delta zg$ represents the maximum $zg$-band change over the ZTF
data time period.

Considering the telescopes we have access to for conducting confirmation 
spectroscopic observations, we also required the $zg$-band magnitudes
predicted to be brighter than 20.5 mag at the times of our observations.
Many CL AGNs located close to our selected candidates
in Figure~\ref{fig:select} were not selected and observed because of 
this requirement. In total, we selected and observed 34 RCL-AGN candidates. 
Their multi-band light curves are shown in Appendix
Figures.~\ref{fig:crcl} and \ref{fig:nrcl}.

\subsection{Spectroscopic identification}

We collected available spectra for the 34 candidates from SDSS, LAMOST,
and DESI. We also obtained new spectra for the candidates with the Lijiang
2.4-m Telescope (LJT; \citealt{wbf+19}), the 3.6-m Devasthal Optical
Telescope (DOT; \citealt{okk+19}), and the 2-m Himalayan Chandra Telescope
(HCT; \citealt{csp02}). Equipped with these three telescopes, the Yunnan Faint
Object Spectrograph and Camera (YFOSC),
the ARIES-Devasthal Faint Object Spectrograph and Camera (ADFOSC), and
the Hanle Faint Object Spectrograph and Camera (HFOSC) were respectively 
used for our observations. The grism and slit
settings were chosen according to the redshift $z$ of each target, such 
that the main Balmer-line regions, especially H$\beta$, were covered. 
Arc-lamp spectra,
including HeNe lamps for the LJT observations and FeNeAr lamps for the DOT 
and HCT observations, were taken for wavelength calibration. Also, a variety
of standard stars were spectroscopically observed during the same nights
of the target observations for flux calibration.
The observational information is provided in Appendix Table~\ref{tab:obs}.

The raw spectra were reduced with standard procedures in \textsc{IRAF}. The
main steps included bias subtraction, flat-field correction, one-dimensional
spectral extraction, wavelength calibration, and flux calibration. 
The obtained spectra were compared with the archival SDSS, LAMOST, and DESI
spectra to determine whether a second CL transition had occurred.

For the spectroscopic identification, we followed a procedure similar to that
used by \citet{dzg+25}. We first fitted the available spectra of each candidate
with \texttt{QSOFITMORE} \citep{qsofitmore,fwj+22}, a quasar spectral fitting 
code developed from \texttt{PyQSOFit} \citep{gsw18}. The identification
was based on the broad H$\beta$ component, whose appearance or disappearance 
was used to trace repeating transitions. The broad H$\alpha$ component
was inspected when available, but
was not included in the quantitative determination.

We quantified the change of broad H$\beta$ between the bright and dim
spectral states. A relative variation factor $R_{\rm s}$ is defined as
\begin{equation}
	R_{\rm s} = \frac{S_{\rm b}-S_{\rm d}}{S_{\rm b}},
\end{equation}
where $S_{\rm b}$ and $S_{\rm d}$ are the integrated broad H$\beta$ flux
in the bright and dim state, respectively. Sources with $R_{\rm s}>0.3$ were
considered as having a CL transition. This quantitative step helped exclude
sources with only weak or marginal broad-line changes.

Finally, we visually inspected all multi-epoch spectra. This step was 
necessary
because automatic fitting could be affected by spectral quality, continuum
subtraction, and blending with nearby narrow lines. During the inspection, we
required the change of broad H$\beta$ to be clear and consistent with an
on--off--on or off--on--off sequence. We also checked whether the spectra in
the dim state still showed reliable narrow emission lines, such as [O III],
to ensure that the apparent disappearance of broad H$\beta$ was not caused by
poor data quality. Sources with uncertain fits or ambiguous H$\beta$ changes
were not included in the confirmed RCL-AGNs. Through this process, we 
identified
25 RCL-AGNs from the 34 candidates. The remaining nine sources
were classified as non-confirmed RCL-AGNs.
The multi-epoch spectra used for this identification are presented in 
Appendix Figures~\ref{fig:crcl} and \ref{fig:nrcl}.

\begin{table}[htbp]
	\centering
	\scriptsize
	\caption{Physical properties of the candidates.}
	\begin{tabular}{lcccc}
		\hline
		Source & Mag\_Diff & $dt1$ (day) & $\log(M_{\rm BH}/M_\odot)$ & $\log\lambda_{\rm Edd}$ \\
		\hline
		\multicolumn{5}{c}{RCL} \\
		\hline
		J0759$^*$ & 0.54 & 231.15 & $9.14 \pm 0.07$ & $-$$2.27 \pm 0.08$ \\
		J0804 & 0.35 & 333.57 & $7.84 \pm 0.04$ & $-$$1.52 \pm 0.05$ \\
		J0831 & 0.59 & 1382.40 & $8.94 \pm 0.25$ & $-$$2.02 \pm 0.33$ \\
		J0836 & 0.26 & 148.99 & $8.39 \pm 0.22$ & $-$$1.56 \pm 0.29$ \\
		J0912 & $-$1.62 & 1604.00 & $8.53 \pm 0.03$ & $-$$1.44 \pm 0.04$ \\
		J0936 & $-$0.92 & 436.49 & $8.81 \pm 0.24$ & $-$$1.15 \pm 0.32$ \\
		J1021 & $-$0.61 & 236.63 & $8.39 \pm 0.26$ & $-$$1.57 \pm 0.35$ \\
		J1040 & $-$0.30 & 2465.03 & $8.68 \pm 0.10$ & $-$$2.13 \pm 0.12$ \\
		J1047 & $-$0.43 & 3007.31 & $7.69 \pm 0.07$ & $-$$1.50 \pm 0.07$ \\
		J1104$^*$ & 0.28 & 39.45 & $8.40 \pm 0.06$ & $-$$2.23 \pm 0.06$ \\
		J1110 & 0.60 & 738.50 & $8.05 \pm 0.11$ & $-$$1.44 \pm 0.13$ \\
		J1113 & 0.42 & 274.47 & $7.80 \pm 0.24$ & $-$$1.55 \pm 0.33$ \\
		J1210 & $-$0.80 & 357.34 & $8.66 \pm 0.03$ & $-$$1.30 \pm 0.04$ \\
		J1220 & $-$1.10 & 862.71 & $8.66 \pm 0.02$ & $-$$1.29 \pm 0.03$ \\
		J1309 & $-$0.35 & 488.16 & $8.20 \pm 0.05$ & $-$$1.36 \pm 0.06$ \\
		J1326 & 1.26 & 990.73 & $8.26 \pm 0.20$ & $-$$1.53 \pm 0.25$ \\
		J1330 & $-$1.09 & 307.42 & $7.69 \pm 0.12$ & $-$$1.02 \pm 0.13$ \\
		J1430 & 0.15 & 158.32 & $8.71 \pm 0.03$ & $-$$2.16 \pm 0.04$ \\
		J1525 & $-$0.41 & 808.87 & $8.57 \pm 0.03$ & $-$$1.39 \pm 0.04$ \\
		J1535 & 0.78 & 482.31 & $8.02 \pm 0.04$ & $-$$1.47 \pm 0.04$ \\
		J1543 & $-$1.13 & 397.37 & $8.60 \pm 0.05$ & $-$$1.06 \pm 0.05$ \\
		J1617$^*$ & $-$0.09 & 1143.69 & $8.57 \pm 0.03$ & $-$$1.39 \pm 0.04$ \\
		J1632 & $-$0.66 & 663.33 & $8.38 \pm 0.08$ & $-$$1.22 \pm 0.09$ \\
		J1654 & $-$0.32 & 751.42 & $8.81 \pm 0.04$ & $-$$2.08 \pm 0.04$ \\
		J2224 & $-$1.38 & 891.37 & $8.51 \pm 0.26$ & $-$$1.45 \pm 0.36$ \\
		\hline
		\multicolumn{5}{c}{RCL candidate} \\
		\hline
		J0023$^\dagger$ & 1.02 & 770.01 & $8.70 \pm 0.10$ & $-$$2.09 \pm 0.13$ \\
		J0158 & 0.88 & 649.94 & $7.32 \pm 0.05$ & $-$$1.42 \pm 0.05$ \\
		J0213 & $-$0.36 & $--$ & $8.20 \pm 0.08$ & $-$$1.94 \pm 0.09$ \\
		J0726$^*$ & 0.17 & 261.76 & $8.03 \pm 0.09$ & $-$$1.48 \pm 0.10$ \\
		J0853 & 0.10 & 7.94 & $7.95 \pm 0.03$ & $-$$1.48 \pm 0.03$ \\
		J1335 & 0.12 & 29.90 & $7.36 \pm 0.04$ & $-$$1.41 \pm 0.04$ \\
		J1400 & 0.55 & 583.33 & $8.64 \pm 0.06$ & $-$$1.32 \pm 0.07$ \\
		J1422 & 0.14 & 700.04 & $8.64 \pm 0.08$ & $-$$1.31 \pm 0.10$ \\
		J1518 & $-$0.10 & 19.82 & $9.19 \pm 0.14$ & $-$$1.60 \pm 0.16$ \\
		\hline
	\end{tabular}
	\label{tab:sources}
	\begin{flushleft}
		\footnotesize
		\textit{Notes.}
		Mag\_Diff is the magnitude change of a light-curve at
		epochs of the third to the second spectra 
		(positive/negative values indicating fainter/brighter). 
		The quantity $dt1$ is transition timescale estimated from 
the $zg$-like light curve. 
		Sources marked with $^*$ have also been
		reported as RCL AGNs in
		\citet{wwg+25,gzm+25,dzg+26,gcw+26}. Source J0023 is marked 
		with $^\dagger$, indicating its spectrum was strongly 
contaminated by moonlight; its black hole mass
		and Eddington ratio should be regarded as being indicative 
only. 
	\end{flushleft}
\end{table}

\section{Properties and Multi-band Variability}
\label{sec:prop}

\begin{figure}[htbp]
	\centering
	\includegraphics[width=0.49\textwidth]{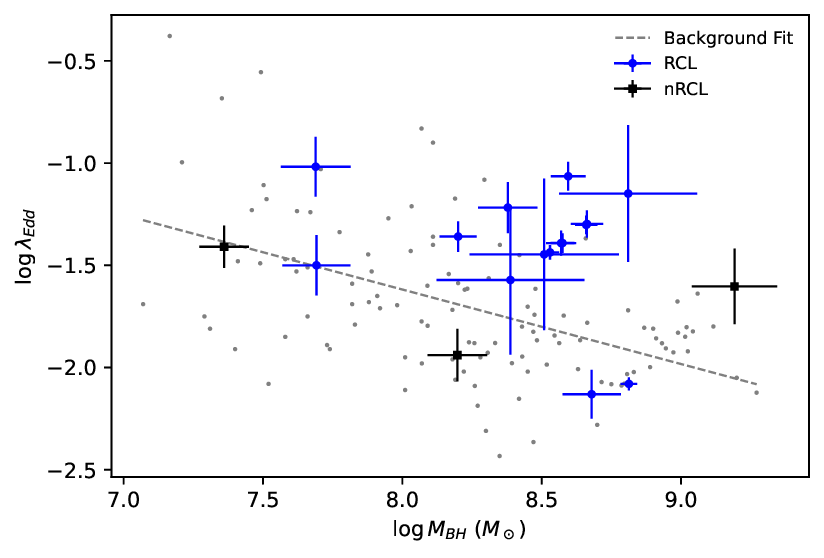}
	\includegraphics[width=0.49\textwidth]{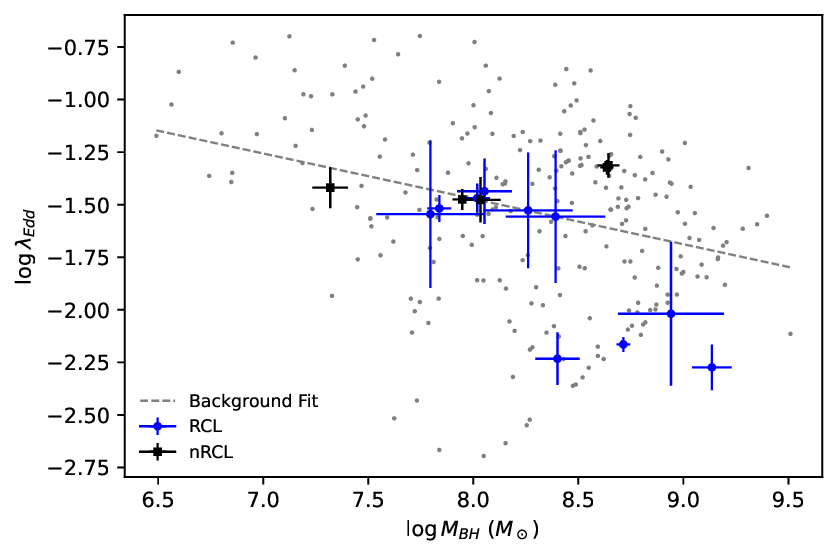}
	\caption{
		Distribution of Eddington ratio versus black hole mass. 
		\textit{Upper panel:} 
		the latest on-state measurements of the on--off--on RCL AGNs
(blue circles) compared with off-state CL AGNs (grey points) from 
the parent samples. 
		\textit{Bottom panel:} the latest off-state measurements of 
		the off--on--off RCL AGNs (blue circles), compared
		with on-state CL AGNs (grey points) from the parent samples. 
		Black squares mark the non-confirmed RCL-AGN candidates 
(source J0023 is not included in the diagram given its contaminated spectrum).
		The dashed lines in both panels show unweighted linear fits 
to the grey points.
	}
	\label{fig:mbh_edd}
\end{figure}

\subsection{Basic properties of the confirmed RCL AGNs}
\label{subsec:basic}

The basic properties of the 25 confirmed  RCL-AGNs are listed in
Table~\ref{tab:sources}, together with those for the
nine non-confirmed RCL (nRCL) candidates. The black hole masses 
$M_{\rm BH}$ and Eddington ratios $\lambda_{\rm Edd}$ given
in the table were derived only from the latest spectra obtained in this work.
The earlier SDSS, LAMOST, and DESI spectra were used to identify the spectral
state changes, but their fitted parameters are not included in the table.

We estimated the black hole masses using single-epoch virial relations, which
can be written in the general form
\begin{equation}
	\log \left(\frac{M_{\rm BH}}{M_{\odot}}\right)
	= a + b\log \left(\frac{\lambda L_{\lambda}}
	{10^{44}~{\rm erg~s^{-1}}}\right)
	+ c\log \left(\frac{\rm FWHM}
	{10^3~{\rm km~s^{-1}}}\right).
\end{equation}
For sources whose latest spectra showed an on state, broad H$\beta$ was used
when it was available, with the calibration of \citet{vp06}. For sources in an
off state, broad H$\beta$ was weak or absent, and we instead used broad
H$\alpha$ when it was covered by the spectra, following the calibration of
\citet{gh05}. For the higher-redshift sources for which H$\alpha$ was
not covered, we used Mg~II with the calibration of \citet{vo09}. 

%In this way, the line used for the mass estimate follows the spectral state and wavelength coverage of the latest spectrum.

The bolometric luminosity $L_{\rm bol}$ was calculated from 
the monochromatic continuum
luminosity using the bolometric corrections of \citet{rls+06},
$L_{\rm bol}=9.26L_{5100}$ for the H$\beta$ and H$\alpha$ estimates, and
$L_{\rm bol}=5.15L_{3000}$ for the Mg~II estimates. The Eddington ratio was
then derived as $\lambda_{\rm Edd}=L_{\rm bol}/L_{\rm Edd}$, where
$L_{\rm Edd}=1.26\times10^{38}(M_{\rm BH}/M_{\odot})~{\rm erg~s^{-1}}$.
The uncertainties were propagated from the fitting errors of the continuum
luminosities and line widths, with an additional systematic term added in
quadrature.

\begin{figure}[htbp]
	\centering
	\includegraphics[width=0.46\textwidth]{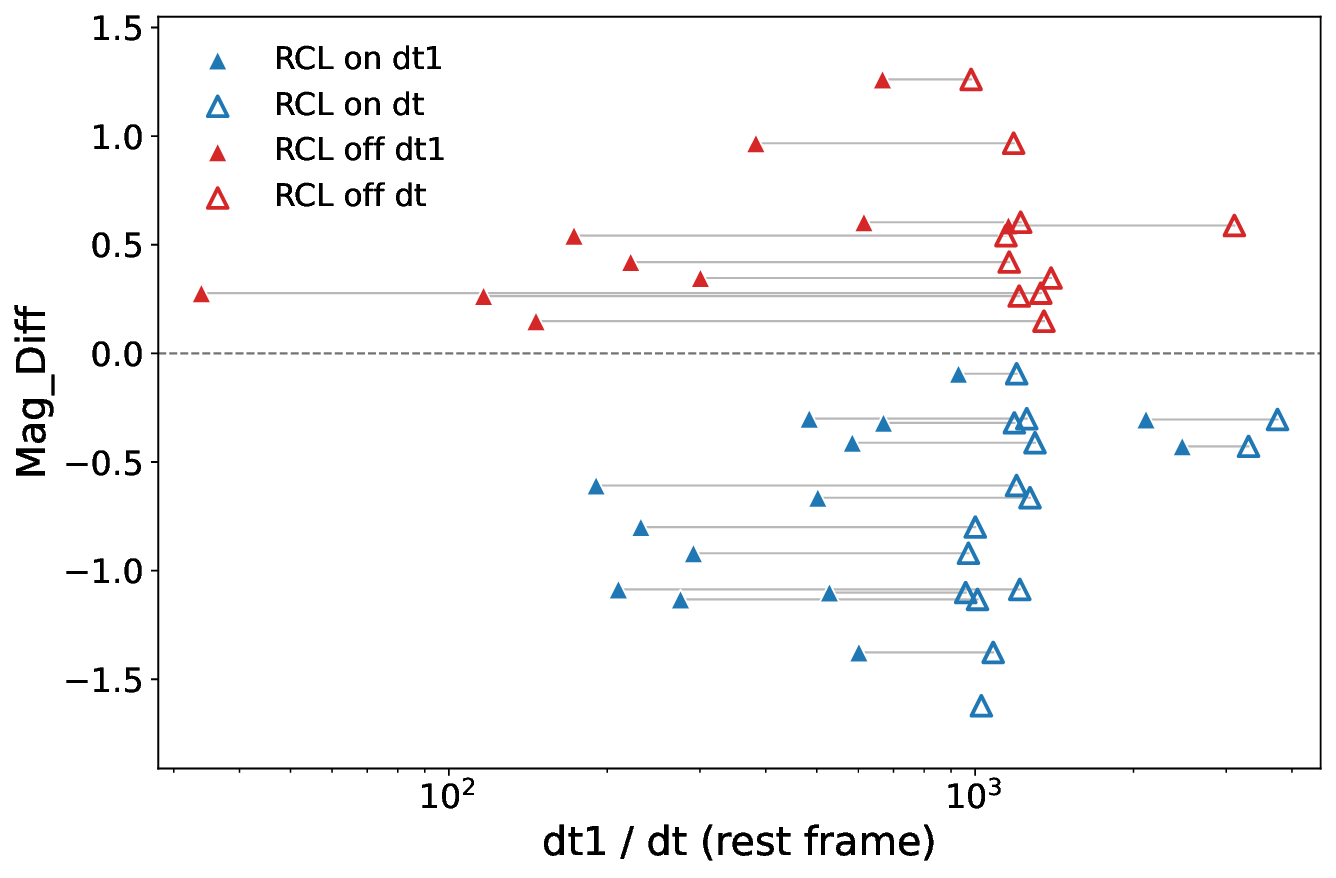}
	\caption{
Optical magnitude change versus the observed and photometric timescales
($dt$ and $dt1$, respectively) for the second CL transition. 
		}\label{fig:fluxdiff}
\end{figure}

We compare these measurements with the CL-AGN populations from
\citet{gzg+25} and \citet{dzg+25} in Figure~\ref{fig:mbh_edd}. Our confirmed
on--off--on and off--on--off sources are with the off-state and on-state
CL AGNs, respectively. Unweighted linear fits to the two states of background 
CL AGNs (the grey dashed line in each panel of Figure~\ref{fig:mbh_edd})
were obtained and are used to guide the comparison.  
As can be seen, the on--off--on sources are mostly located above 
the linear line, indicating Eddington ratios higher 
than the typical off-state CL AGNs. Similarly, the off--on--off sources 
tend to have lower Eddington ratios relative to the on-state CL AGNs.
The trend is consistent with the spectral classifications of the latest
spectra: sources returning to an on state have a stronger accretion-powered
continuum, while those returning to an off state have a weaker one. We note
that the linear fits are not intended to define a universal boundary between
the on and off states. Instead, they help show that the latest 
spectra of our RCL AGNs were shifted in the expected direction relative to the
corresponding parent CL-AGN populations.
%%Unconfirmed candidates are also plotted as black squares.

\subsection{Long-term optical and MIR variability}\label{subsec:lc}

Long-term light curves exhibit the continuum variations, which have been
found to possibly reflect the spectral states (e.g., \citealt{zhu+26}). 
For each RCL AGN, we estimated the $zg$-band magnitudes at the epochs of 
the second and third spectra. When no $zg$-band measurements were available 
near a spectral epoch, we used the
$zg$-like magnitude predicted from the ATLAS $ac$-band light curve
(Section~\ref{sec:sample}). The difference of the magnitude at the later
epoch relative to that at the earlier one, Mag\_Diff, was derived;
the values of the confirmed RCL AGNs and the candidates are listed in 
Table~\ref{tab:sources}.
However, note that there are two sources, J1040 and J1047,
whose second spectral epochs could not be covered even by the ATLAS
data (Figure~\ref{fig:crcl}). For these two, the earliest $zg$-like magnitude 
was taken as
that at the second spectral epoch and the Mag\_Diff value is not reliable.

Two timescales for the second CL transition are determined.
The first, the observed one denoted as $dt$, is the time
interval between the second and third spectroscopic epochs.
Because we selected the candidates from comparing their magnitude 
levels with the values at the second spectral epoch, 
when there were magntidue changes already reaching Mag\_Diff
before $dt$,  the transition could have occurred before the third
spectral epoch (see Figure~\ref{fig:J1654} as an example). 
%%in other words, an AGN already underwent its second transition when the first Mag\_Diff was reached. 
Given this possibility,
we also estimated a photometric timescale, denoted as $dt1$,
for each source. The timescale gives the value when
Mag\_Diff was first reached in the $zg$ light curve of a source. 
The values of $dt1$ and $dt$ are given in Table~\ref{tab:sources} and 
Appendix Table~\ref{tab:obs}, respectively. 
Figure~\ref{fig:fluxdiff} is a diagram of Mag\_Diff versus $dt1$/$dt$,
where the latter are corrected to the rest frame.

The confirmed  RCL AGNs had a broad range of absolute magnitude changes, from
a few tenths to larger than 1 mag, while the latest on-state sources tended
to have larger changes. Most of the transitions had the observed
(rest-frame) timescales of
$\sim$1000--1500\,day, except the two sources J1040 and J1047 mentioned
above and J0831. If we consider the photometric timescales,
the transitions would mostly occur from a few hundreds to one
thousand days (the shortest is $\sim$40\,day, exhibited by J1104).
No clear trend exists between Mag\_Diff and $dt1/dt$, suggesting that 
the transition timescales were not set only by the amplitudes of the flux 
changes.
The MIR light curves also showed long-term variations in many sources. 
Although the MIR cadence was much lower than that of the optical surveys, 
the W1 and W2
variations generally followed the optical brightening or dimming trends. This
behavior indicates that the recent optical changes were linked to changes in
the central continuum, with the MIR emission tracing the response of hot dust.
A detailed discussion of this point and its implication for the RCL behavior 
is given in Section~\ref{sec:dis}.

\subsection{Seven well-sampled RCL-AGNs}\label{sec:seven_rcl}

Among the RCL AGNs, seven sources were covered by single-band photometry at 
all three spectroscopic epochs. Observations of them provide the direct 
comparison between the continuum evolution and the repeated spectral-state 
transitions.  Recent studies of Seyfert 2 AGNs found 
significant optical variability on month-to-year timescales but suppressed 
short-timescale variability, suggesting that the long-term optical changes 
trace coherent changes in the nuclear radiation field \citep{kdy+26}.
We therefore used the low-order polynomial to describe the long-term continuum 
variations that were potentially relevant to the CL transitions.
For all RCL AGNs, we fitted the $zg$-band (including the $zg$-like) light 
curves with a third-order polynomial.
The resulting normalized fits for the RCL AGNs are shown in 
Appendix Figure~\ref{fig:lcfit}.

\begin{figure}[htbp]
	\centering
	\includegraphics[width=0.49\textwidth]{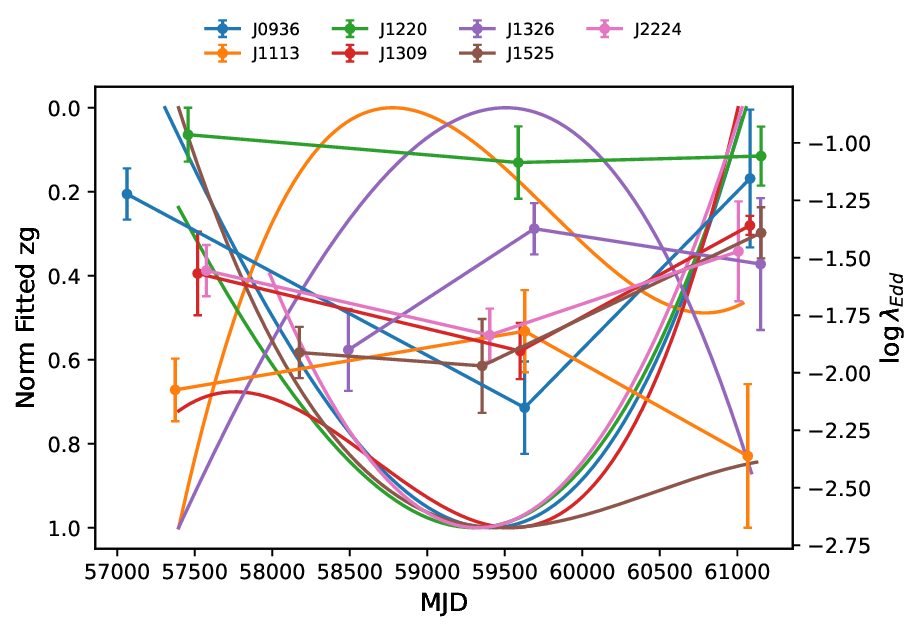}
	\caption{
		Comparison between the normalized fitted $zg$-band light-curve 
trends (solide curves) and the Eddington ratios of the seven well-sampled 
RCL AGNs.  The Eddington ratios measured from the three spectroscopic epochs
are connected by lines of the same color as the trends for a source.  }
	\label{fig:lcedd}
\end{figure}

The fits for the seven well-sampled RCL AGNs are shown in 
Figure~\ref{fig:lcedd}.
The Eddington ratios estimated from the spectra at the three epochs are
also plotted in the figure for comparison.
A clear correspondence is seen between the photometric and spectroscopic 
evolution: all on--off--on sources showed u-shape $zg$-band light curves, 
whereas all off--on--off sources showed inverted, or n-shape, light curves. 
The Eddington ratios varied consistently with the continuum trends, with 
higher and lower accretion states respectively corresponding to 
the photometric brighter and fainter phases. This agreement suggests that the 
RCL transitions are closely connected to recurrent changes in the accretion 
state.

\section{Discussion}\label{sec:dis}

\subsection{Non-confirmed candidates}
\label{sec:un}

In this study, nine candidates were not confirmed as RCL AGNs.
Among them, 
the spectrum of J0023 was strongly contaminated
by moonlight (thus not included in Fig.~\ref{fig:mbh_edd}). 
For J1518, although we classified its latest spectrum as in a non-changing
off state, we cautiously note it as an uncertain case since the H$\beta$ line 
is at the red end of our DOT spectrum, not allowing us to make a clear
classification (Figure~\ref{fig:nrcl}).
Two other candidates, J0158 and J0726, can be considered as special
cases. The first is the nuclear transient
PS16dtm, which has been interpreted as a TDE in a narrow-line Seyfert~1 galaxy
\citep{bnb+17,pli+23}. Its transient broad Balmer emissions may be affected by
TDE-powered reprocessing and need not trace the ordinary AGN BLR (so its
virial parameters should be compared with caution). The second was previously
identified as an on--off--on RCL AGN by \citet{dzg+26}. Broad H$\beta$ did not
disappear in our latest spectrum, and it is retained as an (on) nRCL AGN.
One additional caveat concerns J0213. In
Figure~\ref{fig:select}, its off-state point ($\Delta zg$ close to $-$1)  
was based on the real $zg$ light curve, which did not cover the fast fading 
around its MJD2 (see Figure~\ref{fig:nrcl}).  Using the $zg$-like light curve 
instead gives
$\Delta zg = -0.36$ (Table 1), placing it closer to $\Delta zg =0$ than
its plotted position in Figure~\ref{fig:select} (the upper left nRCL-off
data point).

The remaining non-confirmed candidates provide a useful comparison with the
confirmed RCL AGNs. In Figure~\ref{fig:select}, their $k$ values 
($\lesssim 0.5$) overlap 
with those of a few RCL AGNs. The two non-confirmed (on) sources with 
relatively large $\Delta zg$ (close to 1) are J0023 and J0158, whose special 
circumstances were noted above.
Excluding these two sources, the other non-confirmed ones lie closer to
$\Delta zg=0$ than the confirmed RCL AGNs.
Their continuum variations after MJD2 were therefore
typically smaller.
Consistent with this smaller changes, our spectra indicated that
the previous H$\beta$ state were retained (Appendix Table~\ref{tab:obs}). 
The continuum changes may thus have been
insufficient to move broad H$\beta$ across its source-dependent visibility
threshold (see discussion below in Section~\ref{subsec:blr}).

Figure~\ref{fig:mbh_edd} gives a consistent indication from the inferred
accretion levels. The nRCL (off) sources in the upper panel remain broadly
consistent with the off-state parent population, while the nRCL (on) sources
in the lower panel generally occupy the on-state region. In contrast, the
confirmed RCL AGNs have shifted in the direction expected for their new
spectral states. Although the small sample and the uncertainties of
single-epoch estimates prevent a firm statistical conclusion, the comparison
is consistent with the accretion changes that did not reach the levels required
for another CL transition. 
Continued spectroscopy at near-future light-curve extrema is needed to 
test whether these sources would remain in their present states
or undergo repeating transitions.

\subsection{Evidence for BLR breathing}
\label{subsec:blr}

\begin{figure}[htbp]
	\centering
	\includegraphics[width=0.49\textwidth]{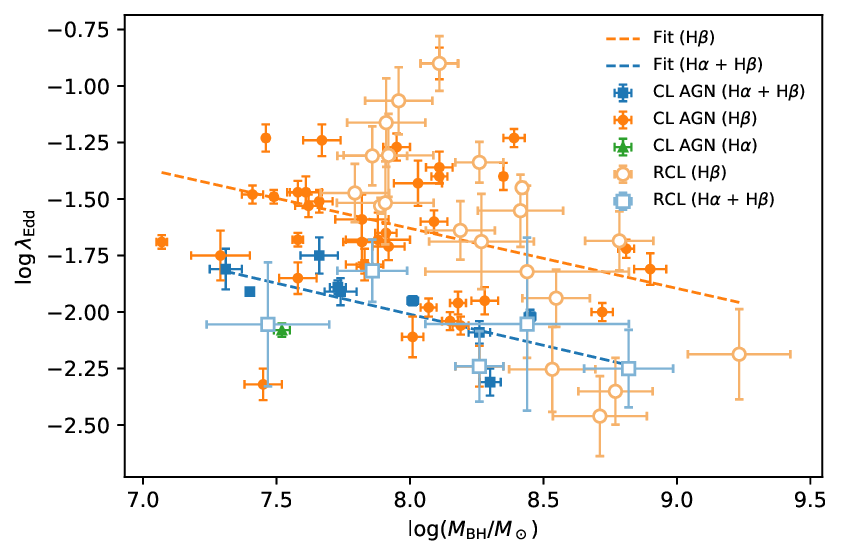}
	\caption{Off-state Eddington ratios and black hole masses for CL and
		RCL AGNs, grouped by the broad lines involved in the transition.
		The filled symbols are CL AGNs from \citet{dzg+25}, where 
		the transition types were classified from the changing broad 
		lines. The open symbols are the RCL AGNs in this work whose 
		off-state spectra cover both H$\alpha$ and H$\beta$. For 
		the RCL AGNs, $M_{\rm BH}$ was fixed to the on-state value of
		the same source. The dashed lines show unweighted linear fits 
		to the H$\beta$ and H$\alpha$+H$\beta$ groups, which combine
		both the CL and RCL sources.}
	\label{fig:breath}
\end{figure}
The variability properties provide an important clue to the nature of the
off state. The optical variability amplitudes and the colour--magnitude
slope $k$ of CLAGNs are closer to those of Type~1 AGNs than to those of
normal Type~2 AGNs (see Figs.~6, 9, and 10 of \citealt{wwy+25}). Shown
in the Appendix Figure~\ref{fig:sig}, the same comparison was conducted;
both the parent CL samples and our RCL sample exhibited a similar behaviour. 
This suggests that
many CL/RCL AGNs in their spectral-off states are not simply classical 
Type~2 AGNs whose nuclei are fully hidden by the torus. Instead, they may be 
connected to the population of `naked' or true Type~2 AGNs, in the sense 
that broad lines can become weak or absent while the nuclear continuum is 
not completely obscured. This view is broadly consistent with models in which
the BLR weakens or disappears below a critical accretion level
\citep[e.g.,][]{eh09,tik+11,eht14}.
Such sources are usually identified with X-ray observations, where
the lack of broad optical lines is accompanied by little or no intrinsic
X-ray absorption \citep[e.g.,][]{haw04,pcb+09,lyp+14}. Although X-ray data
are not available for all sources in our sample, the MIR light curves provide
a complementary test. The WISE W1 and W2 emission is produced by dust
reprocessing of the central UV/optical radiation at radii larger than the
Balmer-line BLR \citep[e.g.,][]{kmy+14}. Therefore, coherent MIR variations
following the optical changes are more naturally explained by intrinsic
changes in the central radiation field than by a single line-of-sight
obscuring cloud \citep[e.g.,][]{swj+17,ygm+23,cas+26}. The correlated
optical and MIR variabilities in our RCL AGNs thus support a picture in which 
the broad lines disappear because the ionizing continuum weakens, rather than 
because the BLR is suddenly hidden from our view.
%%A useful caution comes from SDSS~J1346+1736, for which an optical colour-based obscuration interpretation has been proposed \citep{zxg26}.  However, when the WISE light curves are added, the source also shows strong MIR variability (Fig.~\ref{fig:1346}). This suggests that optical colour changes alone may not be sufficient to distinguish obscuration from intrinsic continuum variations in CLAGN.

Under this intrinsic-variability picture, Figure~\ref{fig:breath} is
plotted to compare
the off-state accretion levels of different line-transition classes. 
We adopt
the H$\beta$-only, H$\alpha$-only, and H$\alpha$+H$\beta$ transition
classification given by \citet{dzg+25} for their CL sample, and classify
our RCL-AGNs in the same way. To avoid wavelength-coverage bias, only 
those with off-state spectra covering both H$\alpha$ and H$\beta$ are 
included.
The dashed lines in the figure show unweighted least-squares fits to 
the combined CL+RCL H$\beta$ and H$\alpha$+H$\beta$ groups. As can be
seen, the H$\beta$-only transitions occur, on average, at higher 
$\lambda_{\rm Edd}$ than those transitions
involving both H$\alpha$ and H$\beta$. 

To verify the significance of this difference, 
we perform a two-sample Kolmogorov-Smirnov (K-S) test on the off-state 
$\log \lambda_{\rm Edd}$ distributions of the H$\beta$-only group (60 sources)
and the H$\alpha$+H$\beta$ group (14 sources). The K-S test yields a maximum 
deviation statistic of $D = 0.667$ with a corresponding $p$-value of 
$2.17 \times 10^{-5}$. 
We also perform a two-sample Anderson-Darling (A-D) test, which yields
a $p$-value of $\le 10^{-3}$ (floored by numerical resolution). These 
results demonstrate that the two distributions are statistically distinct 
at a confidence level exceeding 99.9\%, confirming that broad H$\beta$ can 
become invisible at a relatively higher accretion level, while the
disappearance or significantly weakening of H$\alpha$ requires 
a lower state. This behaviour is consistent with the statistical 
broad-line evolution sequence found by \citet{gfs+25}, who have shown that, 
as AGN luminosity or Eddington ratio decreases, broad H$\beta$ fades first, 
followed by Mg~II and then H$\alpha$.
Their result provides direct empirical support for a line-dependent BLR
visibility threshold. In this framework, our RCL AGNs show the same effect
within repeatedly changing nuclei: as the ionizing continuum weakens, the BLR
does not need to be destroyed, but different broad lines cross their detection
thresholds at different accretion levels. The repeated type changes can
therefore be understood as extreme BLR breathing in a radially stratified and
line-dependent BLR \citep[e.g.,][]{kg04,ch06,bdg+13,gk14,gfs+25}.
As H$\alpha$ and H$\beta$
are both Balmer recombination lines, the difference is not due to different
ionization energies, but to line responsivity, optical depth, emitting radius,
and detectability. Photoionization calculations show that the emissivity and
responsivity of optical recombination lines depend on the incident ionizing
photon flux and on the location of the gas in the BLR \citep{kg04,gk14}. As
the accretion-powered continuum fades, broad H$\beta$ can first fall below the
detection threshold while broad H$\alpha$ remains visible; at lower
states, H$\alpha$ also weakens or disappears. The observed sequence from
H$\beta$-only to H$\alpha$+H$\beta$ transitions therefore does not require
the BLR gas to be repeatedly destroyed and rebuilt. Instead, the same BLR can
cross different line-visibility thresholds as the ionizing continuum rises
and fades. This scenario is compatible with accretion-dependent unification
models, in which the covering factor of circumnuclear material is regulated
by the Eddington ratio \citep{rtk+17,ric26}; our RCL-AGN sample 
shows that line responsivity within the BLR can also play a major role in CL
transitions.

\subsection{Possible driver of the long-term excursions}

The repeating nature of these CL transitions requires a reversible driver
that can operate more than once in the same AGN. The seven well-sampled 
RCL-AGNs provide a clear clue. Their three spectroscopic epochs were
covered by nearly continuous ATLAS light curves, and the spectral states
followed smooth multi-year optical excursions with small-scale AGN variations
superposed (Figure~\ref{fig:lcedd}). The Eddington ratios measured 
at different states varied consistently with the optical flux changes, which
were followed by the MIR flux changes. These facts point to repeated 
changes in the
accretion-powered radiation field, rather than to a single destructive event
or geometrical blocking.
The timescales provide a constraint. The second transition in our RCL-AGN
sample typically occurred over rest-frame timescales of possibly several 
months to a few
years, similar to many CL AGNs \citep[e.g.,][]{lcm+15,nd18,db19}. This is 
much shorter than the viscous timescale of a standard thin disk at 
optical-emitting
radii, making a simple global change in the outer-disk accretion rate
problematic \citep[e.g.,][]{law18,db19}. However, the smooth long-term
variation trends of the seven RCL-AGNs are too coherent to be
explained by ordinary short-timescale stochastic flickering. They are
likely better described as finite-duration accretion-power excursions,
during which the ionizing continuum moves the BLR across the H$\beta$ and,
in lower states, H$\alpha$ visibility thresholds.

Several accretion-flow processes may produce such year-long excursions.
Accretion-mode transitions, analogous to state changes in X-ray binaries, can
alter the inner disk and the soft ionizing continuum without requiring the
whole optical disk to evolve on the standard viscous timescale
\citep{nd18,rae+19,hll+26}. Heating or cooling fronts related to inner-disk 
torque changes or
thermal instabilities offer another route, suggested to explain rapid 
CL-quasar variability \citep{rfg+18}. Magnetically supported disks can 
shorten the inflow time \citep{db19}, while radiation-pressure instability 
in an inner unstable disk
region can produce multiple outbursts and is therefore relevant for repeating
events \citep{scb+20,hys+26}. These mechanisms differ in detail, but all can 
produce
temporary changes in the inner accretion power and hence in the ionizing
continuum on the required timescales.

\begin{figure}[htbp]
	\centering
	\includegraphics[width=0.49\textwidth]{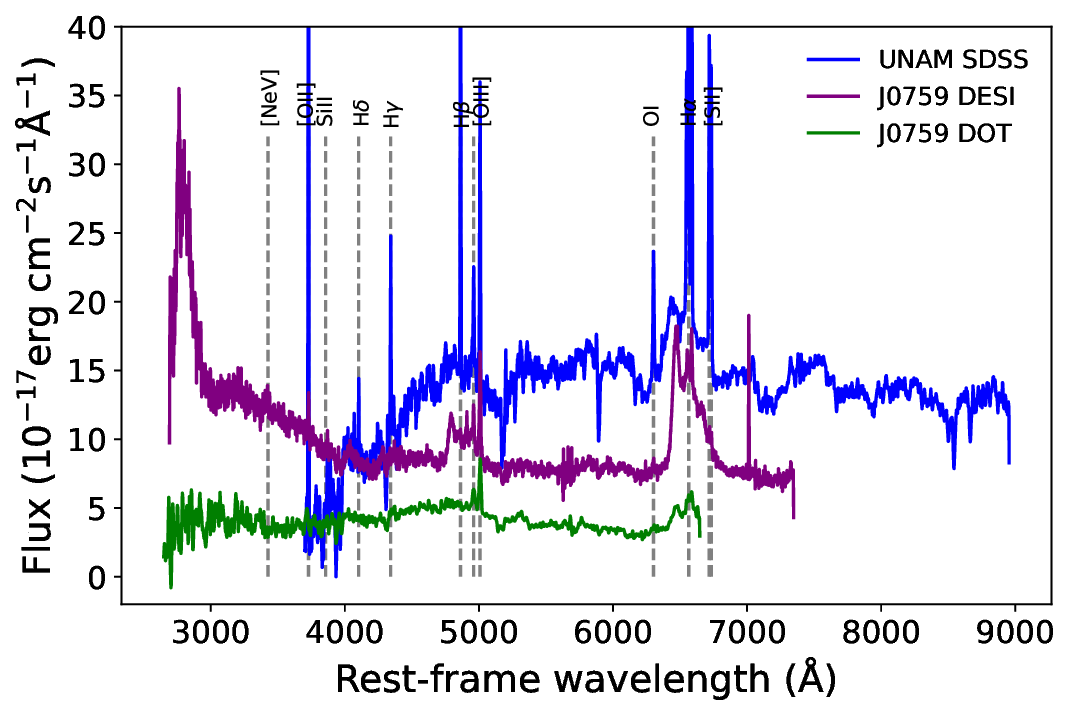}
	\caption{Rest-frame spectra comparing J0759 with UNAM-KIAS 613. The SDSS spectrum of UNAM-KIAS 613 has been divided by 5 and shifted downward by 12 for visual comparison. }	
	\label{fig:0759_spec2}
\end{figure}

Among our RCL AGNs, we also note that J0759 may provide an interesting case 
in which the long-term accretion-power
excursion was accompanied by a transient or persistent wind-like kinematic
component. This source has been studied in detail by \citet{gzm+25}, who
identified it as a recurring CL AGN with multiple optical and MIR flares and
argued that its RCL events are inconsistent with simple obscuration and were
more likely driven by changes in the accretion rate. They further decomposed
the asymmetric broad H$\alpha$ profile into a blue wing component and a main
core component, and found similar asymmetric velocity structures in Mg~II,
H$\beta$, and H$\alpha$. These suggest that the displaced broad-line 
component was connected to a real BLR kinematic structure rather than to 
line blending alone.
Our spectra show the same key feature (Figure~\ref{fig:0759_spec2}): the DESI
spectrum contains a strong asymmetric broad component in the H$\alpha$ region,
and the blue-shifted component is also clearly seen around H$\beta$, where
contamination from the H$\alpha$+[N II] complex is avoided. 
A useful comparison
is UNAM-KIAS 613, for which \citet{cmh+26} reported transient blue- and
red-shifted broad H$\alpha$ components and interpreted them as a short-lived
bipolar outflow or an additional kinematic component superposed on the normal
BLR emission. The presence of such a component in J0759 does not require a
separate origin from BLR breathing. A rise in the ionizing continuum can
increase the BLR emissivity and change the effective line-emitting radius,
while the same enhanced radiation field may also strengthen a disk wind or
accelerate line-emitting gas. Therefore, the J0759 case suggests that in some 
CL events, the observed broad-line changes may include both the breathing
response of the BLR and a temporary change in the gas kinematics.

%\subsection{Alternative scenarios}
RCL events do not necessarily require a single trigger. Several
models proposed for CL AGNs may also operate in RCL AGNs if the inner accretion
flow can be reconfigured more than once \citep{kgm+26}. For example, 
Lense-Thirring--driven disk tearing
and subsequent reassembly can rapidly change the inner accretion power and
the illumination of the BLR \citep{nkp+12,klw+25}, while magnetic-flux
inversion, as proposed for 1ES~1927+654, provides another way to produce a
rapid UV/optical outburst through an inner-disk magnetic reconfiguration
\citep{sbd21}. In this regard, although TDEs in AGNs may explain some 
individual CL-like flares \citep[e.g.,][]{bnb+17,tam+19}, ordinary TDEs 
are less natural as the common explanation for RCL events with correlated
optical--MIR variability and line-dependent $\lambda_{\rm Edd}$ thresholds.
Thus, the most general interpretation is that RCL events trace repeated,
finite-duration changes in the central accretion power, while the specific
trigger may vary from source to source, which remains to be further 
investigated.

\section{Conclusion}
We analyzed the long-term optical/MIR light curves of known CL-AGN samples 
to search for RCL-AGN candidates and confirmed 25 RCL AGNs out of 34 candidates.
Among the confirmed, 22 were discovered by us, which are the largest newly
discovered sample (to the best of our knowledge).

The sample of RCL AGNs showed the following properties:
\begin{enumerate}
\item The latest spectra of the confirmed RCL AGNs showed Eddington-ratio
changes in the expected direction. Sources returning to an on state generally
moved to higher accretion levels, while sources returning to an off state moved
to lower levels. 
\item  The observed timescales for the second CL transition mostly were
3--4 years, and the estimated photometric timescales for the CL transition
could be as short as several months.
%The MIR variations generally follow the optical changes, supporting an intrinsic change in the central radiation field rather than simple line-of-sight obscuration.
\item Seven well-sampled RCL AGNs showed a close connection between their
spectral states and multi-year optical excursions, with Eddington ratios
varying consistently with the trends of the continuum. 
\item Similar to CL AGNs, the RCL sources also had their H$\beta$-only
transitions occurring at higher Eddington ratios than transitions involving both
H$\alpha$ and H$\beta$. This property supports a line-dependent BLR visibility 
threshold, in consistency with extreme BLR breathing.
\end{enumerate}

This study of RCL AGNs suggests that RCL events are produced by reversible
changes in the central ionizing luminosity, while different accretion-flow
processes may act as the underlying triggers. Future high-cadence spectroscopy,
together with optical, MIR, UV, and X-ray monitoring, will be essential for
more completely
understanding how the CL transitions are related to physical changes in an AGN
and how different broad
lines respond to recurrent accretion-power excursions of the AGN.

\section*{Data availability}
The derived data underlying this article are available from the corresponding 
authors on reasonable request.

\begin{acknowledgements} 
	
	We acknowledge the support of the staff of the Lijiang 2.4m telescope. Funding for the telescope has been provided by Chinese Academy of Sciences and the People's Government of Yunnan Province.\\	
	This work is based on observations obtained at the 3.6m Devasthal Optical Telescope (DOT), which is a National Facility run and managed by Aryabhatta Research Institute of Observational Sciences (ARIES), an autonomous Institute under the Department of Science and Technology, Government of India.\\
	This study makes use of data obtained from the 2-m Himalayan Chandra Telescope (HCT). We thank the staff at IAO, Hanle, and CREST, Hosakote, operated by the Indian Institute of Astrophysics, Bengaluru (India), for their support in facilitating these observations.\\
	This work was based on observations obtained with the Samuel Oschin Telescope 48-inch and the 60-inch Telescope at the Palomar Observatory as part of the Zwicky Transient Facility project. ZTF is supported by the National Science Foundation under Grant No. AST-2034437 and a collaboration including Caltech, IPAC, the Weizmann Institute for Science, the Oskar Klein Center at Stockholm University, the University of Maryland, Deutsches Elektronen-Synchrotron and Humboldt University, the TANGO Consortium of Taiwan, the University of Wisconsin at Milwaukee, Trinity College Dublin, Lawrence Livermore National Laboratories, and IN2P3, France. Operations are conducted by COO, IPAC, and UW.\\
	Funding for the Sloan Digital Sky Survey V has been provided by the Alfred P. Sloan Foundation, the Heising-Simons Foundation, the National Science Foundation, and the Participating Institutions. SDSS acknowledges support and resources from the Center for High-Performance Computing at the University of Utah. SDSS telescopes are located at Apache Point Observatory, funded by the Astrophysical Research Consortium and operated by New Mexico State University, and at Las Campanas Observatory, operated by the Carnegie Institution for Science. The SDSS web site is \url{www.sdss.org}.\\
	SDSS is managed by the Astrophysical Research Consortium for the Participating Institutions of the SDSS Collaboration, including the Carnegie Institution for Science, Chilean National Time Allocation Committee (CNTAC) ratified researchers, Caltech, the Gotham Participation Group, Harvard University, Heidelberg University, The Flatiron Institute, The Johns Hopkins University, L'Ecole polytechnique f\'{e}d\'{e}rale de Lausanne (EPFL), Leibniz-Institut f\"{u}r Astrophysik Potsdam (AIP), Max-Planck-Institut f\"{u}r Astronomie (MPIA Heidelberg), Max-Planck-Institut f\"{u}r Extraterrestrische Physik (MPE), Nanjing University, National Astronomical Observatories of China (NAOC), New Mexico State University, The Ohio State University, Pennsylvania State University, Smithsonian Astrophysical Observatory, Space Telescope Science Institute (STScI), the Stellar Astrophysics Participation Group, Universidad Nacional Aut\'{o}noma de M\'{e}xico, University of Arizona, University of Colorado Boulder, University of Illinois at Urbana-Champaign, University of Toronto, University of Utah, University of Virginia, Yale University, and Yunnan University.  \\	
	This publication makes use of data products from the Wide-field Infrared Survey Wise, which is a joint project of the University of California, Los Angeles, and the Jet Propulsion Laboratory/California Institute of Technology, funded by the National Aeronautics and Space Administration. This publication also makes use of data products from NEOWISE, which is a project of the Jet Propulsion Laboratory/California Institute of Technology, funded by the Planetary Science Division of the National Aeronautics and Space Administration.\\	
	This work has made use of data from the Asteroid Terrestrial-impact Last Alert System (ATLAS) project. 
	The Asteroid Terrestrial-impact Last Alert System (ATLAS) project is primarily funded to search for near earth asteroids through NASA grants NN12AR55G, 80NSSC18K0284, and 80NSSC18K1575; byproducts of the NEO search include images and catalogs from the survey area. 
	This work was partially funded by Kepler/K2 grant J1944/80NSSC19K0112 and HST GO-15889, and STFC grants ST/T000198/1 and ST/S006109/1. 
	The ATLAS science products have been made possible through the contributions of the University of Hawaii Institute for Astronomy, the Queen’s University Belfast, the Space Telescope Science Institute, the South African Astronomical Observatory, and The Millennium Institute of Astrophysics (MAS), Chile.	
	This work made use of the data from LAMOST (Large Sky Area Multi-Object Fiber Spectroscopic Telescope, also known as the Guoshoujing Telescope) (https://cstr.cn/31118.02.LAMOST). LAMOST is a Chinese national mega-science facility, operated by National Astronomical Observatories, Chinese Academy of Sciences.\\	
	This material is based upon work supported by the U.S. Department of Energy (DOE), Office of Science, Office of High-Energy Physics, under Contract No. DE–AC02–05CH11231, and by the National Energy Research Scientific Computing Center, a DOE Office of Science User Facility under the same contract. Additional support for DESI was provided by the U.S. National Science Foundation (NSF), Division of Astronomical Sciences under Contract No. AST-0950945 to the NSF’s National Optical-Infrared Astronomy Research Laboratory; the Science and Technology Facilities Council of the United Kingdom; the Gordon and Betty Moore Foundation; the Heising-Simons Foundation; the French Alternative Energies and Atomic Energy Commission (CEA); the National Council of Science and Technology of Mexico (CONACYT); the Ministry of Science and Innovation of Spain (MICINN), and by the DESI Member Institutions.\\
	Guoshoujing Telescope (the Large Sky Area Multi-Object Fiber Spectroscopic Telescope, LAMOST) is a National Major Scientific and Technological Infrastructure of China, operated and managed by the National Astronomical Observatories, Chinese Academy of Sciences.
	
This research is supported by the National Natural Science Foundation of
	China (12273033) and the Xingdian Talent Support Project of
	the Yunnan Province (XDYC-YLXZ-2023-0016).
	
\end{acknowledgements}

\appendix

\restartappendixnumbering
\section{Information for the RCL-AGN candidates}
In Table~\ref{tab:obs}, detailed information for the 34 selected candidates
is given. We also include our spectroscopic observational information and
the related light-curve variation information (used to derive properties
of the sources in the main text) in the table.

\begin{table*}
%\begin{sidewaystable*}[p]
	\centering
	\scriptsize
	\caption{Information for the 34 candidates.}
	\label{tab:obs}
	\begin{tabular}{lccccccccccccc}
		\hline
		Name & R.A. & Decl. & $z$ & MJD3 & State & Telescope & Standard & $t_{\rm exp}$ (s) & LC MJD2 & LC MJD3 & Mag2 & Mag3 & $dt$ (day) \\
		\hline
		\multicolumn{14}{c}{RCL} \\
		\hline
		J0759$^*$ & 119.94888 & 11.41870 & 0.34 & 61136 & off & DOT & Feige 66 & 3600 & 59622 & 61119$^a$ & 19.12 & 19.67 & 1530 \\
		J0804 & 121.22590 & 21.60752 & 0.11 & 61083 & off & LJT & Feige 66 & 3000 & 59532 & 61057$^a$ & 18.88 & 19.23 & 1547 \\
		J0831 & 127.88440 & 36.77150 & 0.20 & 61082 & off & LJT & G163-50 & 3000 & 57395$^a$ & 61057$^a$ & 18.80 & 19.39 & 3715 \\
		J0836 & 129.19525 & 15.21640 & 0.28 & 61061 & off & HCT & HZ 4 & 5400 & 59510 & 61039$^a$ & 18.44 & 18.70 & 1552 \\
		J0912 & 138.13726 & 0.28144 & 0.56 & 61136 & on & DOT & Feige 66 & 3600 & 59527 & 61144$^a$ & 20.67 & 19.05 & 1604 \\
		J0936 & 144.17900 & 55.18870 & 0.50 & 61082 & on & LJT & G163-50 & 3600 & 59630 & 61058$^a$ & 19.91 & 18.99 & 1454 \\
		J1021 & 155.31182 & 5.29313 & 0.24 & 61083 & on & LJT & Feige 66 & 3600 & 59624 & 61065$^a$ & 19.93 & 19.32 & 1489 \\
		J1040 & 160.18080 & 27.58860 & 0.17 & 61067 & on & LJT & Feige 66 & 3600 & 57396$^a$ & 61058$^a$ & 19.74 & 19.43 & 4383 \\
		J1047 & 161.77150 & 54.73500 & 0.21 & 61061 & on & HCT & HZ 4 & 2400 & 57397$^a$ & 61038$^a$ & 19.11 & 18.69 & 4018 \\
		J1104$^*$ & 166.09671 & 63.71816 & 0.17 & 61153 & off & DOT & HZ 44 & 3600 & 59621 & 61092$^a$ & 19.80 & 20.08 & 1553 \\
		J1110 & 167.60601 & $-$0.05947 & 0.22 & 61082 & off & LJT & G163-50 & 7200 & 59623 & 61065$^a$ & 19.12 & 19.72 & 1477 \\
		J1113 & 168.37368 & 53.22744 & 0.24 & 61067 & off & LJT & Feige 66 & 3600 & 59629 & 61038$^a$ & 19.63 & 20.05 & 1438 \\
		J1210 & 182.63880 & $-$1.29880 & 0.54 & 61150 & on & DOT & Feige 66 & 3600 & 59623 & 61125$^a$ & 19.86 & 19.06 & 1544 \\
		J1220 & 185.20005 & 36.46967 & 0.63 & 61153 & on & DOT & HZ 44 & 7200 & 59671$^a$ & 61090$^a$ & 21.18 & 20.08 & 1566 \\
		J1309 & 197.35950 & 53.69183 & 0.18 & 61081 & on & LJT & G163-50 & 3000 & 59683 & 61059$^a$ & 19.18 & 18.83 & 1398 \\
		J1326 & 201.64203 & 31.24318 & 0.49 & 61150 & off & DOT & Feige 66 & 7000 & 59689 & 61097$^a$ & 18.84 & 20.10 & 1460 \\
		J1330 & 202.68510 & 7.44130 & 0.46 & 61150 & on & DOT & HZ 44 & 3600 & 59312 & 61125$^a$ & 20.32 & 19.23 & 1780 \\
		J1430 & 217.56690 & 23.06235 & 0.08 & 61067 & off & LJT & Feige 66 & 1800 & 59600 & 61008$^a$ & 17.33 & 17.47 & 1461 \\
		J1525 & 231.32320 & 40.23270 & 0.38 & 61153 & on & DOT & HZ 44 & 7200 & 59353 & 61124$^a$ & 20.68 & 20.27 & 1799 \\
		J1535 & 233.75947 & 34.92733 & 0.13 & 61066 & off & LJT & G163-50 & 2400 & 59679 & 61058$^a$ & 17.99 & 18.77 & 1387 \\
		J1543 & 235.91460 & 31.79208 & 0.44 & 61153 & on & DOT & HZ 44 & 1980 & 59700 & 61098$^a$ & 20.50 & 19.36 & 1457 \\
		J1617$^*$ & 244.29759 & 6.64264 & 0.23 & 61150 & on & DOT & HZ 44 & 1800 & 59673 & 61125$^a$ & 18.62 & 18.53 & 1475 \\
		J1632 & 248.23804 & 24.05535 & 0.32 & 61082 & on & LJT & G163-50 & 3600 & 59404 & 61058$^a$ & 20.16 & 19.49 & 1678 \\
		J1654 & 253.64931 & 33.04762 & 0.12 & 61067 & on & LJT & Feige 66 & 1800 & 59735 & 61058$^a$ & 17.30 & 16.98 & 1332 \\
		J2224 & 336.11631 & 27.29065 & 0.48 & 61007 & on & DOT & Feige 110 & 6000 & 59404 & 60969 & 20.59 & 19.21 & 1604 \\
		\hline
		\multicolumn{14}{c}{RCL candidate} \\
		\hline
		J0023 & 5.85873 & 28.35356 & 0.24 & 61067 & on & LJT & Feige 66 & 2400 & 59503 & 61030$^a$ & 17.62 & 18.65 & 1564 \\
		J0158 & 29.51979 & $-$0.87275 & 0.08 & 61082 & on & LJT & G163-50 & 3000 & 59540 & 61060$^a$ & 17.93 & 18.81 & 1513 \\
		J0213 & 33.49910 & 0.70740 & 0.18 & 61061 & off & HCT & HZ 4 & 4800 & 57744$^a$ & 61030$^a$ & 18.39 & 18.03 & 3315 \\
		J0726$^*$ & 111.73365 & 41.02669 & 0.13 & 61081 & on & LJT & G163-50 & 1800 & 59622 & 61058$^a$ & 18.02 & 18.18 & 1462 \\
		J0853 & 133.49610 & 21.46760 & 0.08 & 61082 & on & LJT & G163-50 & 1500 & 58461 & 61057$^a$ & 17.02 & 17.12 & 2617 \\
		J1335 & 203.96660 & 54.74730 & 0.11 & 61061 & off & HCT & HZ 4 & 2400 & 59638 & 61038$^a$ & 18.26 & 18.38 & 1421 \\
		J1400 & 210.06941 & $-$1.13949 & 0.35 & 61150 & on & DOT & HZ 44 & 2400 & 59623 & 61125$^a$ & 17.92 & 18.46 & 1537 \\
		J1422 & 215.72282 & 47.54785 & 0.20 & 61081 & on & LJT & G163-50 & 2200 & 59356 & 61059$^a$ & 18.15 & 18.28 & 1725 \\
		J1518 & 229.52241 & 32.67494 & 0.76 & 61154 & uncertain & DOT & HZ 44 & 3000 & 59671 & 61098$^a$ & 19.49 & 19.38 & 1489 \\
		\hline
	\end{tabular}
	\begin{flushleft}
		\footnotesize
		\textit{Notes.}
Column MJD3 gives the epoch of our spectroscopic observation, State indicates 
whether broad H$\boldsymbol{\beta}$ is present (on) or absent/weak (off) in 
the third spectrum, LC MJD2 and LC MJD3 are the light-curve epochs matched 
to the second and third spectra respectively, Mag2 and Mag3 are 
the corresponding magnitudes at the two epochs, and $dt$ is the time interval 
between the second and third epochs. The exposure times are the total 
integrations. The superscript ``a'' indicates that the corresponding magnitude 
was obtained from the predicted $zg$-like light curve.
Sources marked with $^*$ have also been reported as RCL AGNs in
		\citet{wwg+25,gzm+25,dzg+26,gcw+26}.
	\end{flushleft}
%\end{sidewaystable*}
\end{table*}

\section{Multi-band light curves and multi-epoch spectra}

The multi-band light curves and spectra (from the archives and our observations)
of the confirmed RCL AGNs and nRCL candidates are shown in
Figures~\ref{fig:crcl} and \ref{fig:nrcl}, respectively.

%\begin{figure*}[htbp]
	%\centering
	%\includegraphics[width=0.49\textwidth]{J1346.eps}
%	%\caption{Long-term optical and MIR light curves of SDSS~J1346+1736. This source was interpreted by \citet{zxg26} as a possible obscuration-driven CLAGN based on its optical colour variability and Balmer-line properties. The additional WISE light curves show strong MIR variability during the same long-term evolution, providing an extra constraint on a pure moving-cloud obscuration scenario.}
%\label{fig:1346}
%\end{figure*}

\begin{figure*}[htbp]
	\centering
	\includegraphics[width=0.49\textwidth]{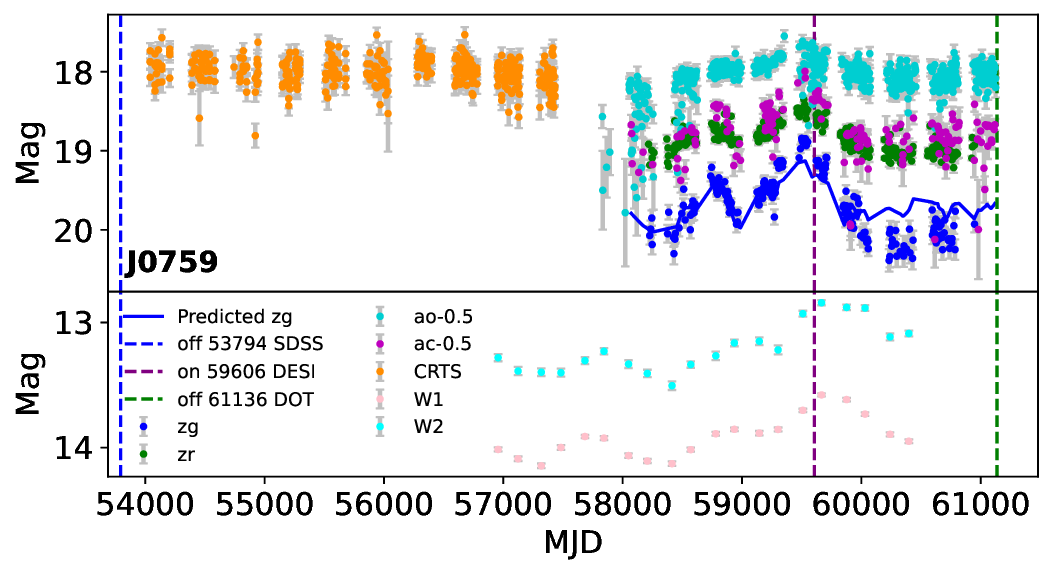}
	\includegraphics[width=0.48\textwidth]{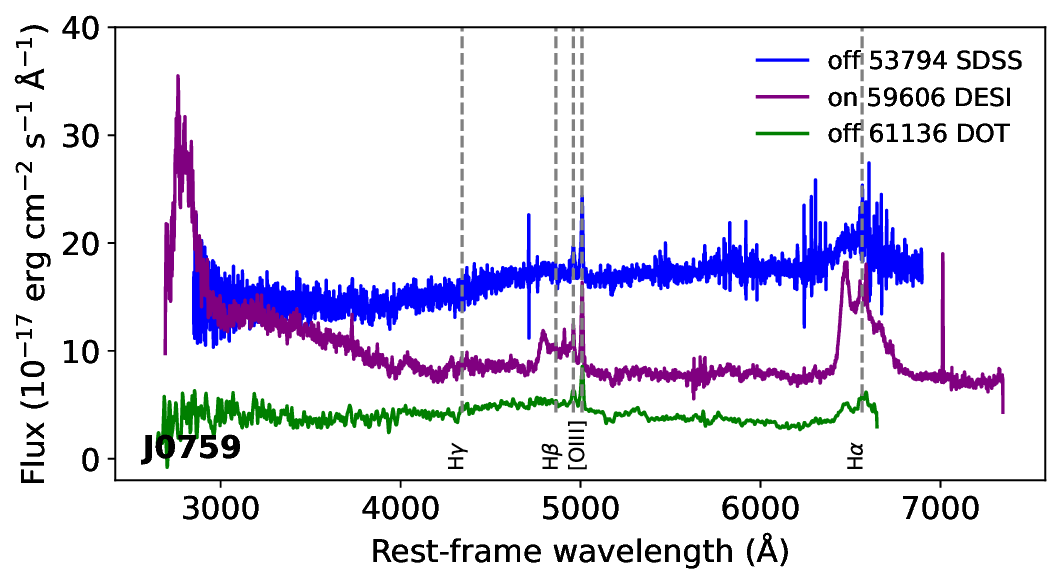}
	\includegraphics[width=0.5\textwidth]{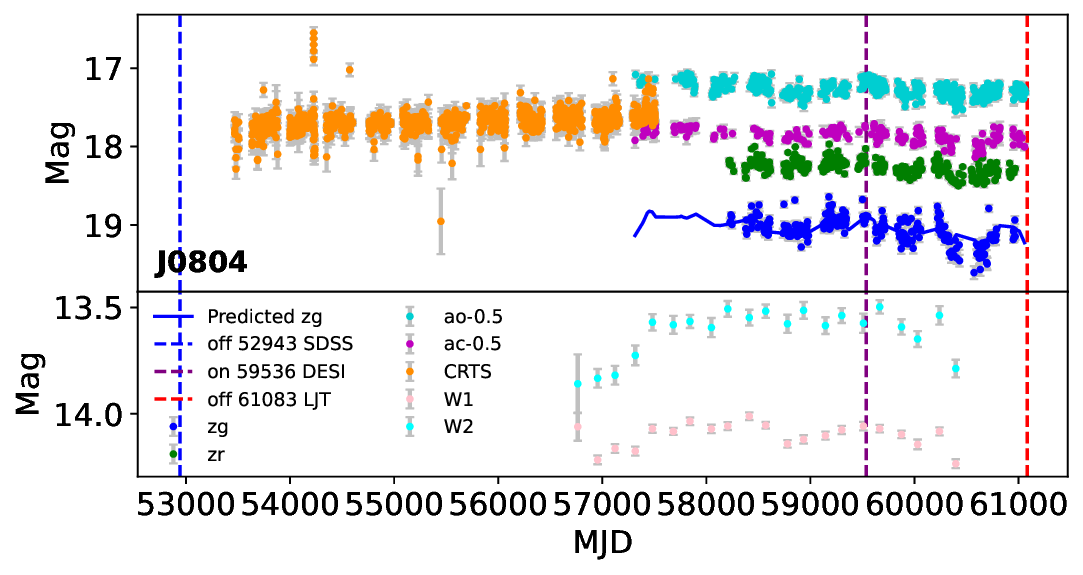}
	\includegraphics[width=0.48\textwidth]{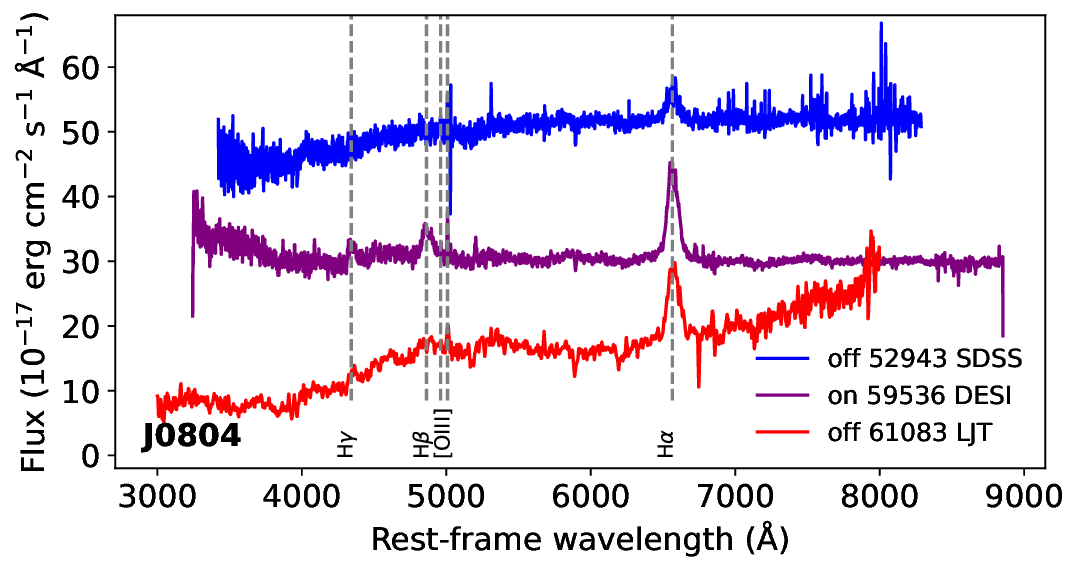}
	\includegraphics[width=0.5\textwidth]{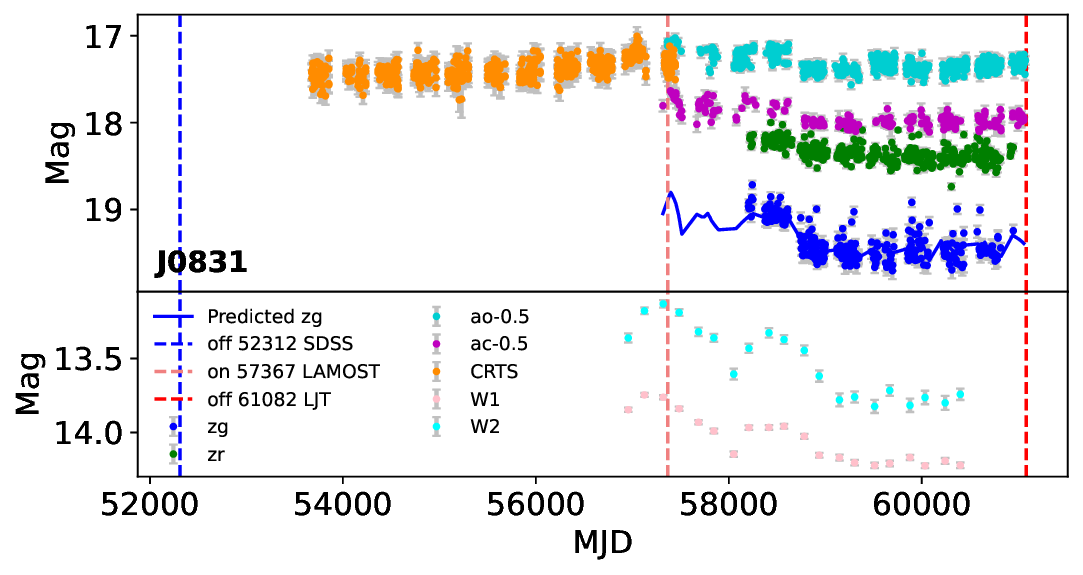}
	\includegraphics[width=0.48\textwidth]{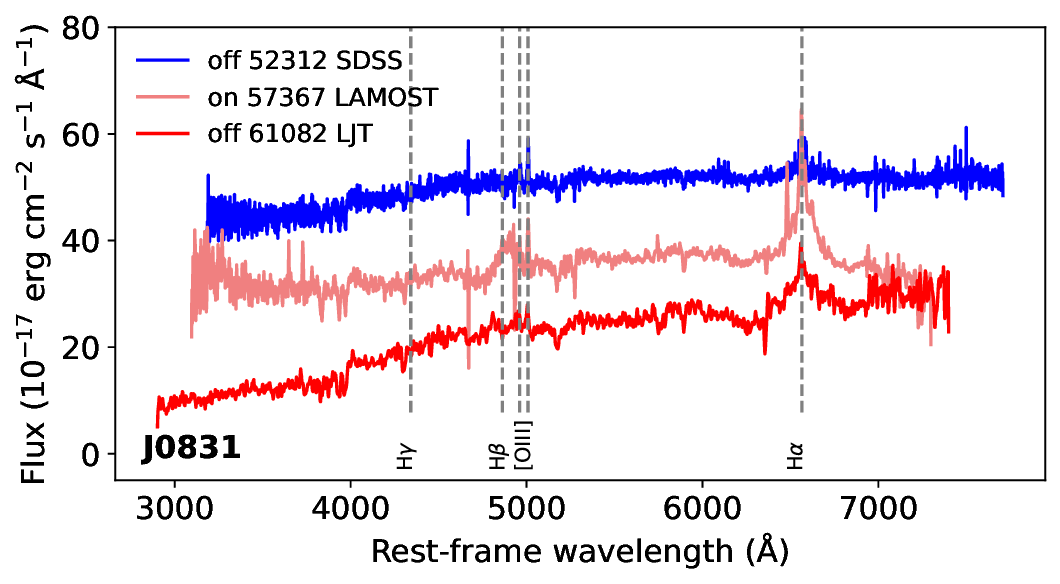}
	\includegraphics[width=0.49\textwidth]{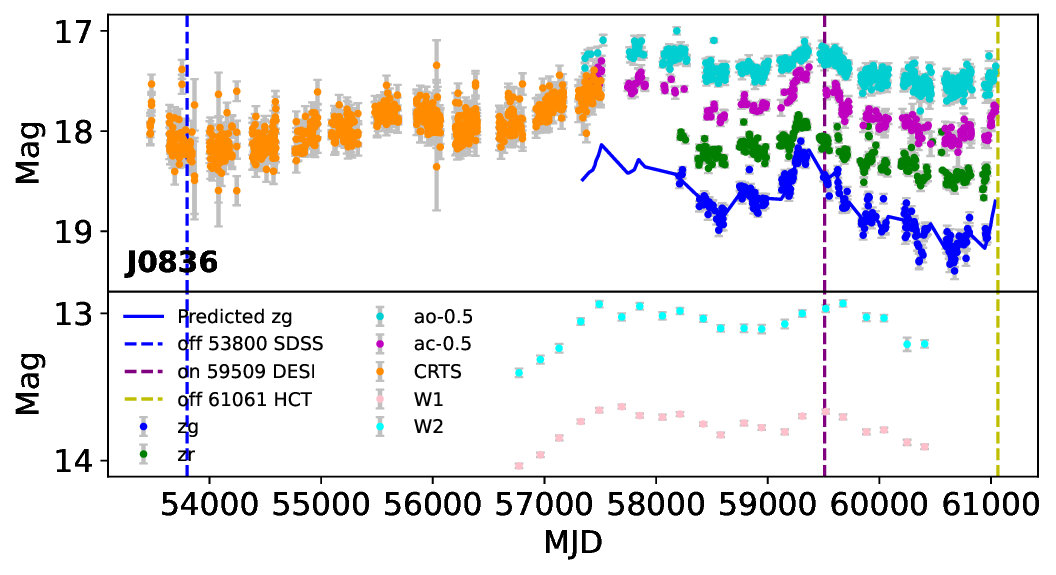}
	\includegraphics[width=0.48\textwidth]{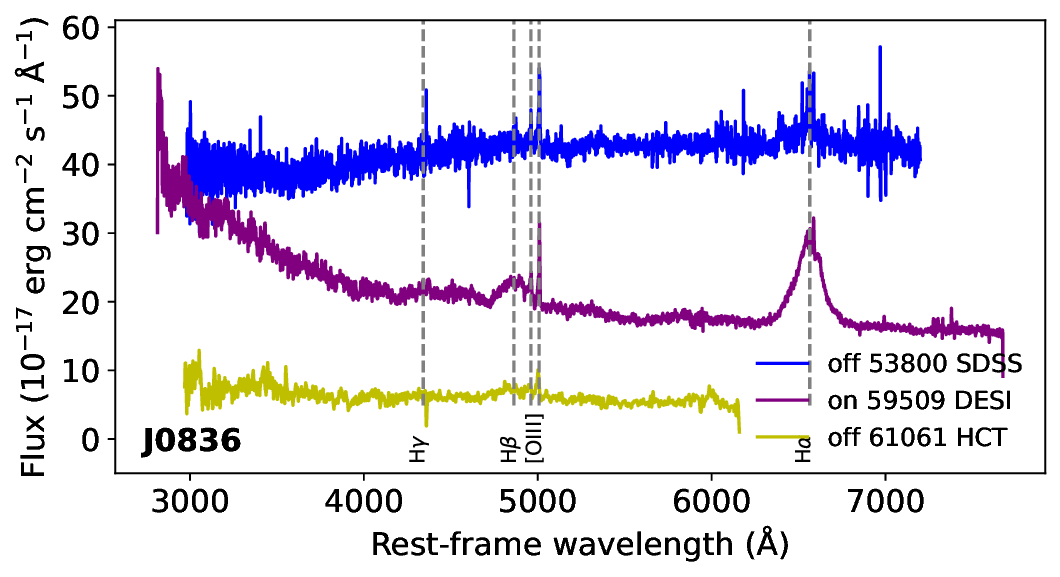}
	\includegraphics[width=0.49\textwidth]{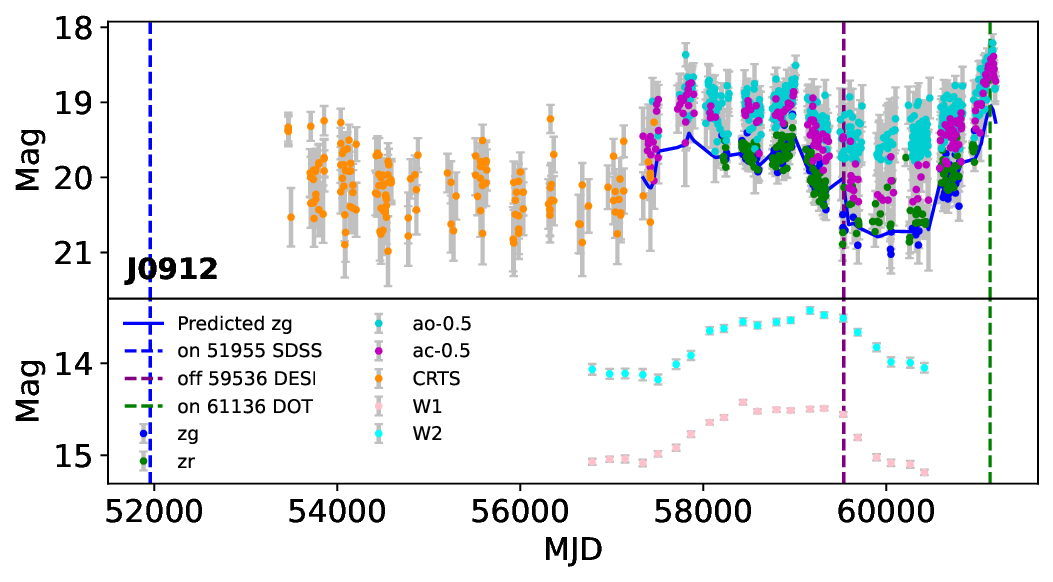}
	\includegraphics[width=0.48\textwidth]{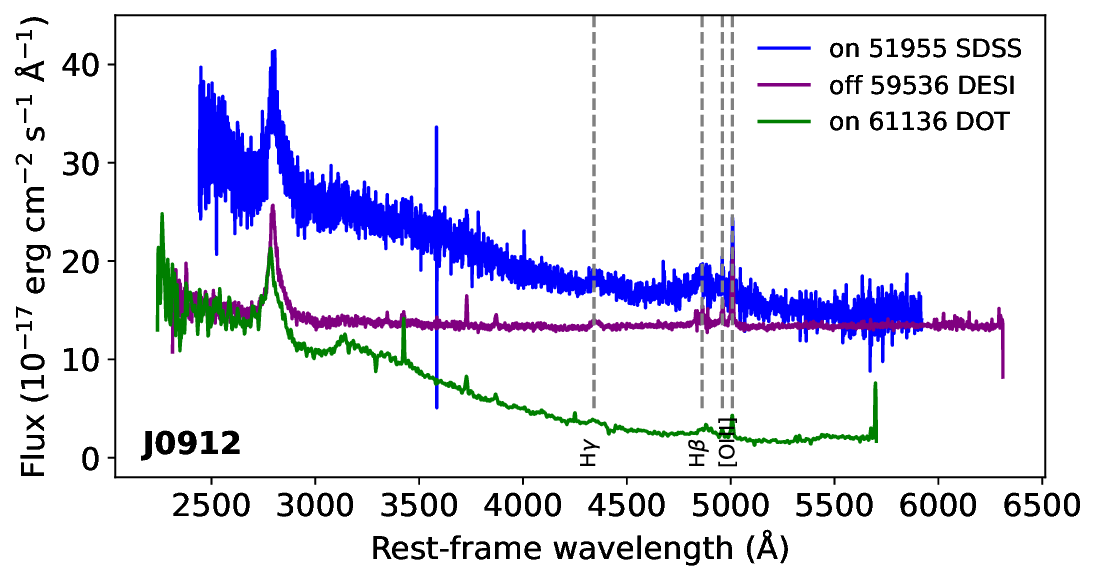}
	\caption{Light curves ({\it left}) and spectra ({\it right}) of 
	25 confirmed RCL AGNs. In the left panels, vertical dashed lines 
	mark the epochs of the spectra shown in the right panel. Main emission
	lines of the spectra are marked with vertical dashed lines. 
	For clarity, some spectra have been smoothed and vertically shifted.}
	\label{fig:crcl}
	\end{figure*}
\begin{figure*}[htbp]
	\figurenum{B1}
	\centering
	\includegraphics[width=0.5\textwidth]{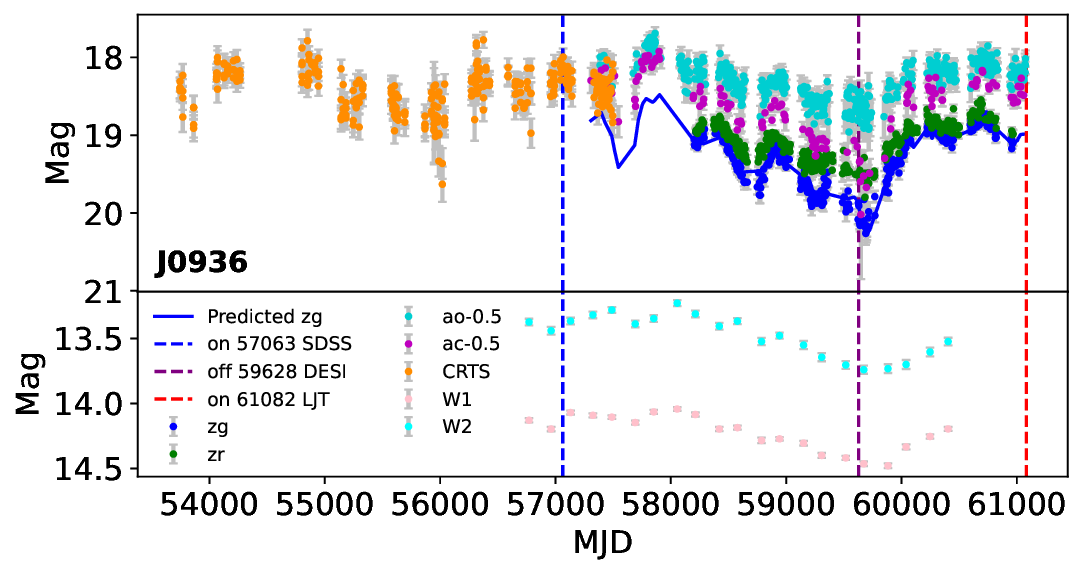}
	\includegraphics[width=0.48\textwidth]{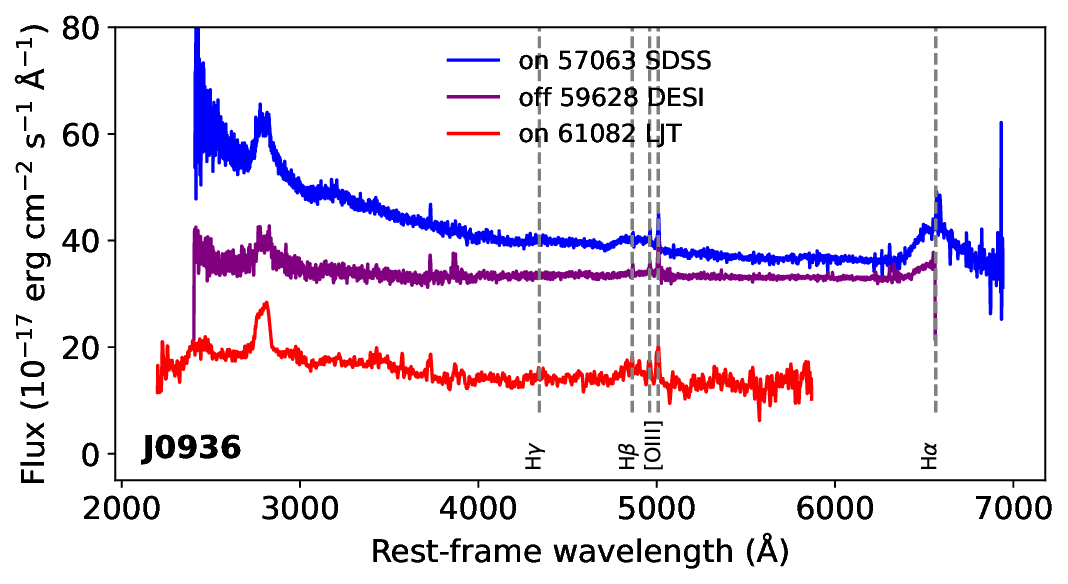}
	\includegraphics[width=0.5\textwidth]{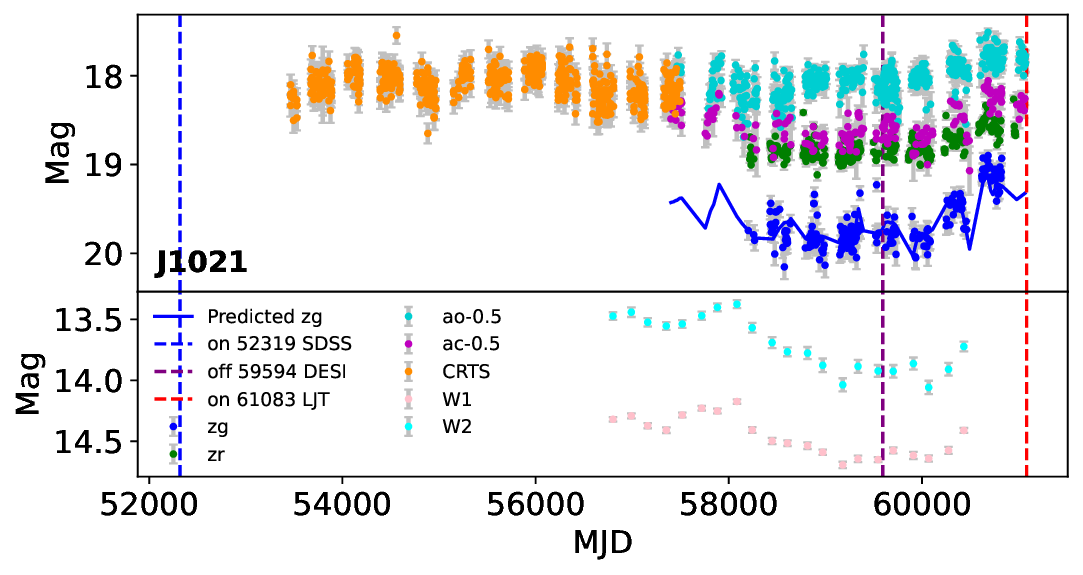}
	\includegraphics[width=0.48\textwidth]{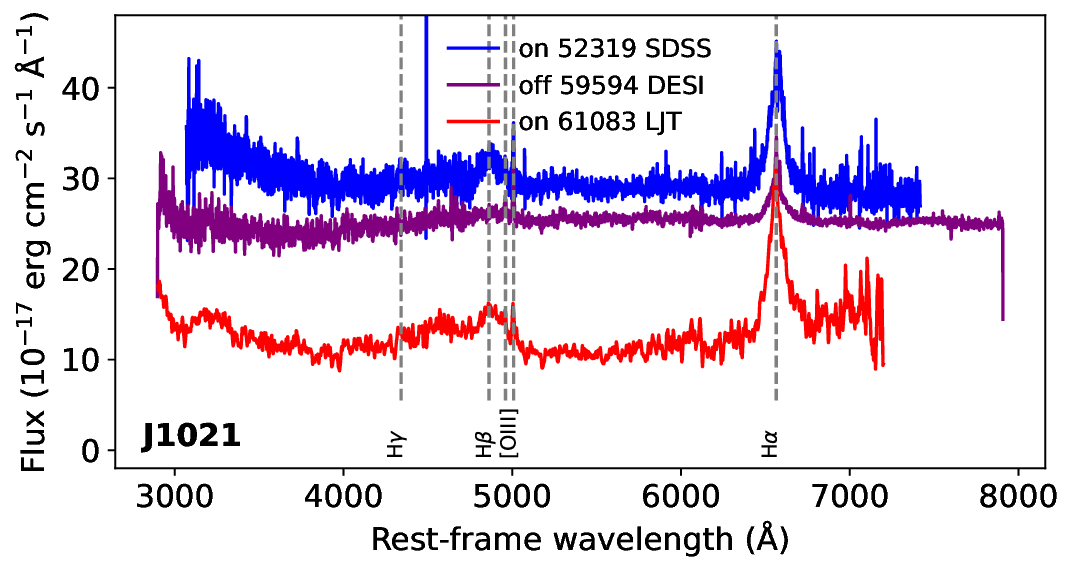}
	\includegraphics[width=0.5\textwidth]{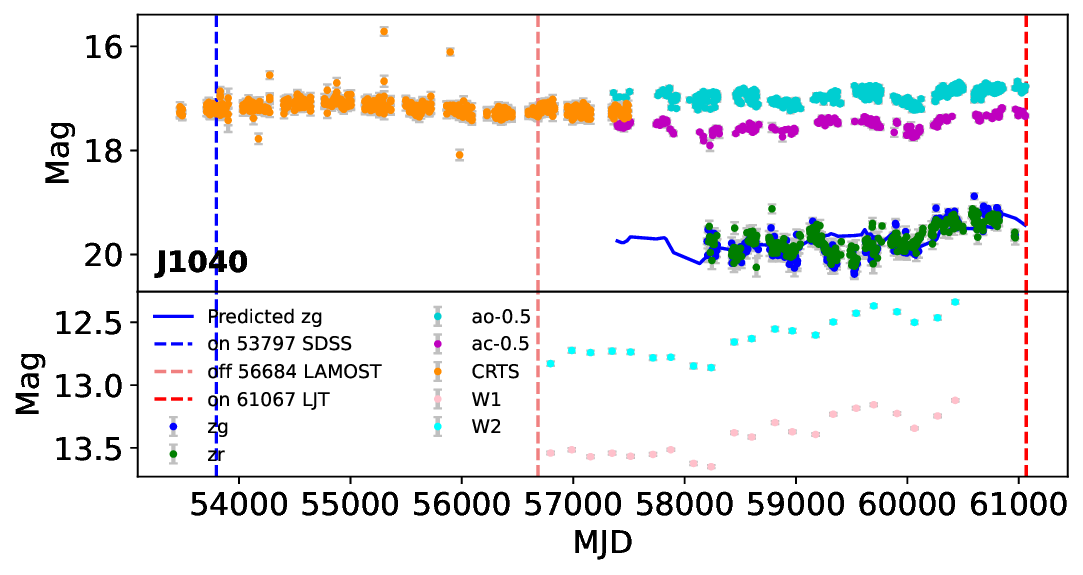}
	\includegraphics[width=0.48\textwidth]{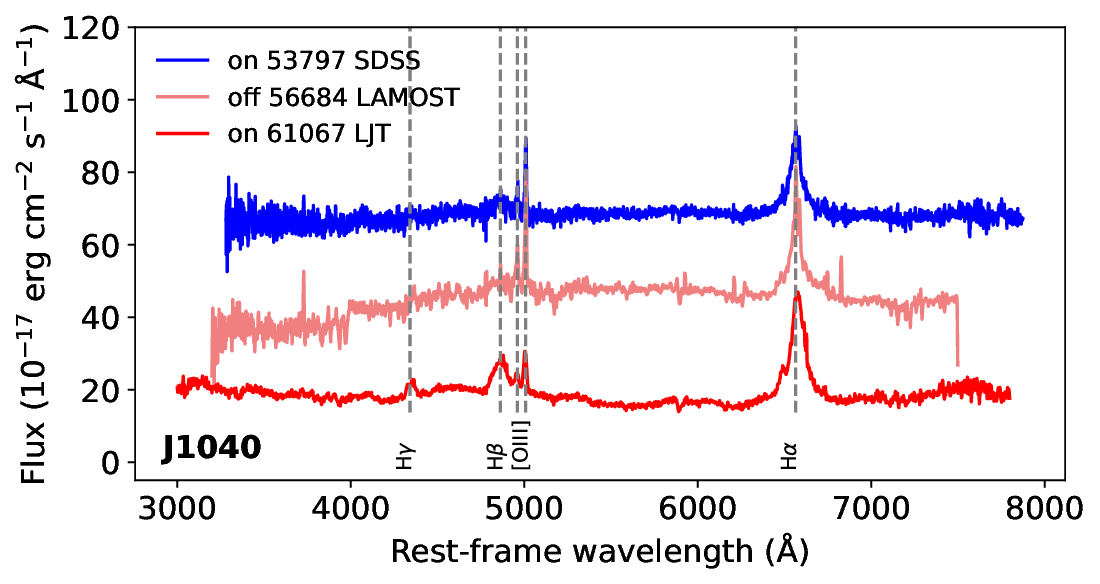}
	\includegraphics[width=0.5\textwidth]{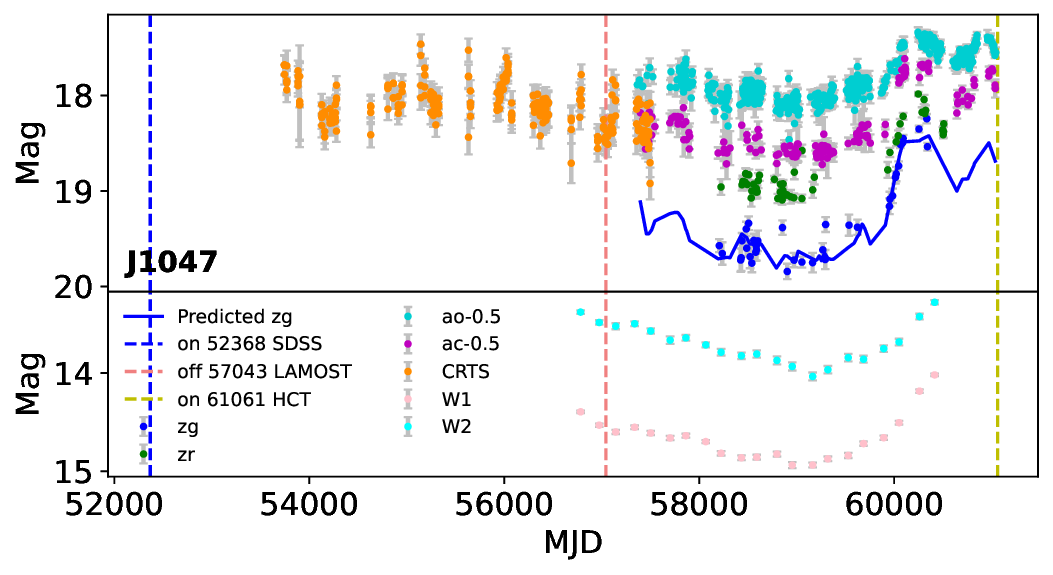}
	\includegraphics[width=0.48\textwidth]{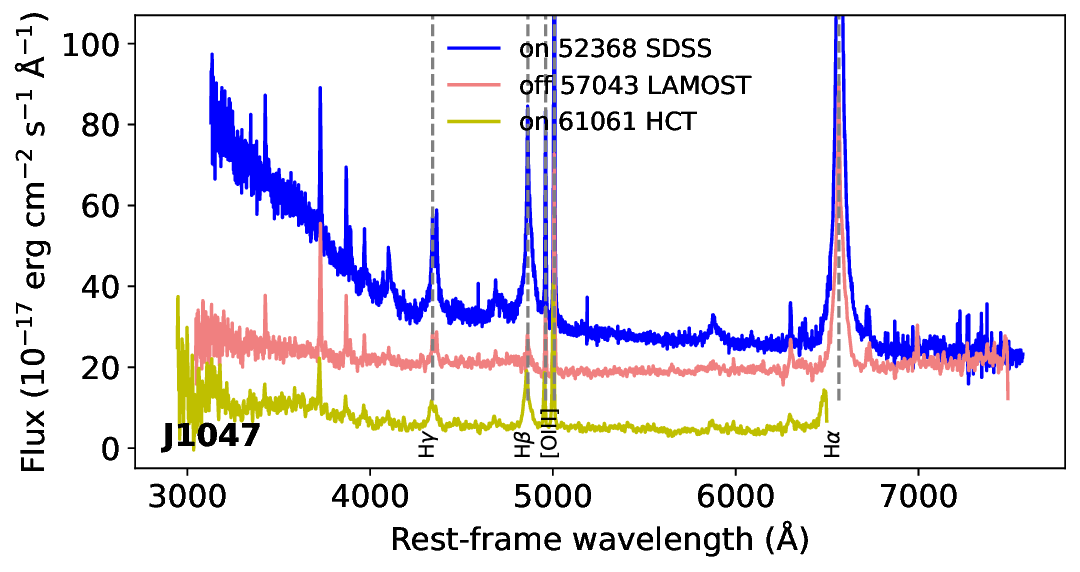}
	\includegraphics[width=0.5\textwidth]{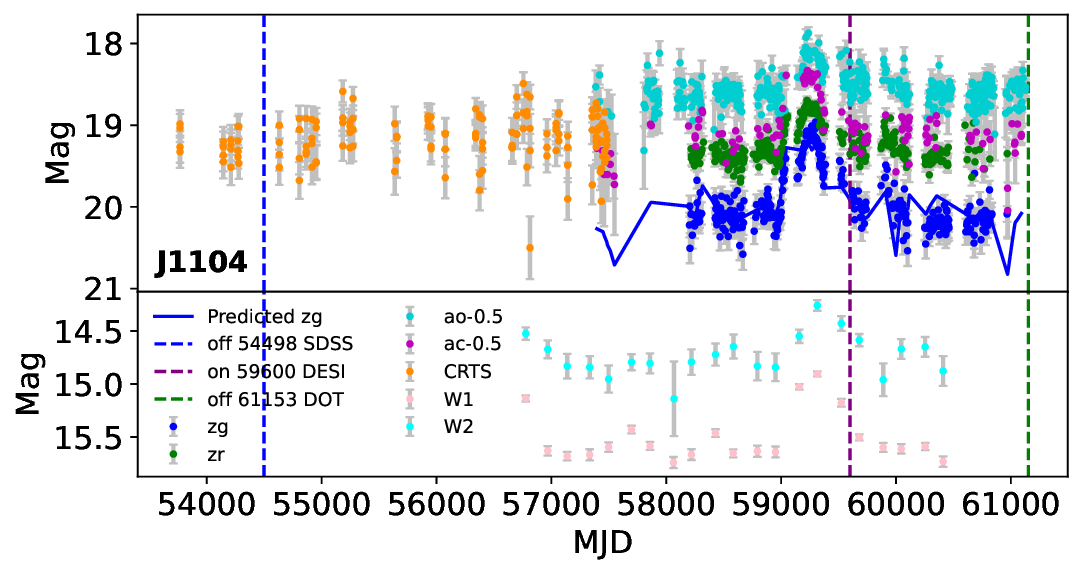}
	\includegraphics[width=0.48\textwidth]{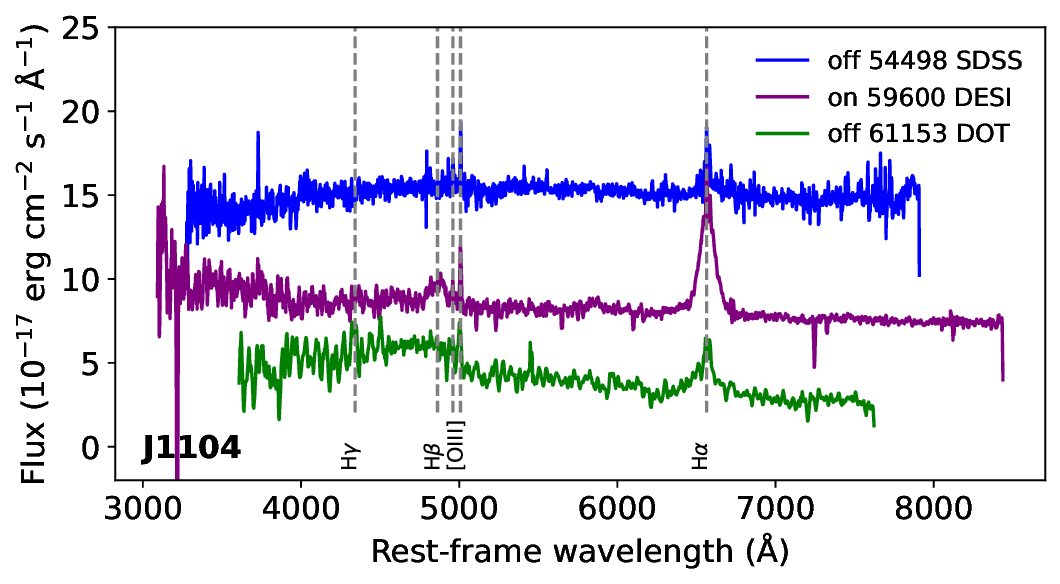}
	\addtocounter{figure}{-1}
	\caption{Continued.}
	\end{figure*}
\begin{figure*}[htbp]
	\figurenum{B1}
	\centering	
	\includegraphics[width=0.5\textwidth]{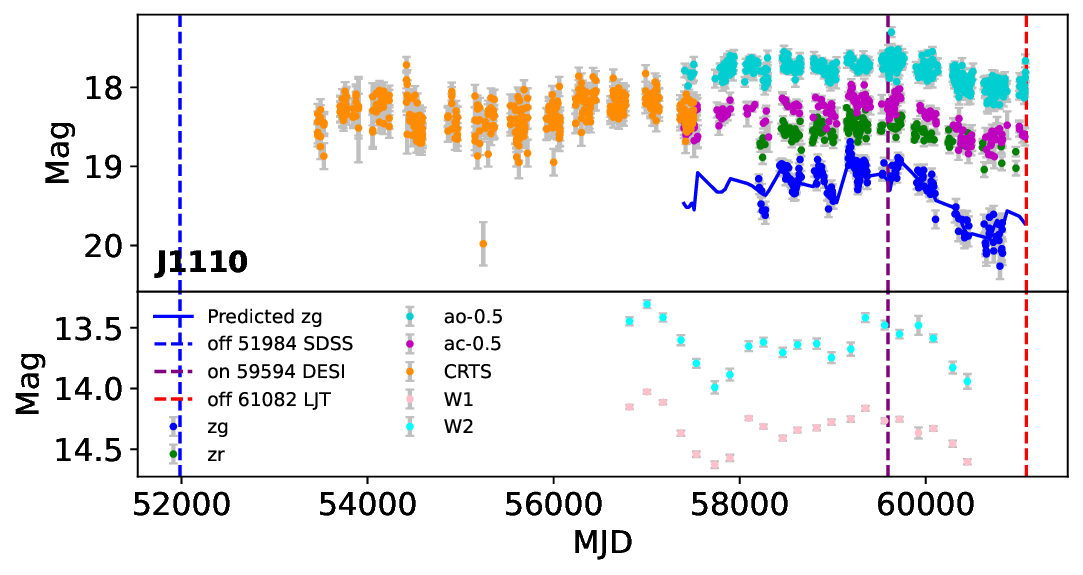}
	\includegraphics[width=0.48\textwidth]{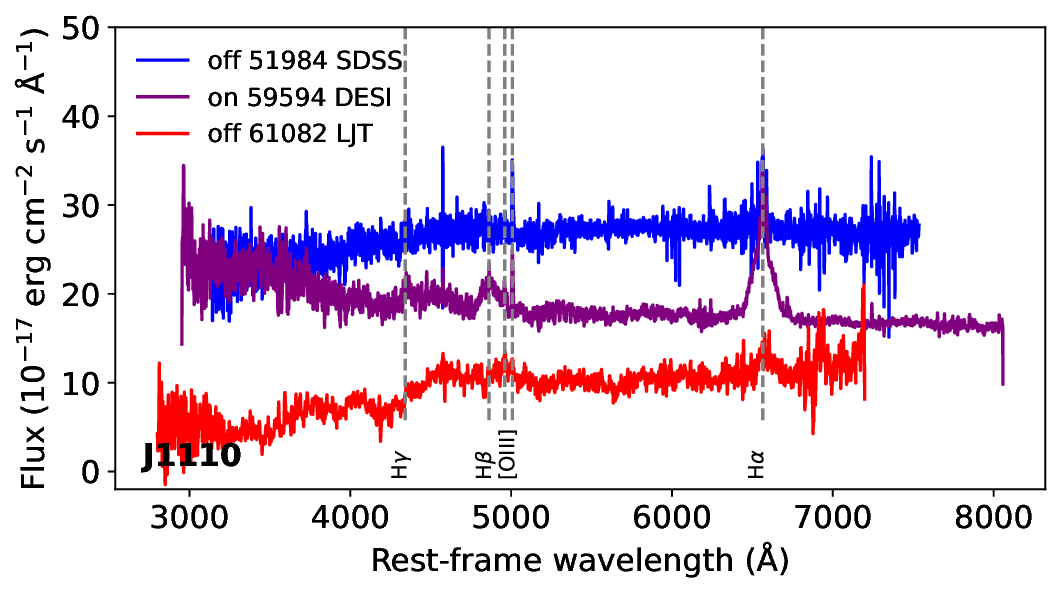}
	\includegraphics[width=0.5\textwidth]{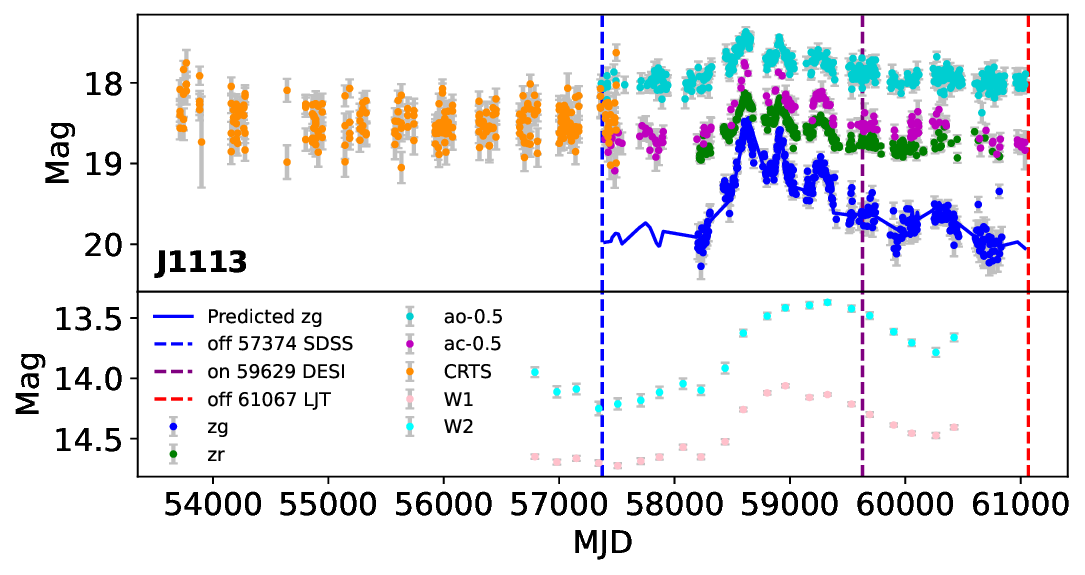}
	\includegraphics[width=0.48\textwidth]{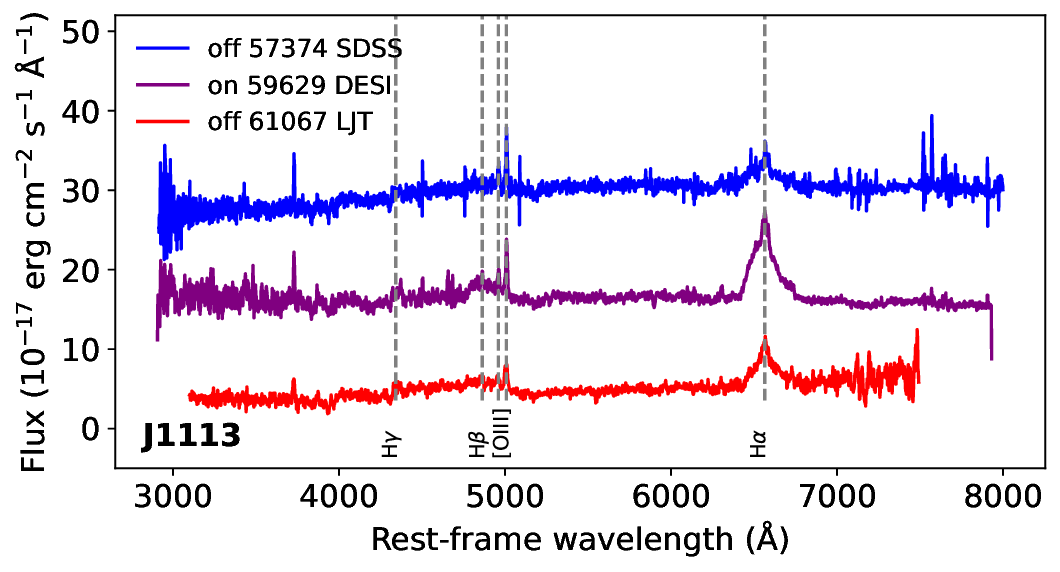}
	\includegraphics[width=0.5\textwidth]{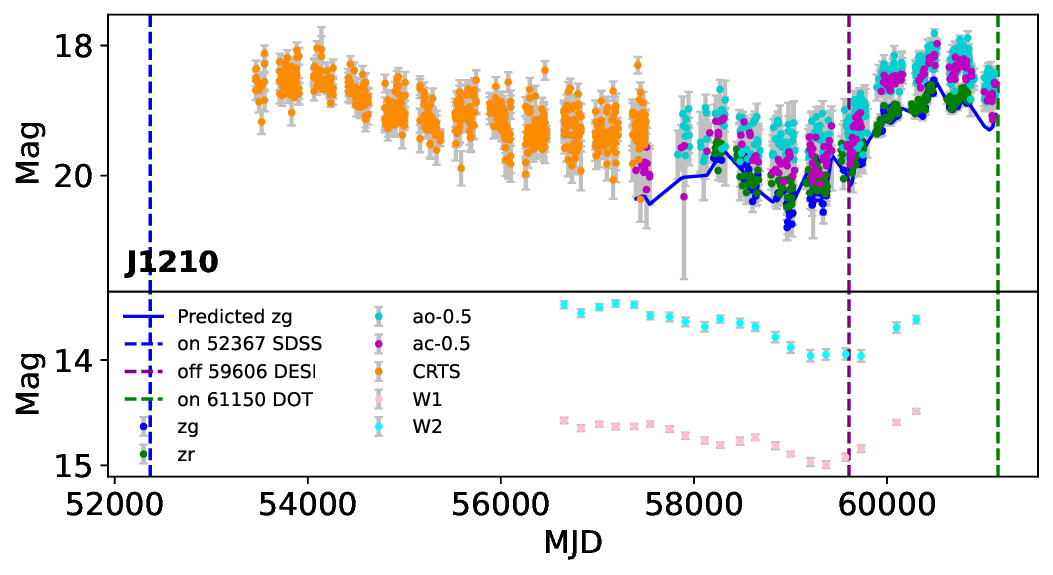}
	\includegraphics[width=0.48\textwidth]{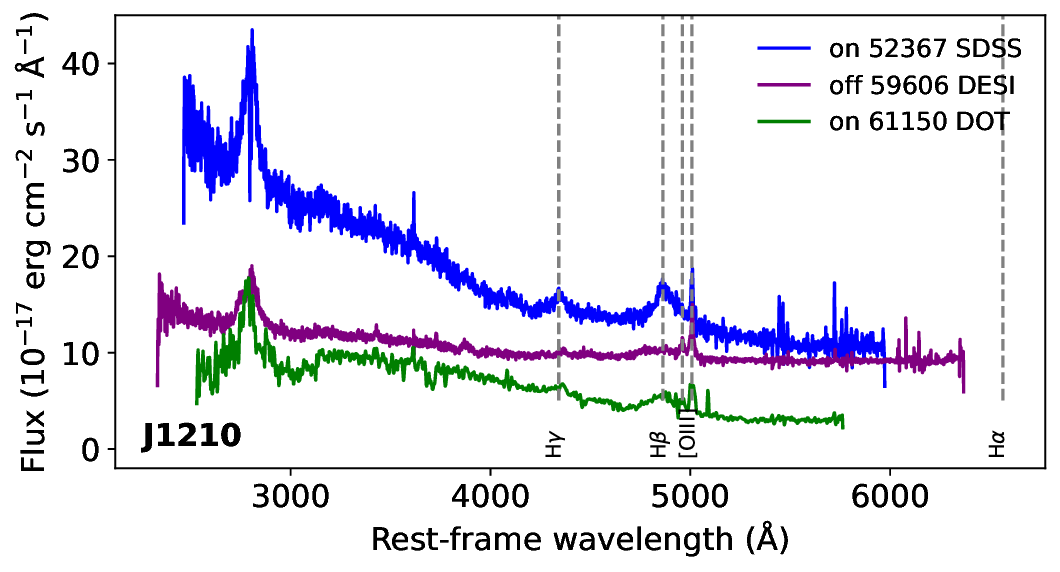}
	\includegraphics[width=0.5\textwidth]{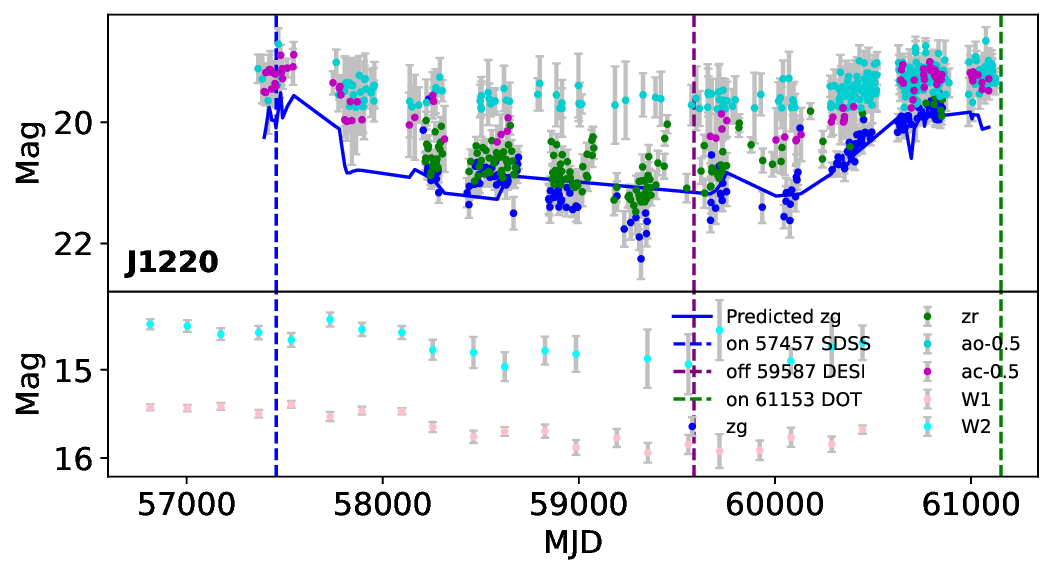}
	\includegraphics[width=0.48\textwidth]{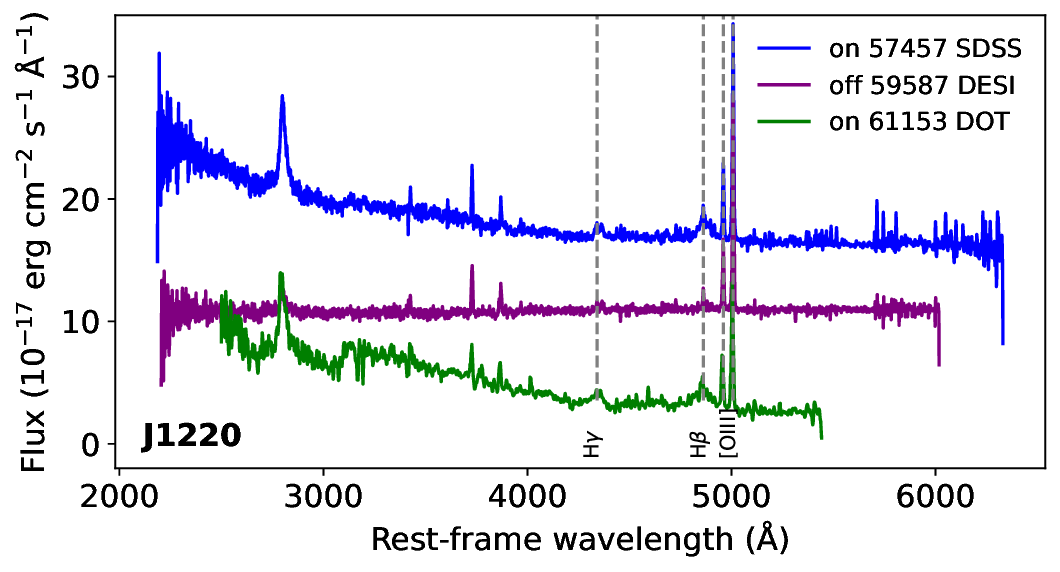}
	\includegraphics[width=0.5\textwidth]{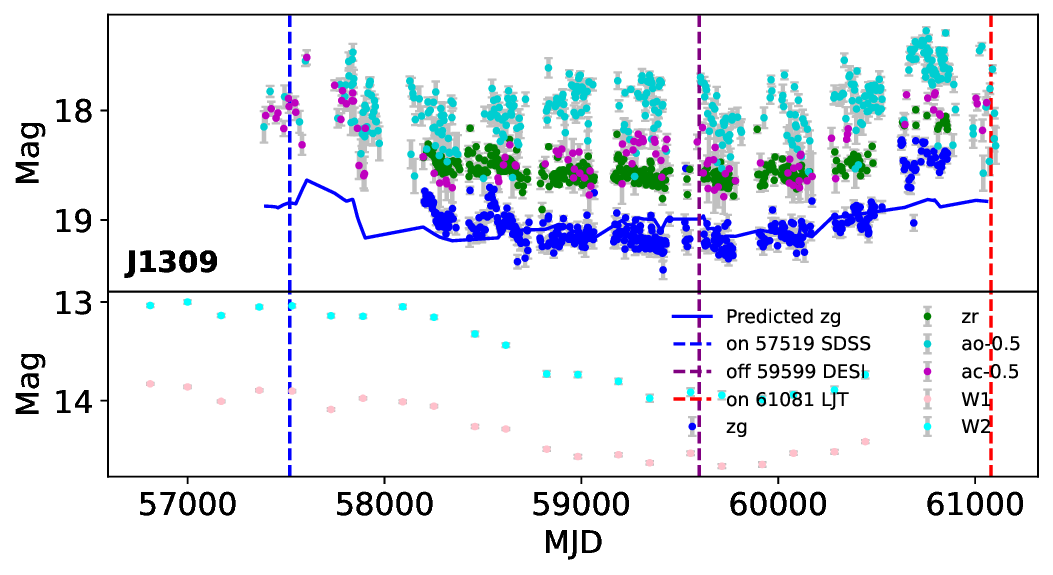}
	\includegraphics[width=0.48\textwidth]{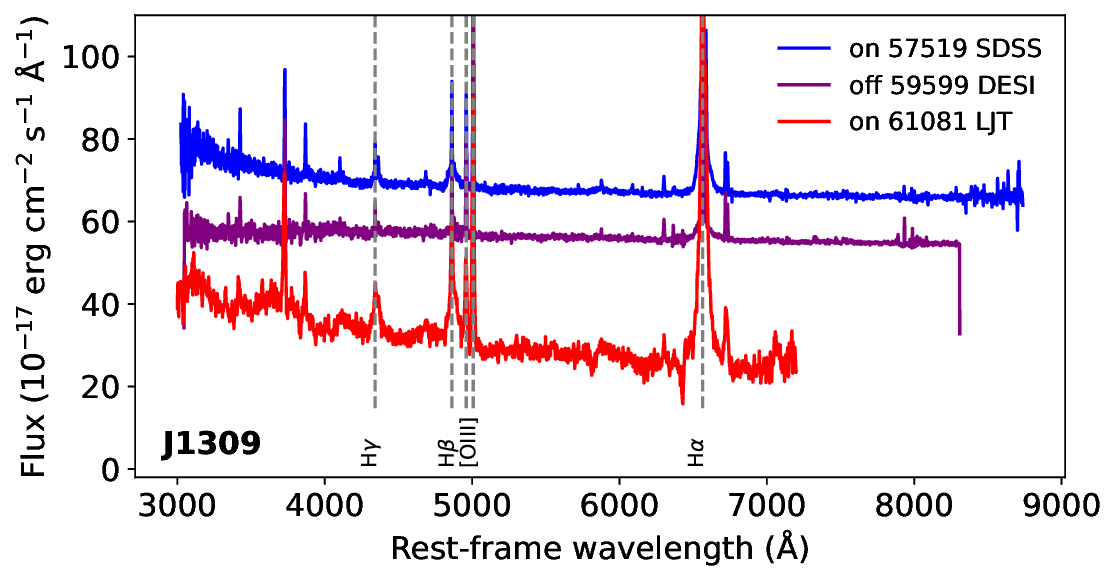}
	\addtocounter{figure}{-1}
	\caption{Continued.}
	\end{figure*}
\begin{figure*}[htbp]	
	\figurenum{B1}
	\centering	
	\includegraphics[width=0.5\textwidth]{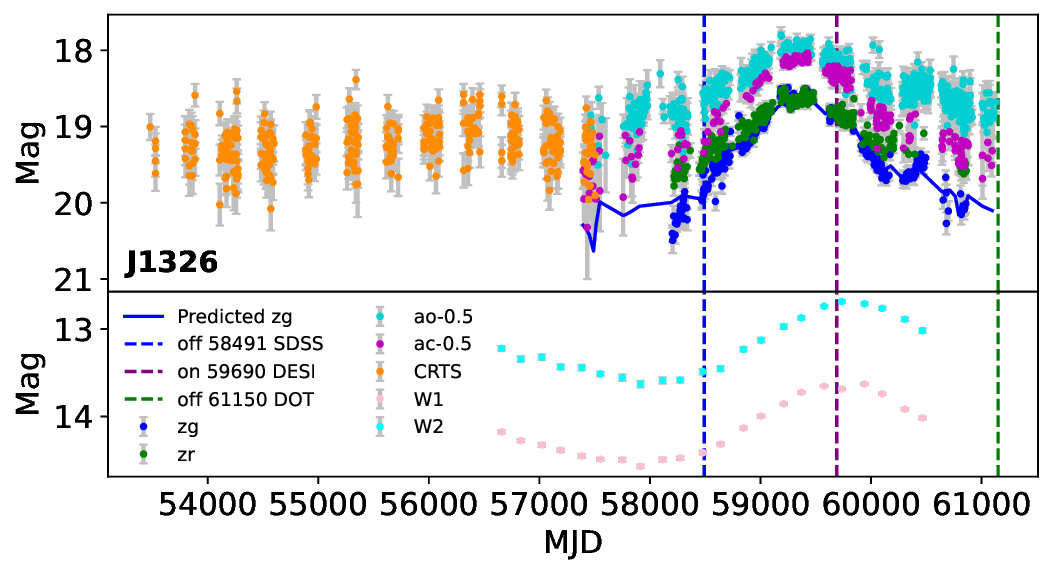}
	\includegraphics[width=0.48\textwidth]{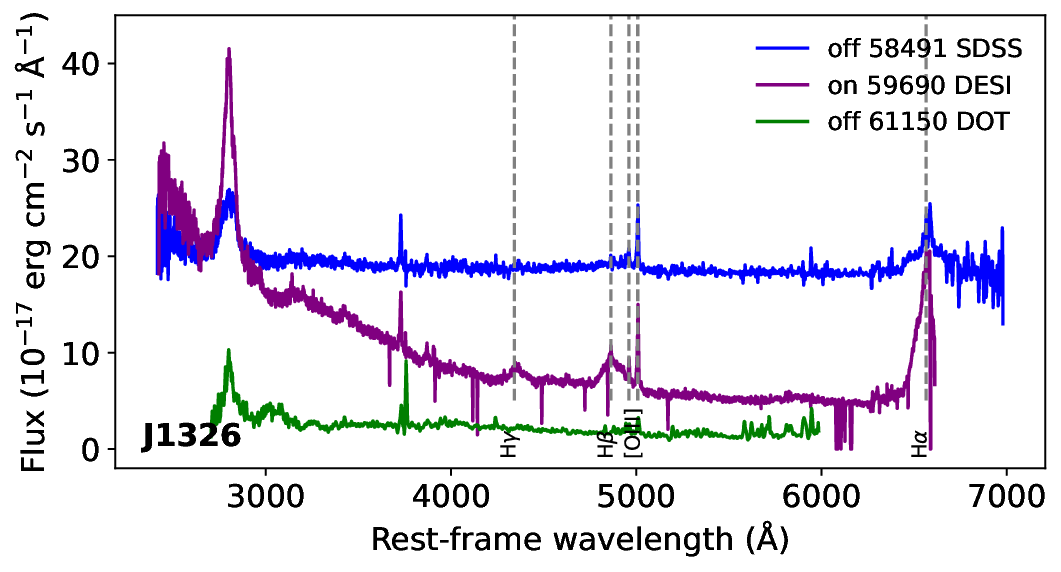}
	\includegraphics[width=0.5\textwidth]{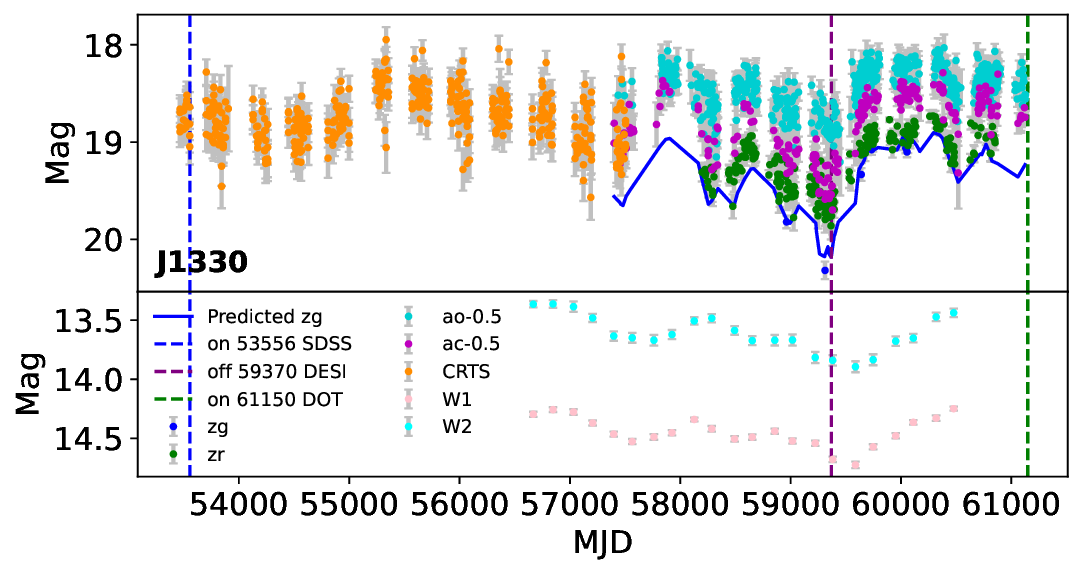}
	\includegraphics[width=0.48\textwidth]{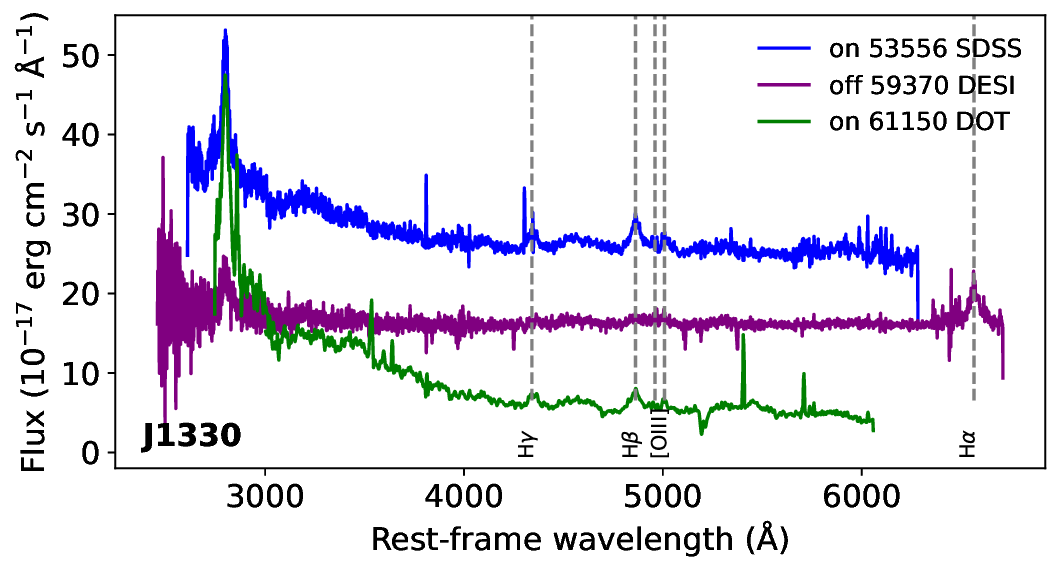}
	\includegraphics[width=0.5\textwidth]{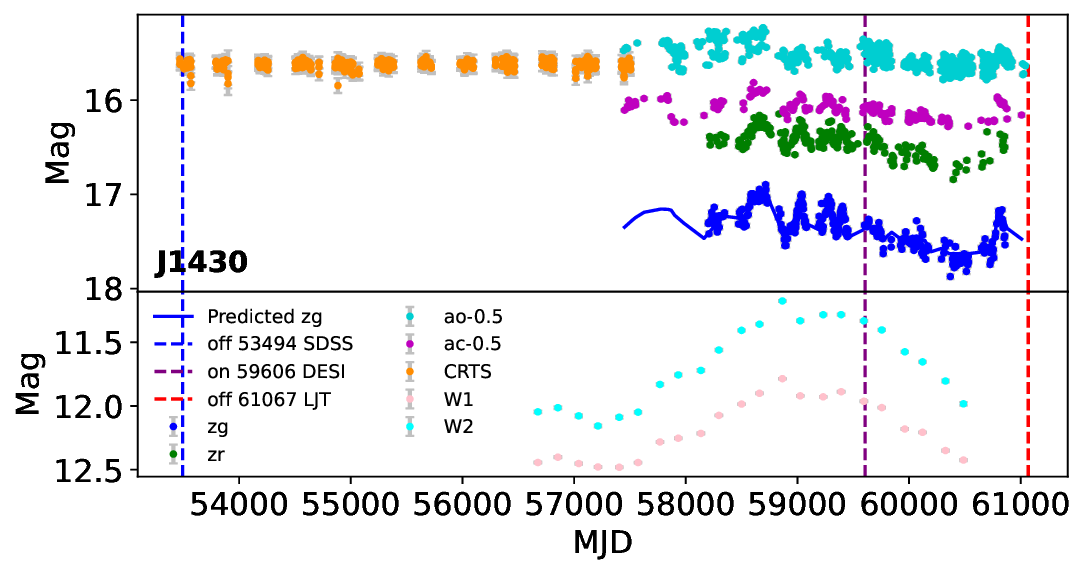}
	\includegraphics[width=0.48\textwidth]{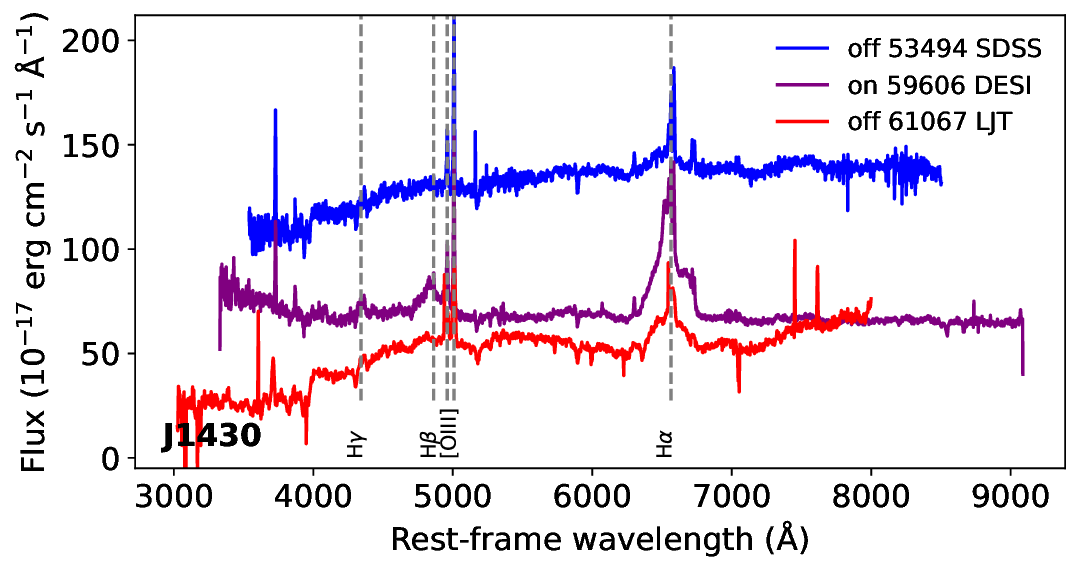}
	\includegraphics[width=0.5\textwidth]{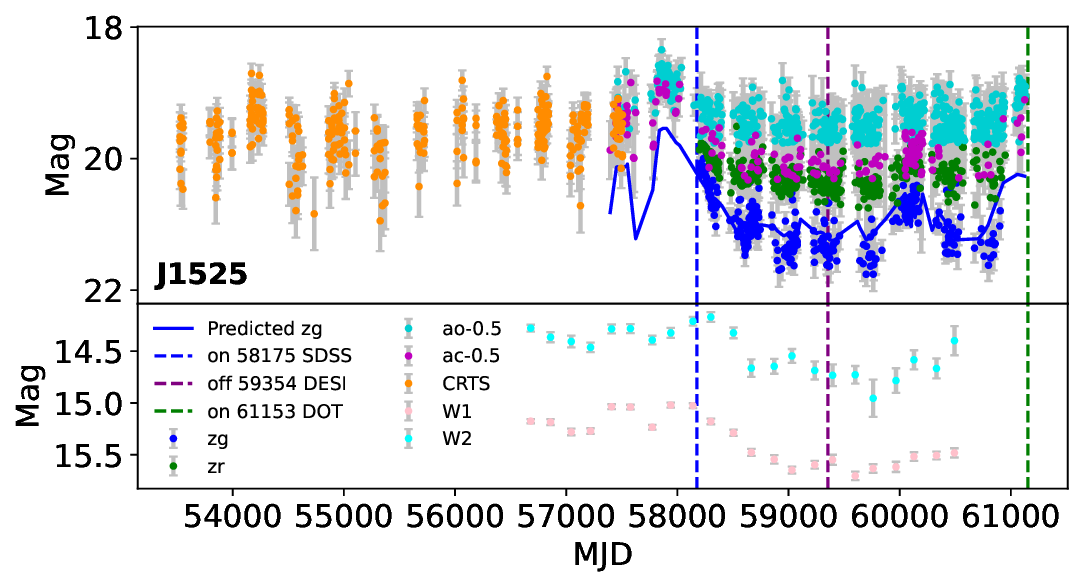}
	\includegraphics[width=0.48\textwidth]{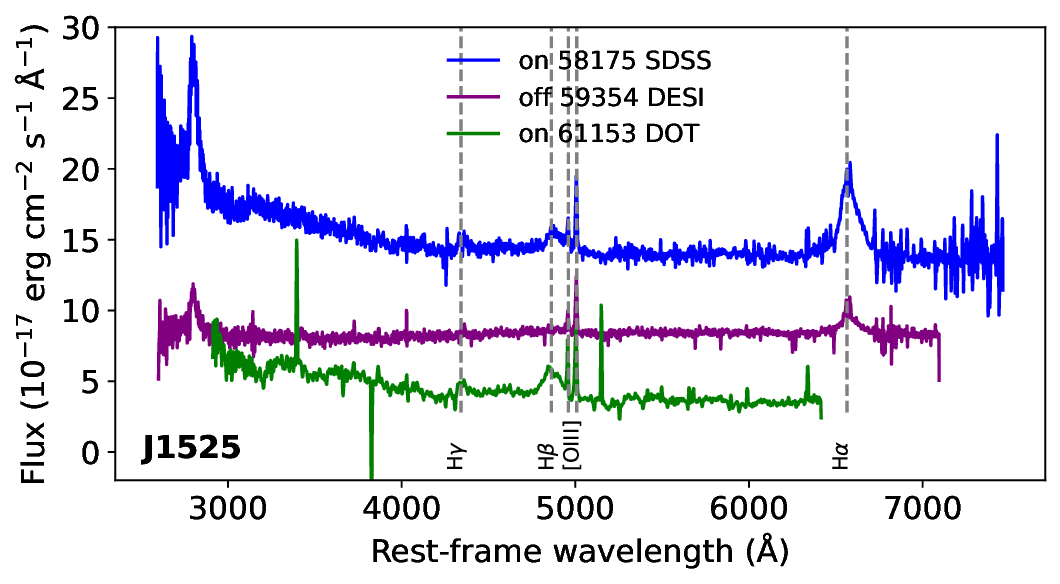}
	\includegraphics[width=0.5\textwidth]{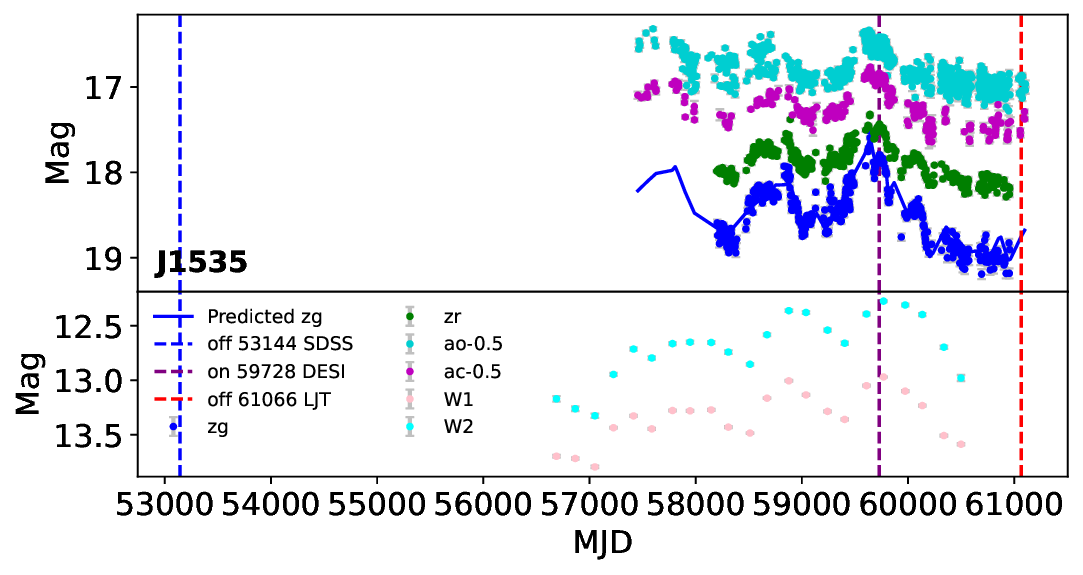}
	\includegraphics[width=0.48\textwidth]{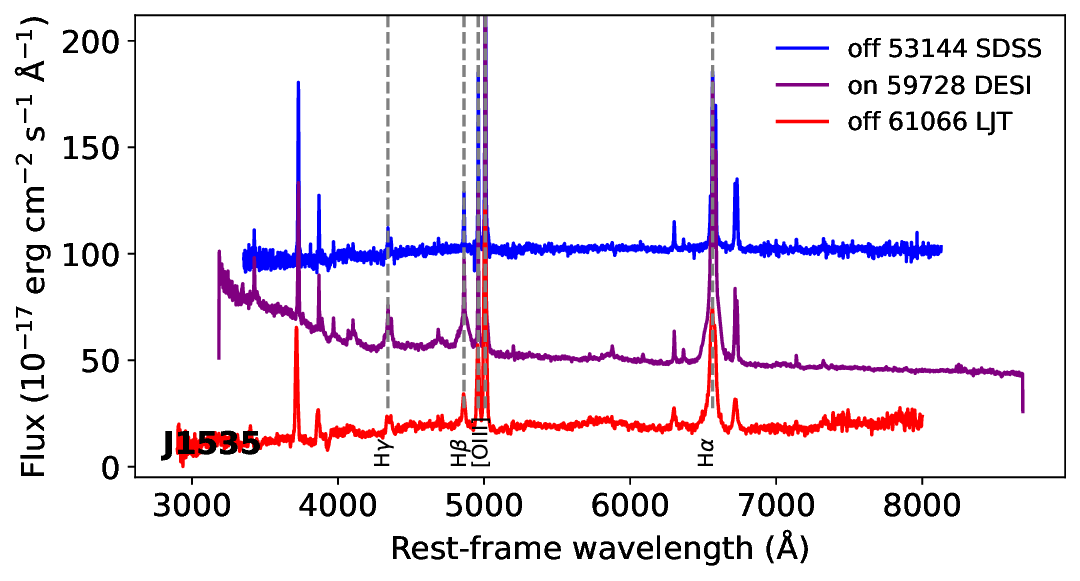}
	\addtocounter{figure}{-1}
	\caption{Continued.}
\end{figure*}
\begin{figure*}[htbp]	
	\figurenum{B1}
	\centering	
	\includegraphics[width=0.5\textwidth]{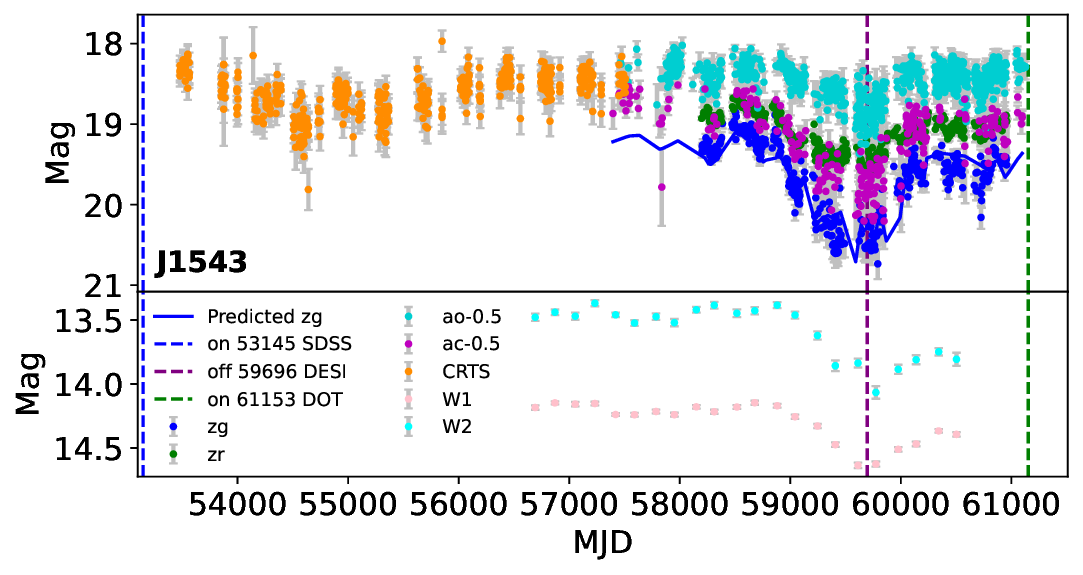}
	\includegraphics[width=0.48\textwidth]{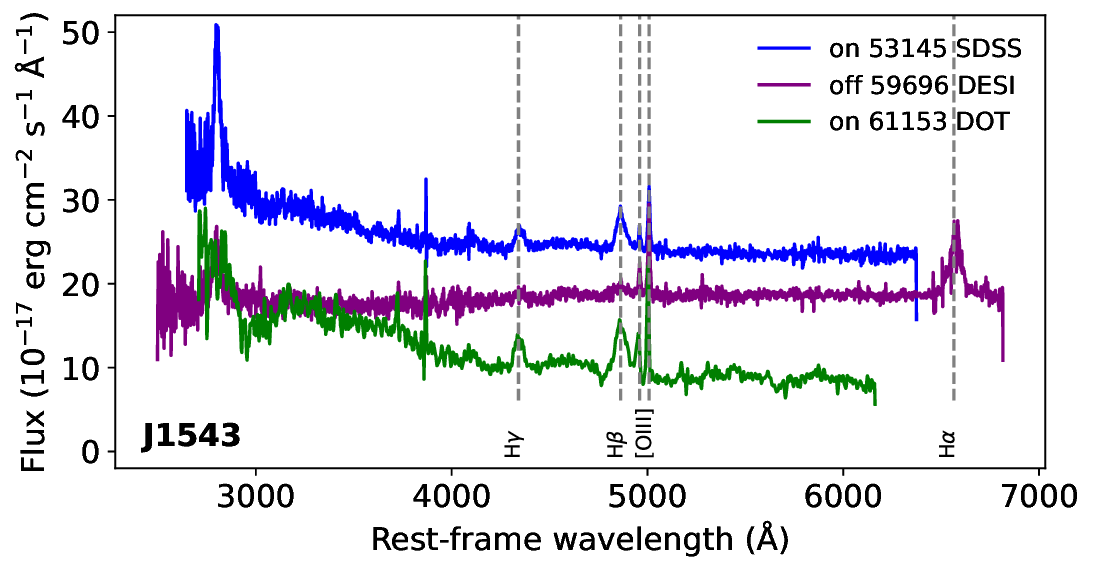}
	\includegraphics[width=0.5\textwidth]{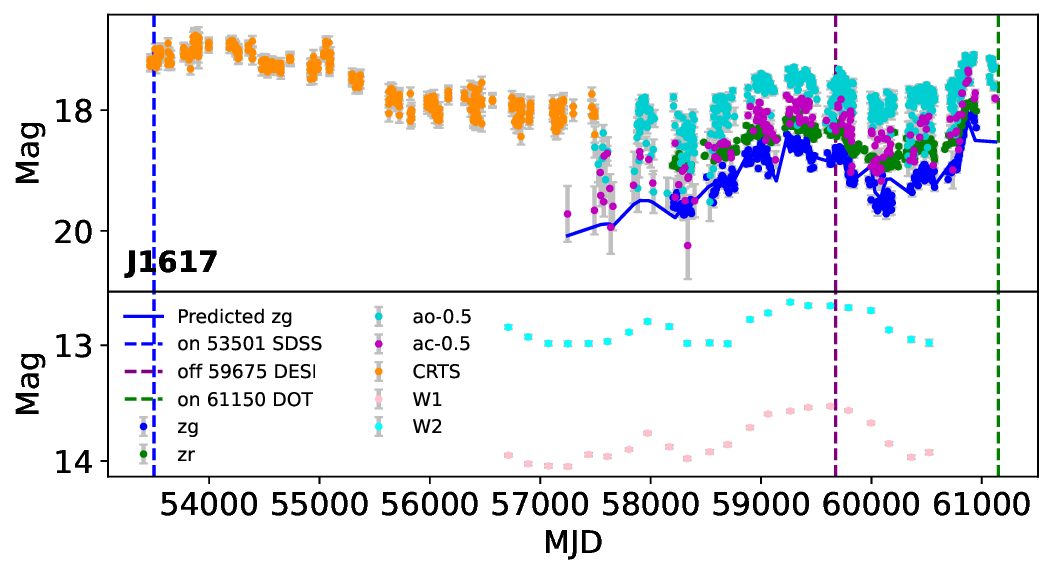}
	\includegraphics[width=0.48\textwidth]{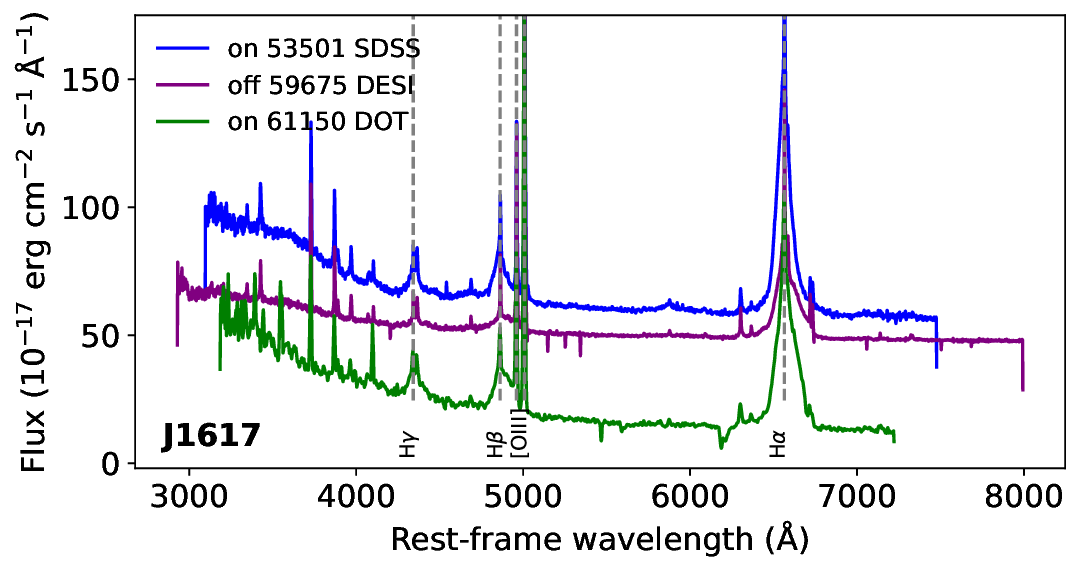}
	\includegraphics[width=0.5\textwidth]{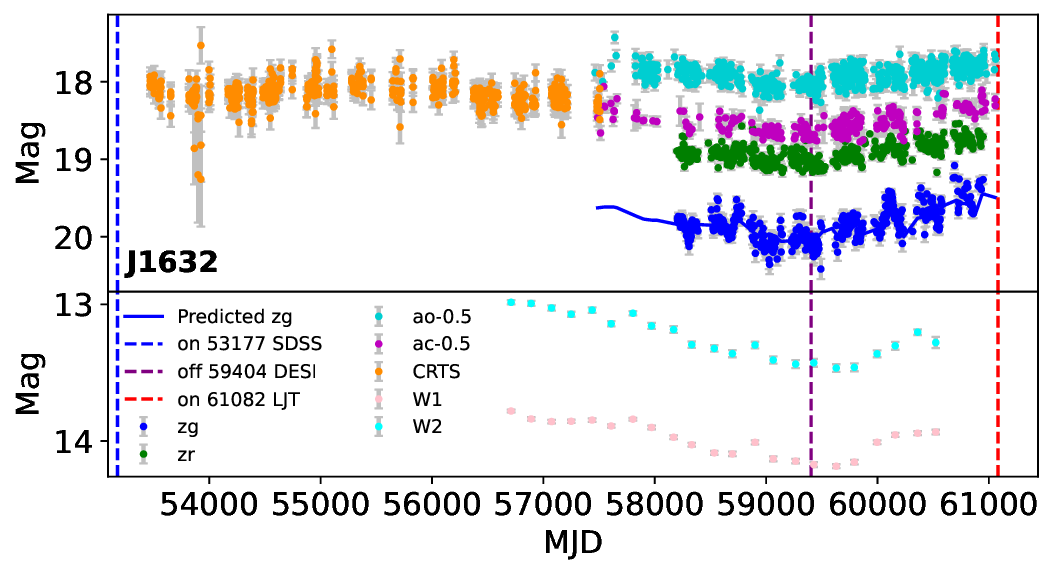}
	\includegraphics[width=0.48\textwidth]{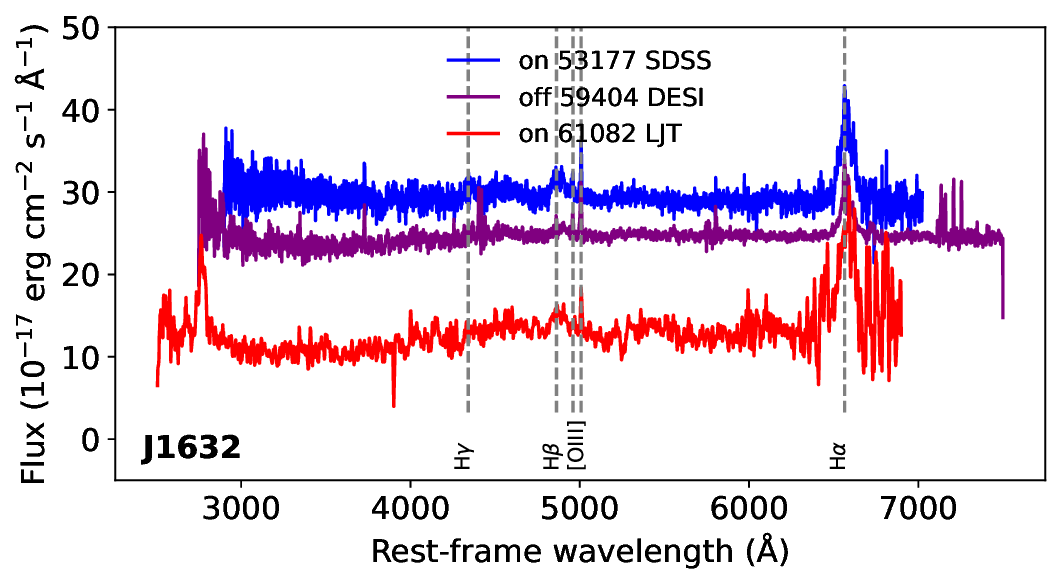}
	\includegraphics[width=0.5\textwidth]{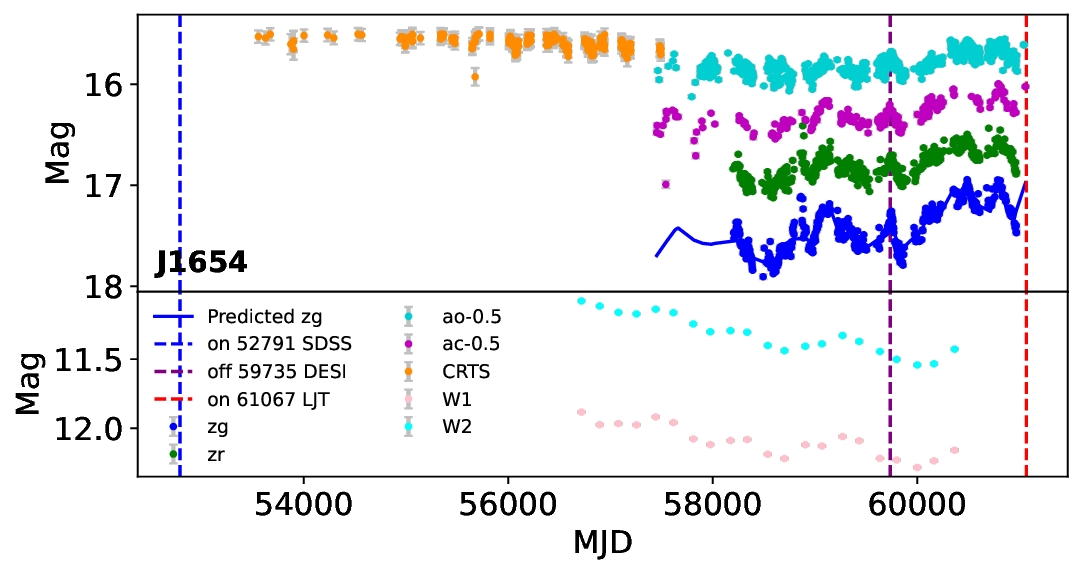}
	\includegraphics[width=0.48\textwidth]{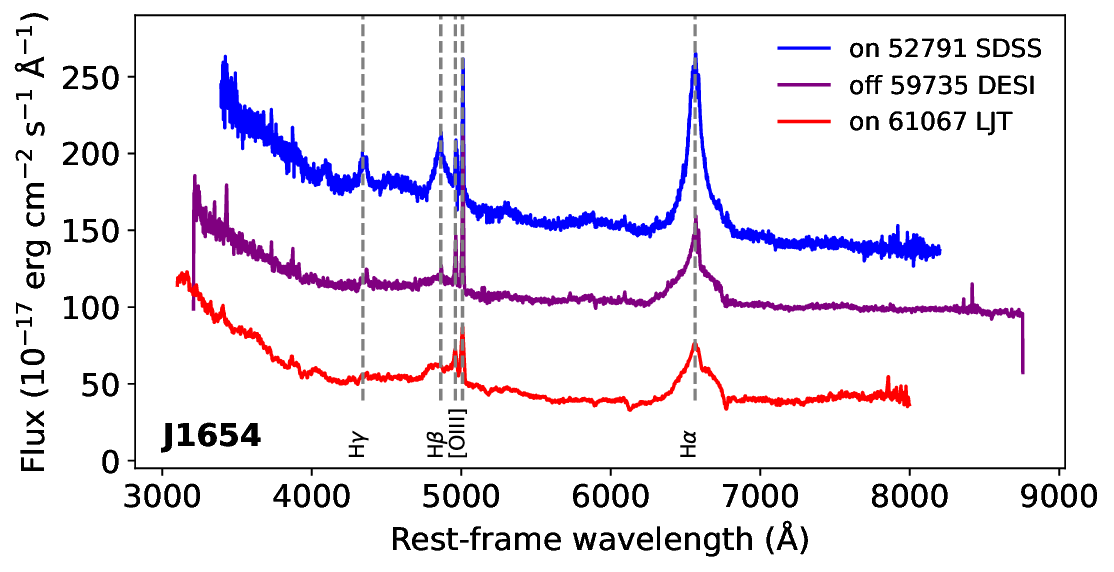}
	\includegraphics[width=0.5\textwidth]{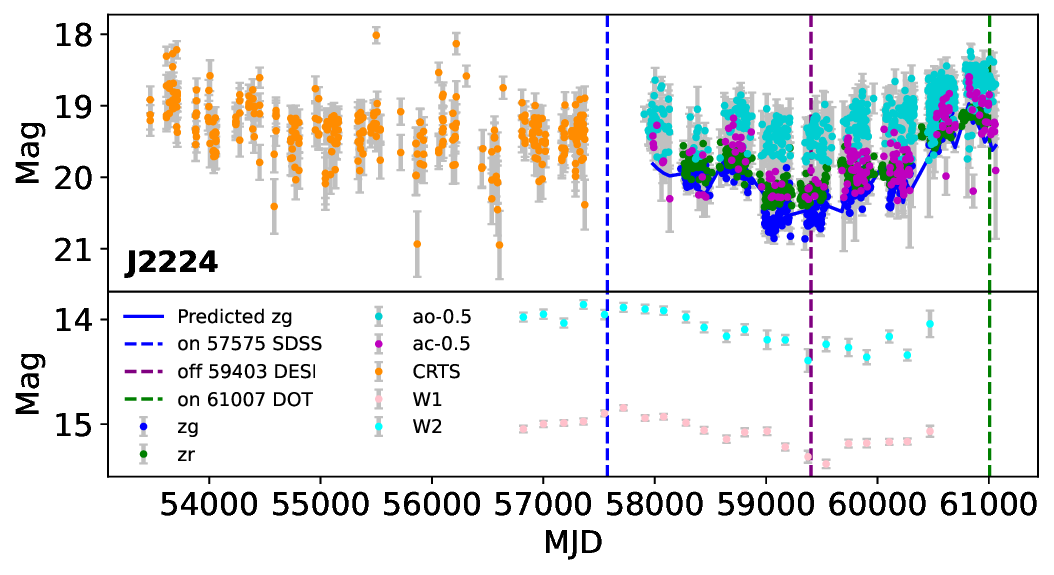}
	\includegraphics[width=0.48\textwidth]{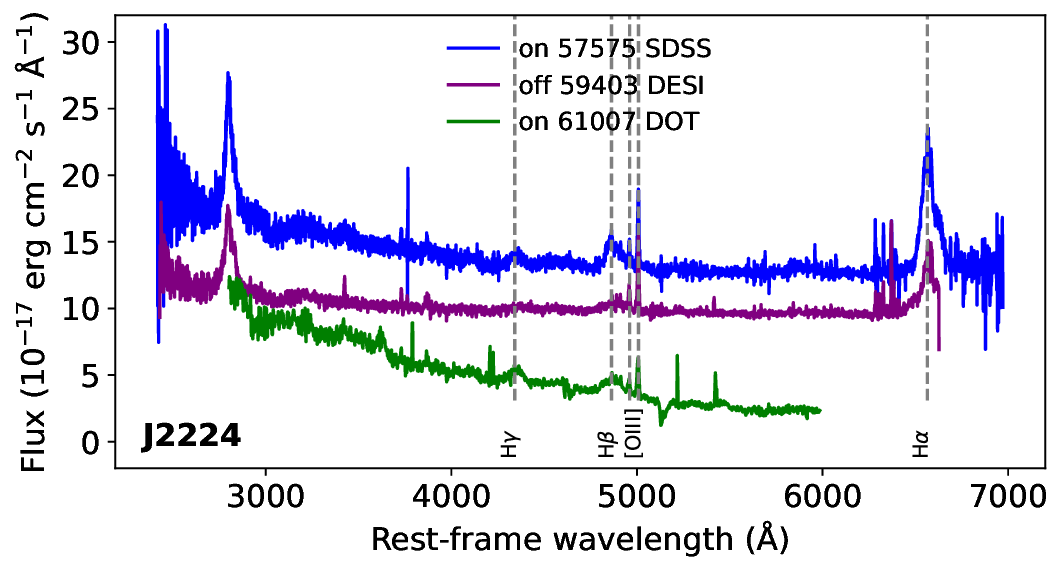}
	\addtocounter{figure}{-1}
	\caption{Continued.}
\end{figure*}

\begin{figure*}[htbp]
	\centering
	\includegraphics[width=0.5\textwidth]{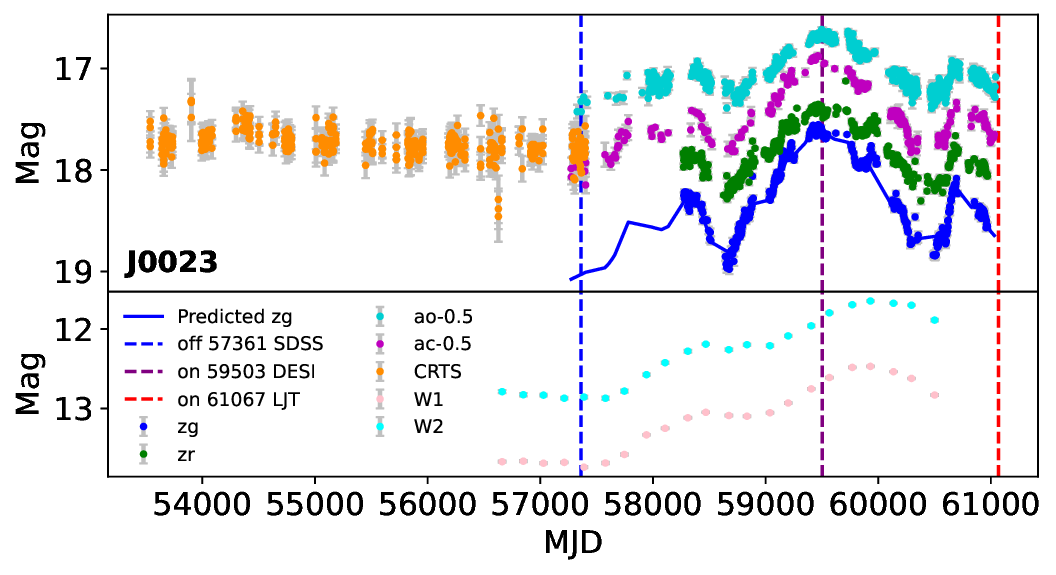}
	\includegraphics[width=0.48\textwidth]{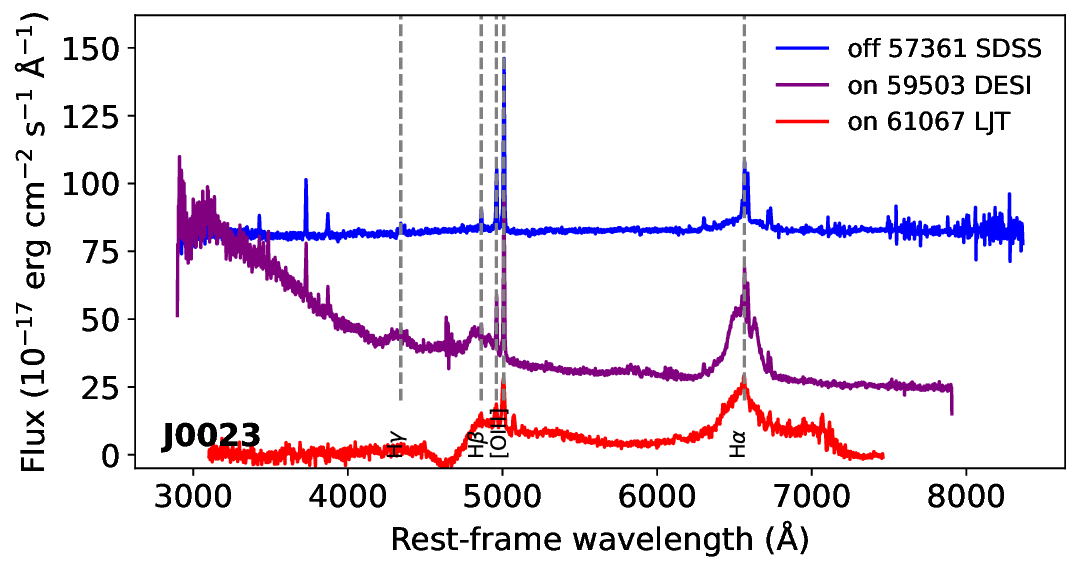}
	\includegraphics[width=0.5\textwidth]{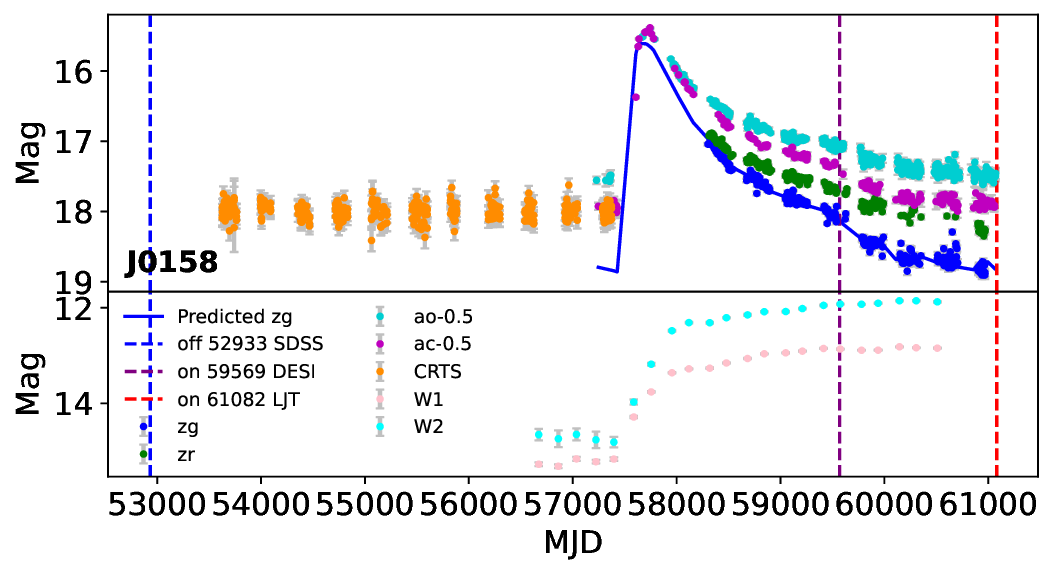}
	\includegraphics[width=0.48\textwidth]{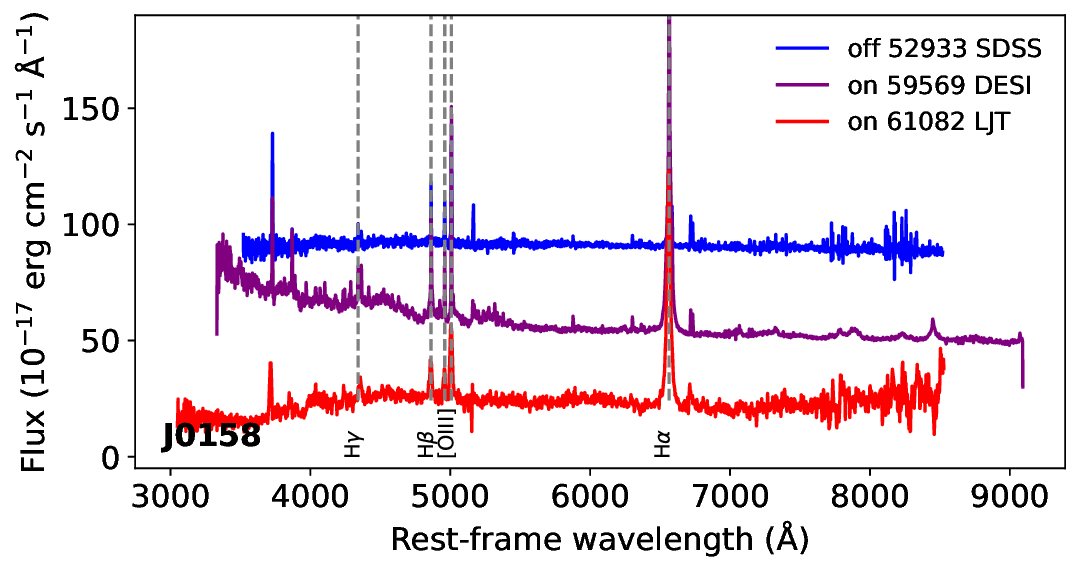}
	\includegraphics[width=0.5\textwidth]{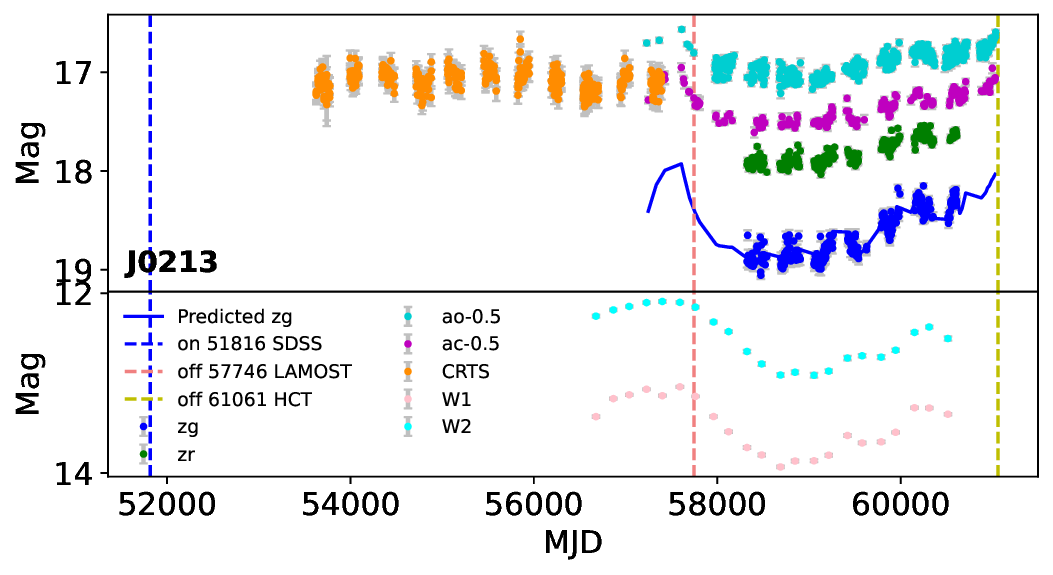}
	\includegraphics[width=0.48\textwidth]{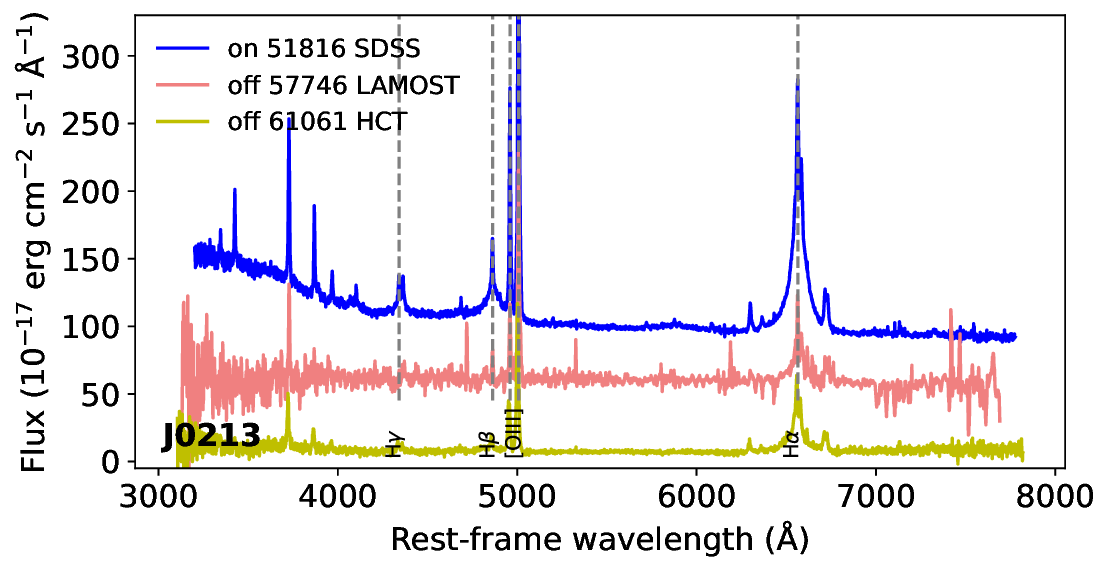}
	\includegraphics[width=0.5\textwidth]{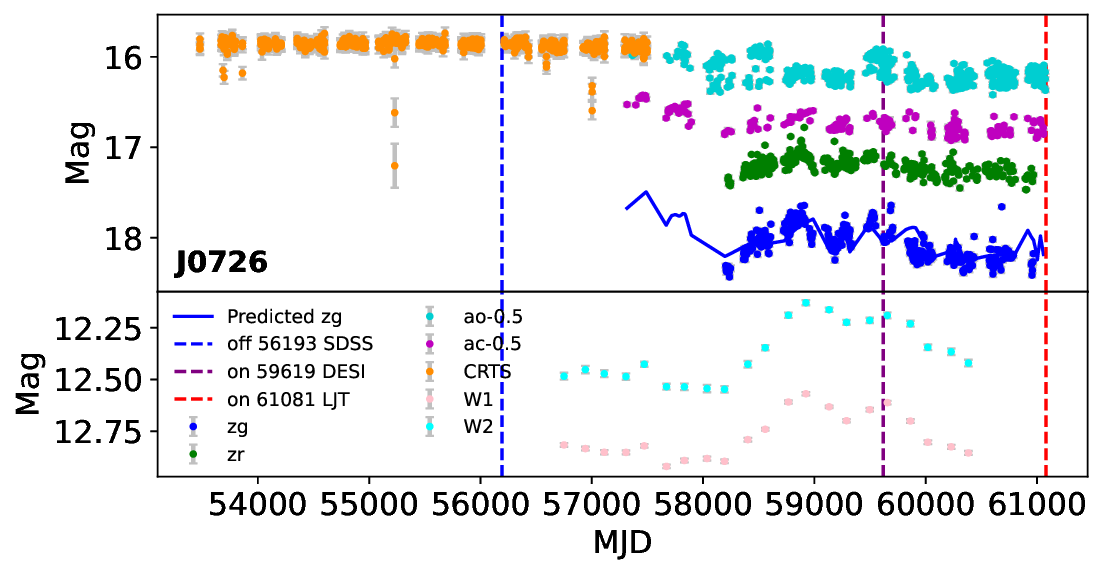}
	\includegraphics[width=0.48\textwidth]{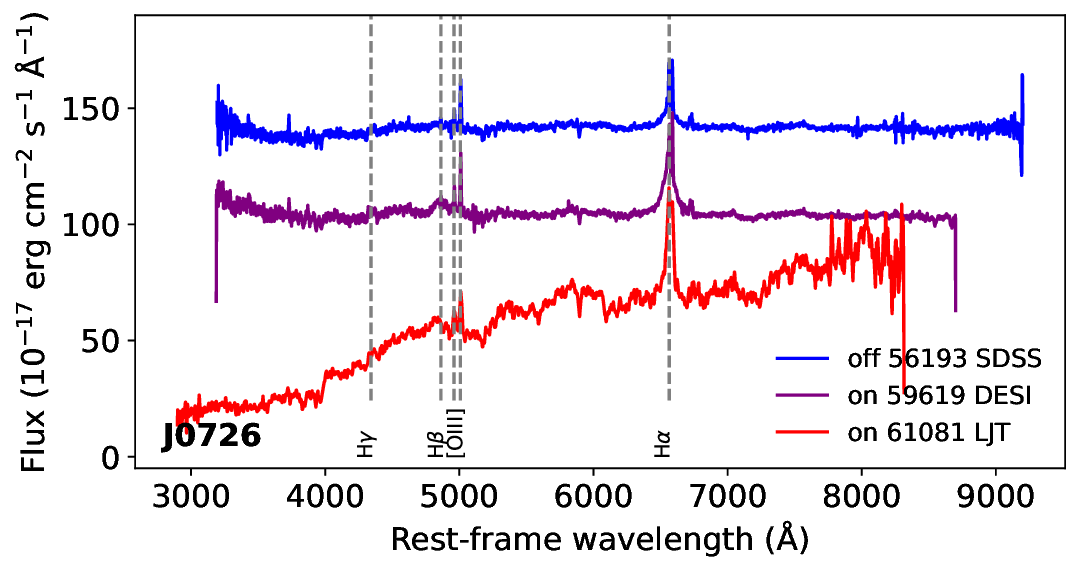}
	\includegraphics[width=0.5\textwidth]{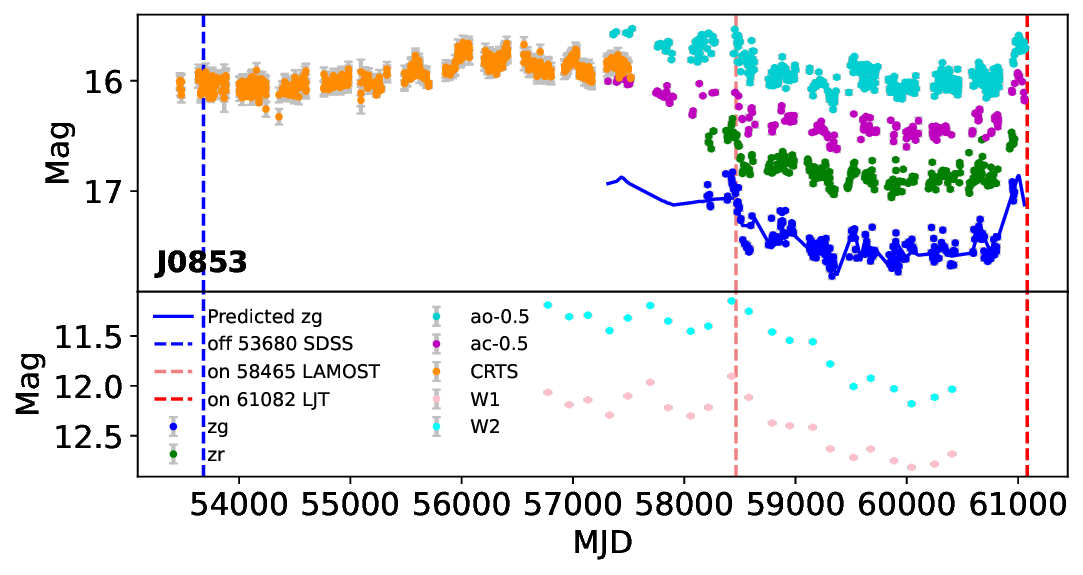}
	\includegraphics[width=0.48\textwidth]{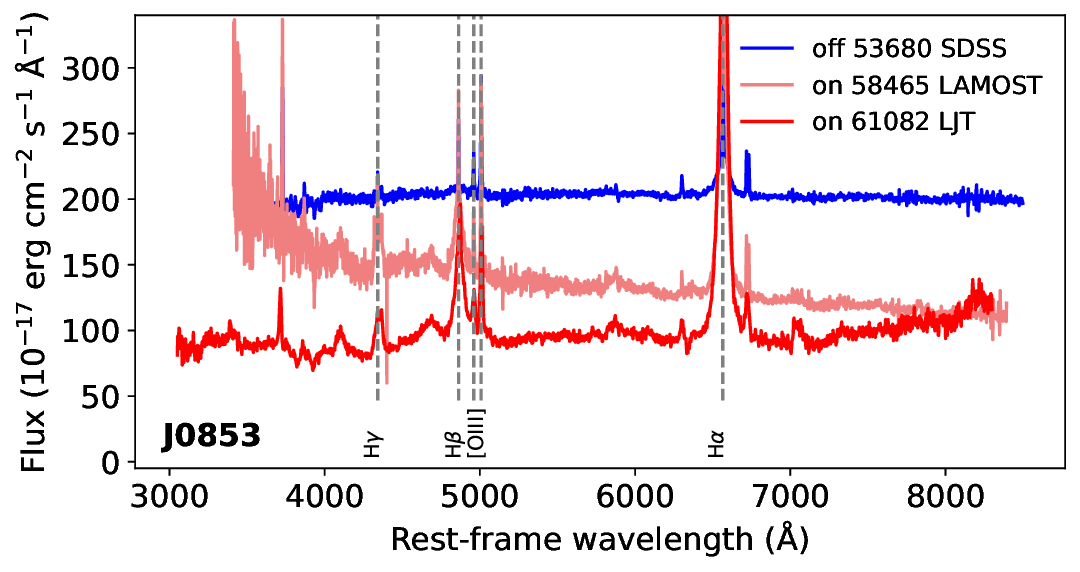}
	\caption{Same as Appendix Fig.~\ref{fig:crcl} for the nine
		unconfirmed candidates.}
	\label{fig:nrcl}
\end{figure*}
\begin{figure*}[htbp]
	\figurenum{B2}
	\centering	
	\includegraphics[width=0.5\textwidth]{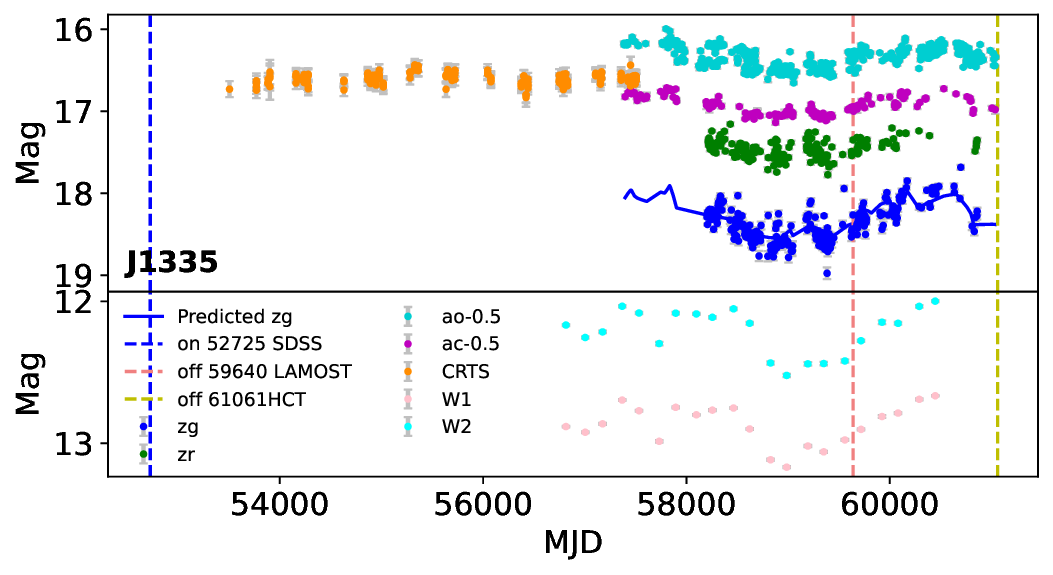}
	\includegraphics[width=0.48\textwidth]{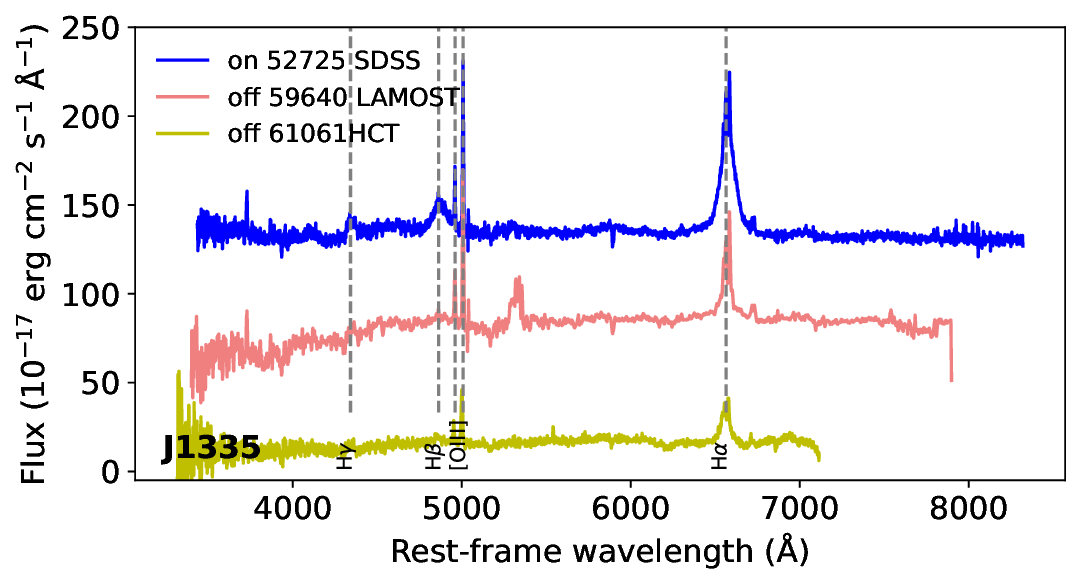}
	\includegraphics[width=0.5\textwidth]{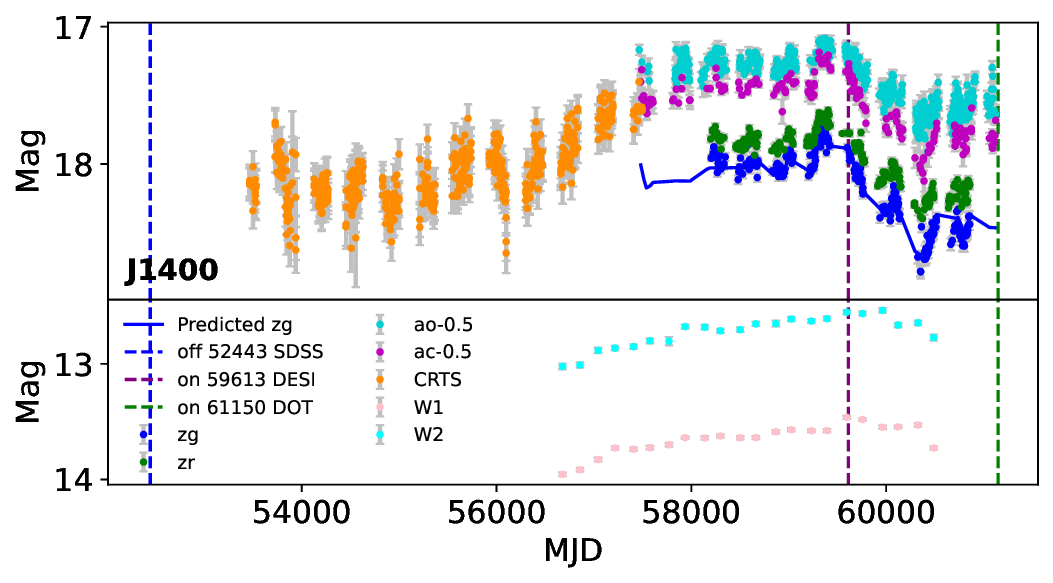}
	\includegraphics[width=0.48\textwidth]{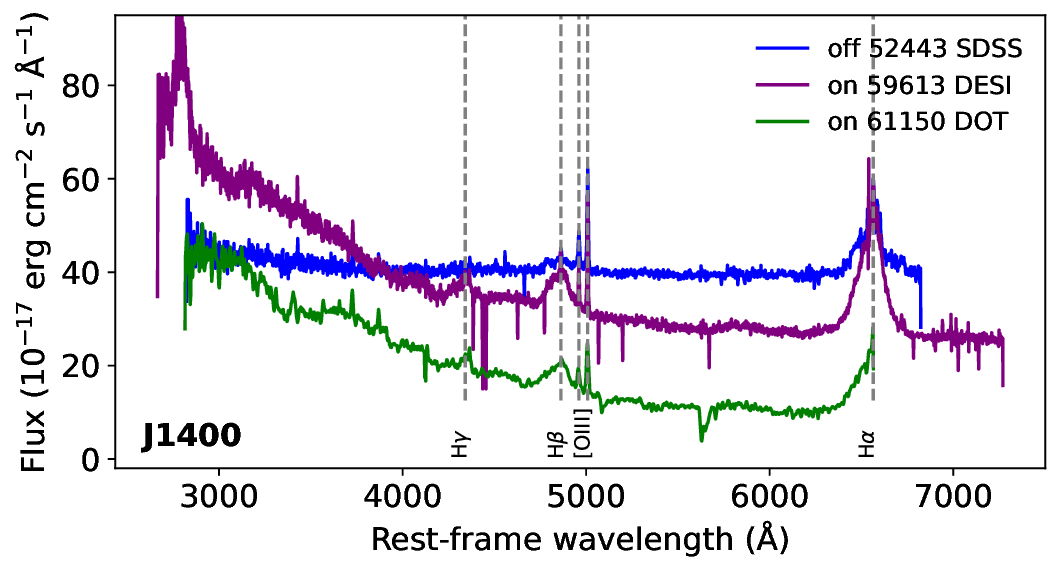}
	\includegraphics[width=0.5\textwidth]{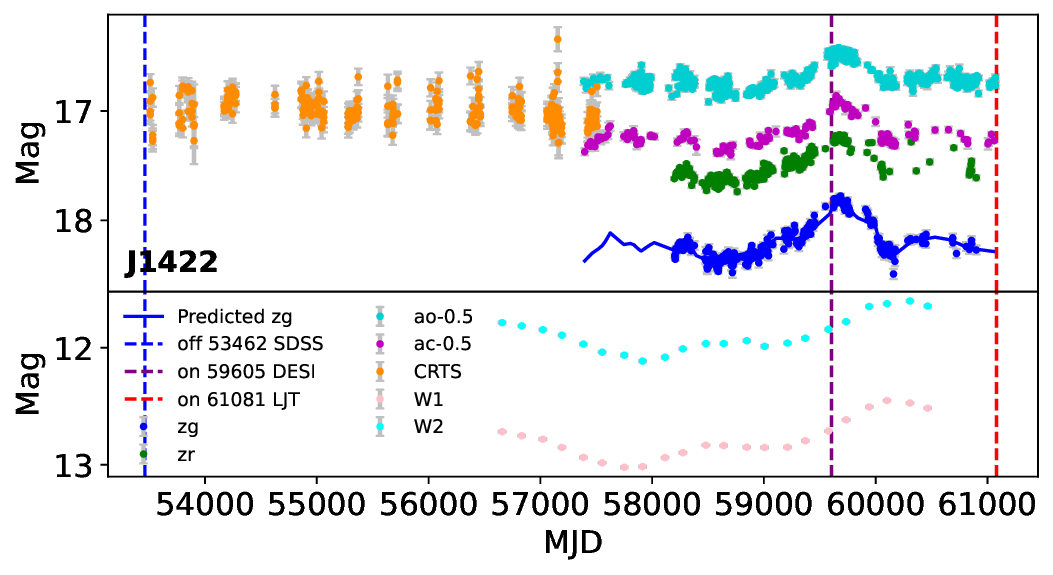}
	\includegraphics[width=0.48\textwidth]{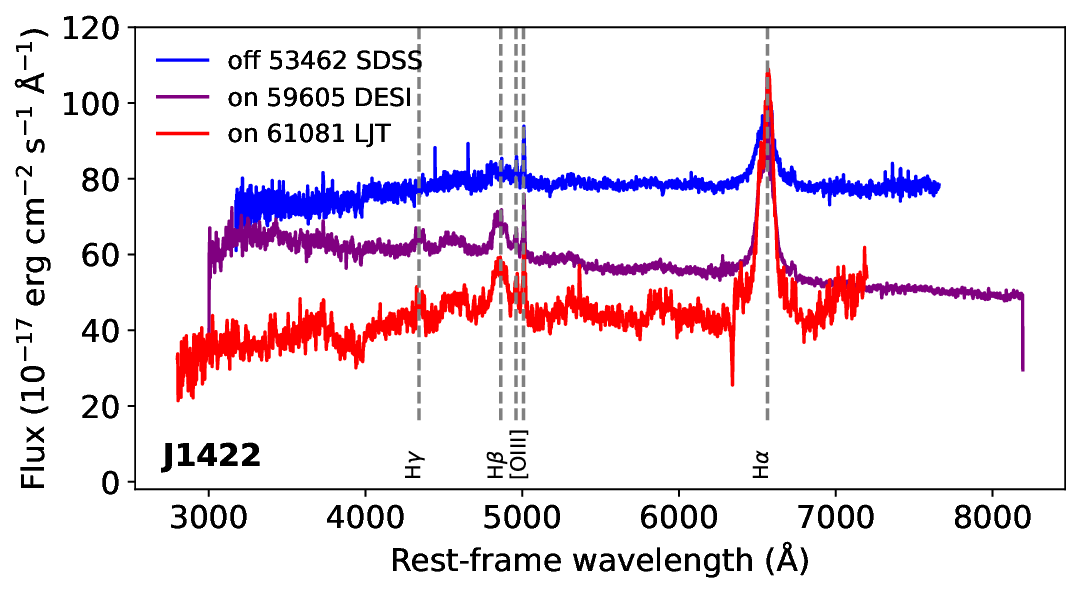}
	\includegraphics[width=0.5\textwidth]{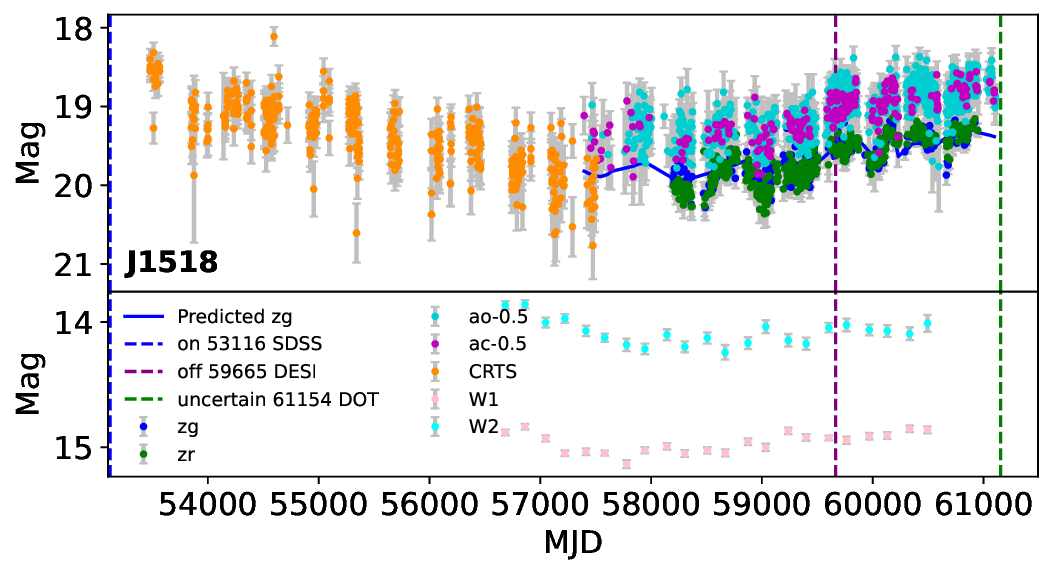}
	\includegraphics[width=0.48\textwidth]{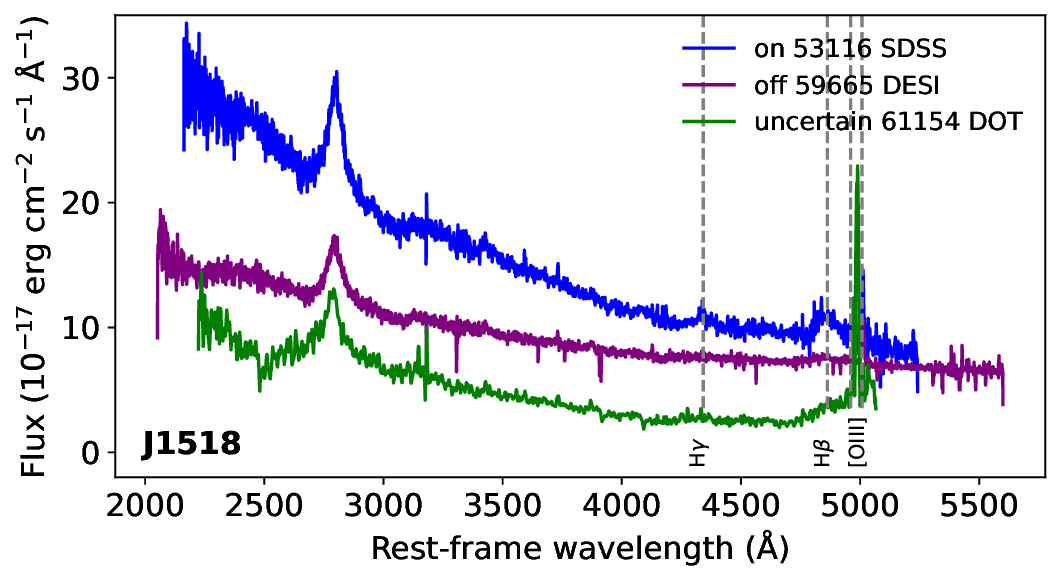}
	\addtocounter{figure}{-1}
	\caption{Continued.}
\end{figure*}

\section{Polynomial fits}
\begin{figure}[htbp]
	\centering
	\includegraphics[width=0.7\textwidth]{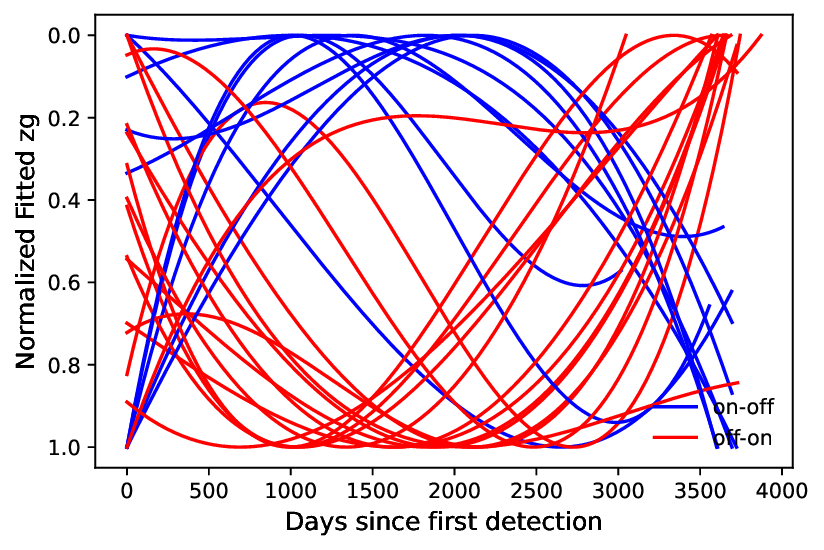}
	\caption{
	Normalized third-order polynomial fits to the $zg$-band light curves 
of the RCL AGNs.  Blue and red curves represent on--off and off--on transition 
cases, respectively. 
	}
	\label{fig:lcfit}
\end{figure}

In order to describe the long-term optical variation trends of the RCL AGNs,
we used a third-order polynomial to fit their $zg$-band light curves. The
obtained fits (normalized) are shown in Figure~\ref{fig:lcfit}.

\section{Distributions of the variability parameters}
\begin{figure*}[htbp]
	\centering
	%\includegraphics[width=0.49\textwidth]{on_log_sig_qso_log_sig_var.eps}
	%\includegraphics[width=0.49\textwidth]{off_log_sig_qso_log_sig_var.eps}
	%\caption{Variability-parameter distributions of the CL parent samples and our repeating CL-AGN candidates. The left panel shows sources selected from known turn-on CL-AGNs, and the right panel shows sources selected from known turn-off CL-AGNs. The background Type~1 and Type~2 AGNs are SDSS DR16 sources with S/N$>10$. Both the CL parent samples and our RCL sources are located closer to the Type~1 AGN distribution than to the normal Type~2 population, suggesting that their off states are not simply classical obscured Type~2 states.}	
\includegraphics[width=0.7\textwidth]{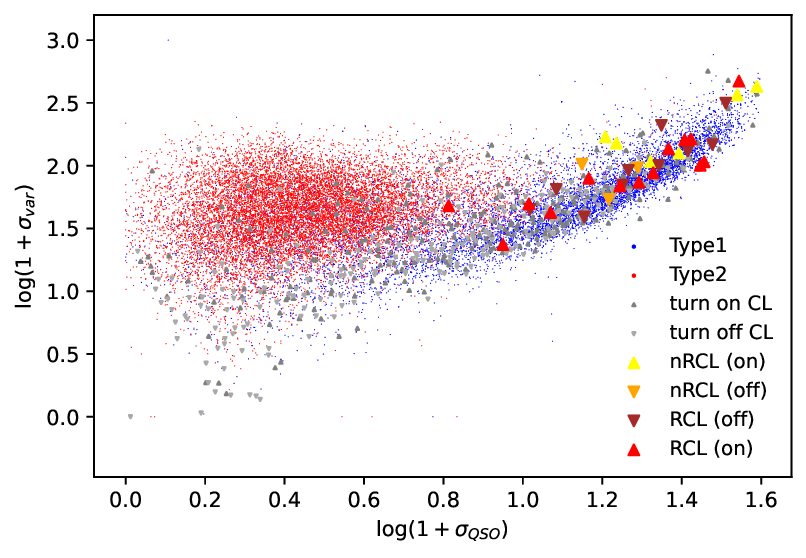}
	\caption{Distributions of the variability parameters for the CL parent 
	samples and our RCL-AGN candidates. The symbol definitions are the 
	same as in Figure~\ref{fig:select}. The background Type~1 and Type~2 
	AGNs are SDSS DR16 sources with S/N$>10$. }
	\label{fig:sig}
\end{figure*}

Using the metrics developed by \citet{bb11}, in which $\sigma_{\rm QSO}$ is the 
best-fit Damped Random Walk (DRW) variability amplitude under the quasar model 
hypothesis and $\sigma_{\rm var}$ represents the overall empirical standard 
deviation of a light curve, we obtained the variability parameters for
the sources in this study (cf., Figure~\ref{fig:select}). The distributions of
the variability parameters are shown in Figure~\ref{fig:sig}.
Both the CL parent samples and our RCL candidates are located closer to 
the Type~1 AGNs than to the Type~2 population, suggesting that the off states
of the CL AGNs are different from the classical obscured Type~2 states.

\bibliography{rcl}{}
\bibliographystyle{aasjournal}
%\end{CJK*}
\end{document}